\documentclass[11pt]{article}
\usepackage{amsfonts}
\usepackage{amssymb}
\usepackage{graphicx}
\usepackage{amsmath}
\usepackage{makeidx}
\usepackage{indentfirst}
\usepackage[T1]{fontenc}
\usepackage[utf8]{inputenc}

\newcounter{resultnum}[section]
\newcounter{conclusionnum}[section]
\newcounter{conditionnum}[section]
\newcounter{conjecturenum}[section]
\newcounter{examplenum}[section]
\newcounter{exercisenum}[section]
\newcounter{lemmanum}[section]
\newcounter{notationnum}[section]
\newcounter{theoremnum}[section]
\newcounter{definitionnum}[section]
\newcounter{corollarynum}[section]
\newcounter{remarknum}[section]
\newcounter{propositionnum}[section]
\newcounter{acknowledgementnum}[section]
\newcounter{algorithmnum}[section]
\newcounter{axiomnum}[section]
\newcounter{casenum}[section]
\newcounter{claimnum}[section]
\newcounter{summarynum}[section]
\newcounter{problemnum}[section]
\begin{document}

\title{Kaluza--Klein gravity \& cosmology emerging from G. Perelman's
entropy functionals and quantum geometric information flows}
\date{December 10, 2020}
\author{ \vspace{.1 in} {\textbf{Iuliana Bubuianu}}\thanks{%
email: iulia.bubu@gmail.com } \\
%EndAName
{\small \textit{Radio Ia\c{s}i, \ 44 Lasc\v{a}r Catargi street, Ia\c{s}i, \
700107, Romania}} \vspace{.1 in} \\
\vspace{.1 in} \textbf{Sergiu I. Vacaru} \thanks{
emails: sergiu.vacaru@gmail.com ;\newline
\textit{Address for post correspondence in 2019-2020 as a visitor senior
researcher at YF CNU Ukraine is:\ } Yu. Gagarin street, nr. 37, ap. 3,
Chernivtsi, Ukraine, 58008} \\
{\small \textit{Physics Department, California State University at Fresno,
Fresno, CA 93740, USA; and }}\\
{\small \textit{Dep. Theoretical Physics and Computer Modelling, 101
Storozhynetska street, Chernivtsi, 58029, Ukraine}} \vspace{.1 in} \\
and \vspace{.1 in} \\
{\ \textbf{El\c{s}en Veli Veliev} \vspace{.1 in} }\thanks{%
email: elsen@kocaeli.edu.tr and elsenveli@hotmail.com } \\
{\small \textit{Department of Physics,\ Kocaeli University, 41380, Izmit,
Turkey}} }
\maketitle

\begin{abstract}
%%%%%%%
We elaborate on quantum geometric information flows, QGIFs, and emergent (modified) Einstein-Maxwell and Kaluza--Klein, KK, theories formulated in Lagrange-Hamilton and general covariant variables. There are considered nonholonomic deformations of Grigory Perelman's F- and W-functionals (originally postulated for Riemannian metrics) for describing relativistic
geometric flows, gravity and matter field interactions, and associated statistical thermodynamic systems. We argue that the concept of Perelman W-entropy presents more general and alternative possibilities to characterize geometric flow evolution, GIF, and gravity models than the Bekenstein--Hawking and another area--holographic type entropies. Formulating the theory of QGIFs, a set of fundamental geometric, probability and quantum concepts, and methods of computation, are reconsidered for curved spacetime and (relativistic) phase spaces. Such generalized metric--affine spaces are modelled as nonholonomic Lorentz manifolds, (co)
tangent Lorentz bundles and associated vector bundles. Using geometric and entropic and thermodynamic values, we define QGIF versions of the von Neumann entropy, relative and conditional entropy, mutual information etc. There are analyzed certain important inequalities and possible applications of G. Perelman and related entanglement and R\'{e}nyi entropies to theories of KK QGIFs and emergent gravitational and electromagnetic interactions. New
classes of exact cosmological solutions for GIFs and respective quasiperiodic evolution scenarios are elaborated. We show how classical and quantum thermodynamic values can be computed for cosmological quasiperiodic solutions and speculate how such constructions can be used for explaining structure formation in dark energy and dark matter physics.

\vskip5pt


\vskip5pt

PACS2010:\ 02.40.-k, 02.90.+p, 03.67.-a, 04.50.-Cd, 04.90.+e, 05.90.+m

\end{abstract}

\tableofcontents

%\newpage

%\vskip10pt

\section{Introduction}

The goal of this paper is to prove that (modified) Einstein-Maxwell, EM,
theories emerge from models of self-similar generalized geometric flows
(with nonholonomic Ricci solitons) and quantum geometric information flows,
QGIFs, see relevant details in partner works \cite%
{bubuianu19,vacaru19b,vacaru19c,vacaru19d}. The gravitational and
electromagnetic fields are distinguished from other fundamental ones (with
strong and weak quantum interactions) because at long distances such
interactions can be described by well defined relativistic classical
theories. Geometrically, the EM theory (and more general Einstein-Yang-Mills
configurations) can be formulated equivalently as a five dimensional, 5-d,
(and higher dimensions) Kaluza--Klein, KK, theory with corresponding
parametrization of metric, frame and connection structures and assumptions
on compactification on extra-dimension coordinates, for instance, see a
review of results in \cite{kk}. In this work, we do not study non--Abelian
gauge interactions resulting in more sophisticate geometric evolution models
with confinement and nonlinear gauge symmetries.

An important property of the classical EM and KK theories is that they can
be treated as emergent entropic gravity models \cite{vacaru19,vacaru19a}
(for instance, of E. Verlinde type \cite{verlinde10,verlinde16}) in the
theory of Ricci flows \cite{hamilt1,perelman1}. Here we cite D. Friedan \cite%
{friedan2} who considered equivalents of the R. Hamilton equations in
physics before mathematicians elaborated a rigorous geometric analysis and
topology formalism, see \cite{monogrrf1,monogrrf2,monogrrf3} for reviews of
results. Such geometric and non-relativistic physical theories are
characterized by the so-called G. Perelman F- and W-functionals (the last
one is called also as W-entropy) and associated statistical thermodynamical
models. In our works, an approach to modified gravity and information flow
theories was developed in relativistic form and applied in the study of
locally anisotropic stochastic and kinetic processes \cite%
{vacaru2000,vacaru2012} and generalizations for quantum and relativistic/
noncommutative/ supersymmetric theories \cite%
{ruchin13,vacaru09,gheorghiu16,rajpoot17}. Here we cite also a series of
papers on gradient flows of non Abelian gauge fields, conformal and
supersymmetric gauge models \cite%
{narayanan06,luescher09,weisz11,carosso18,bergner18}, see also a work on
Ricci-Yang-Mills geometric flow evolution \cite{streets09}.

The systems of nonlinear partial differential equations, PDEs, describing
relativistic flow evolution and dynamical field equations in modified
gravity theories (MGTs, see reviews \cite{odintsov1,stavr1,bubuianu19}) and
general relativity, GR, can be decoupled and integrated in general forms in
terms of generating functions and sources, and integration functions or
integration constants. We cite \cite%
{vacaru18tc,bubuianu18,bubuianu19,vacaru19a,gheorghiu16} and references
therein for a review of geometric methods for constructing exact and
parametric solutions (on the so-called anholonomic frame deformation method,
AFDM) and applications in modern cosmology and astrophysics. The AFDM is
also important for quantizing such theories because it allows to introduce
Lagrange and/or Hamilton type nonholonomic variables and apply geometric
methods in quantum mechanics, QM, and quantum field theory, QFT. Extending
the main concepts and methods of information theory to classical geometric
information flows, GIFs, and developing the approach for QM models, we can
elaborate on quantum geometric information, QGIF, theories \cite%
{bubuianu19,vacaru19b,vacaru19c,vacaru19d}. In this work, using
corresponding generalizations of nonholonomic geometric and statistical
thermodynamics G. Perelman models for GIFs and QM computation methods with
density matrices, von Neumann entropy and entanglement, we investigate how
emergent EM and KK theories can be derived from QGIFs. It should be noted
that we follow the system of notations from \cite{vacaru19c,vacaru19d}
extended in a form when KK configurations for geometric and physical objects
are described by underlined symbols.

The paper is structured as follows: In section \ref{s2}, we provide
necessary geometric preliminaries and formulate the geometric flow equations
for the EM and KK systems. Such modified Ricci flow equations are written in
general and canonical (with decoupling properties) nonholonomic variables
and using mechanical (Lagrange and Hamilton type) variables. Modified EM and
KK classical field equations are derived as nonholonomic Ricci soliton
configurations for relativistic geometric flows. In section \ref{s3}, we
define nonholonomic versions of G. Perelman F- and W-functionals for EM and
KK systems which allows proofs of respective geometric flow and Ricci
soliton equations. There are constructed associated statistical
thermodynamic models which is used for formulating the geometric flow
information, GIF, theory of EM and KK systems. Section \ref{s4} is devoted
to quantum information models. There are defined quantum versions of
W-entropy and the thermodynamic GIF entropy and respective von Neumann, R%
\'{e}nyi and relative/ conditional entropies. The concepts of geometric flow
and quantum entanglement of EM and KK systems and related quantum geometric
information flow, QGIF, models are elaborated in section \ref{s5}. In
section \ref{s6a}, we apply the anholonomic deformation method, AFDM, and
prove the general decoupling and integrability properties of cosmological
geometric evolution equations. The constructions are considered for the case
of gravitational and electromagnetic fields determined by entropic forces
encoding elastic and quasiperiodic spacetime structures. Then, in section %
\ref{s7}, we summarize the AFDM for constructing cosmological solutions with
GIFs for KK gravity and effective NES. There are studies respective
nonlinear symmetries relating generating functions, pattern forming and
spacetime structures. We show how analogous of Perelman's W-entropy and main
thermodynamics values can be computed for such locally anisotropic
cosmological quasiperiodic configurations and speculate how the respective
statistical and geometric thermodynamic generating function can be used for
constructing QGIF models. Finally, we summarize and conclude the work in
section \ref{s8}.

\section{Relativistic evolution equations for geometric flows of KK systems}

\label{s2} We outline necessary concepts on the geometry of nonholonomic
manifolds and nonlinear connections and introduce fundamental equations for
generalized Ricci flows. For details and proofs and applications in modern
physics, we cite \cite{ruchin13,gheorghiu16,rajpoot17}. The (modified)
gravitational and electromagnetic theories are derived as nonholonomic Ricci
soliton configurations.

\subsection{Geometric preliminaries}

On the geometry of nonholonomic manifolds and bundle spaces enabled with
nonlinear connection structure and related modified gravity theories, we
follow the conventions and method outlined in \cite%
{bubuianu18,vacaru19a,bubuianu19,vacaru19b,vacaru19c}. Readers are
considered to be familiar with geometric methods applied in modern gravity
\cite{misner,kk} and extended to constructions with nonholonomic manifolds,
nonlinear connections, and AFDM.

\subsubsection{Nonlinear connections and adapted variables}

Let us consider a four dimensional, 4-d, spacetime Lorentz manifold $\mathbf{%
V}$ defined by a metric $\mathbf{g}$ of local pseudo-Euclidean signature $%
(+++-).$ We use boldface symbols for spaces with nonholonomic fibred
structure when the geometric objects enabled with (or adapted to) a
nonlinear connection, N-connection, $\mathbf{N}:\ \mathbf{V}=hV\oplus vV$.
This states a conventional 2+2 splitting into horizontal (h), $\dim (hV)=2$,
and vertical (v), $\dim (vV)=2,$ parts, and respective sets of N-linear
adapted frames $\mathbf{e}_{\alpha }=(\mathbf{e}_{i},e_{a})$ and dual frames
$\mathbf{e}^{\beta }=(e^{j},\mathbf{e}^{b}).$\footnote{%
Such frames define respective nonholonomic frame structures on the tangent
bundle $T\mathbf{V,}$ where $i,j,..=1,2$ are h-indices and $a,b,...=3,4$ are
(co) v-indices for certain general Greek indices $\alpha ,\beta ,...$. In a
similar form, all geometric constructions can be performed for a
nonholonomic splitting on $\ ^{\shortmid }\mathbf{N}:\mathbf{V}^{\ast
}=hV\oplus \ ^{\shortmid }cvV,$ with a conventional fiber and enabled with a
dual metric $\ ^{\shortmid }\mathbf{g}$. We can introduce Lagrange-Hamilton
variables, see details in \cite{bubuianu19,vacaru19b,vacaru19c} and
references therein, determined by a prescribed Lagrange generating function $%
L$ on $\mathbf{V}$ and related via a Legendre transform a Hamilton
generating function, $H$ on $\mathbf{V}^{\ast }$, inducing respective
canonical geometric mechanical data $(\widetilde{\mathbf{g}},\widetilde{%
\mathbf{N}})$ ) and/ or $(\ ^{\shortmid }\widetilde{\mathbf{g}},\
^{\shortmid }\widetilde{\mathbf{N}})$. Local coordinates are denoted: $%
u^{\alpha }=(x^{i},y^{a})$ (in brief, $u=(x,y)$) on $\mathbf{V;}$ and,
respectively, $\ ^{\shortmid }u^{\alpha }=(x^{i},p_{a})$ (in brief, $\
^{\shortmid }u=(x,p)$) on $\mathbf{V}^{\ast }.$ To generate exact generic
off-diagonal solutions depending, in principle, on all spacetime coordinates
is necessary to consider nonholonomic splitting of type 2+2 of the total
spaces. In another turns, splitting of type 3+1 are more convenient to
elaborating on thermodynamics, kinetic, stochastic and QM models. Such
double fibrations requests different systems of notations for coordinates
and indices of geometric objects. Readers are recommended to see \ partner
works \cite{bubuianu19,vacaru19b,vacaru19c,vacaru19d} and \cite%
{vacaru18tc,bubuianu19,bubuianu18} for details on nonholonomic differential
geometry, geometric flows and applications in classical and quantum
information theory.} We shall consider coefficient formulas for respective
N-connections and adapted frames in section \ref{sshamvar}. Diadic
N-connection splittings will be prescribed in such forms when it is possible
to define certain general decoupling and integration of physically important
systems of nonlinear PDEs or elaborated analogous mechanical and/or
thermodynamic models. Having elaborated in N-adapted variables on explicit
classes of such solutions and/or analogous models all geometric/ physical
objects can be re-written equivalently in arbitrary frames.

A distinguished connection, d-connection, $\mathbf{D}=(h\mathbf{D},v\mathbf{D%
})$ is a linear connection preserving under parallelism a N-connection
splitting. It defines a distortion d-tensor $\mathbf{Z}$ (it is a
distinguished tensor, d-, adapted to a N-connection) from the Levi-Civita,
LC, connection $\nabla ,$ when $\mathbf{D}=\nabla +$ $\mathbf{Z.}$\footnote{%
There are preferred linear and d-connection structures which can be uniquely
defined by geometric data $\mathbf{g}$ and/or $\mathbf{N}$ following certain
geometric principles. For instance, $\nabla $ is a unique metric compatible
linear connection, $\nabla \mathbf{g}=0,$ with zero torsion, $T_{\ \beta
\gamma }^{\alpha }[\nabla ]=0$. The canonical d-connection $\widehat{\mathbf{%
D}}=\nabla +$ $\widehat{\mathbf{Z}}$ is also defined by the same d-metric
structure in a metric compatible form, $\widehat{\mathbf{D}}\mathbf{g}=0,$
with zero torsion on both h- and v-subspaces, but with some (nonholonomic)
nontrivial coefficients relating the h- and v-subspaces. This $\widehat{%
\mathbf{D}}$ is very important because it allows us to prove certain general
decoupling and integrality properties of various classes of modified
Einstein and Ricci flow equations. Such constructions can not be performed
in explicit form if we work only with the LC-connection. Zero torsion
configurations can be always extracted from more general already found
solutions ones bye imposing at the end the condition $\widehat{\mathbf{D}}%
_{|T=0}=\nabla $. This restricts the class of generating/integrating
functions and sources. There is also another very important d-connection
called the Cartan d-connection $\widetilde{\mathbf{D}}$ which is almost
symplectic. It can be defined if we prescribe, for instance, a generating
Lagrange function $L$ on $\mathbf{V}$ when a canonical d-metric ($\mathbf{g}=%
\widetilde{\mathbf{g}}$ up to corresponding frame transforms) can be
constructed as a so-called Sasaki lift for a Hessian metric, $v\widetilde{%
\mathbf{g}},$ and a canonical N-connection $\widetilde{\mathbf{N}}.$ Such a $%
\widetilde{\mathbf{N}}$ is subjected to the condition that corresponding
semi-spray (nonlinear geodesic equations) are equivalent to the
Euler-Lagrange equations for $L.$ The priority of $\widetilde{\mathbf{D}}$
is that it can be used directly for performing deformation quantization or
elaborating on other types of perturbative or nonperturbative quantization.}
The Ricci d-tensor $\mathbf{R}ic=\{\mathbf{R}_{\alpha \beta }\}$ of a
d-connection $\mathbf{D}_{\alpha }$ is defined and computed in standard
form. We write \ $\widehat{\mathbf{R}}ic=\{\widehat{\mathbf{R}}_{\alpha
\beta }\}$ for the canonical d-connection. Similarly, we can label/compute
other type geometric objects, for instance, the scalar curvature $\
_{s}^{\shortmid }\widetilde{R}=\ ^{\shortmid }\mathbf{g}^{\alpha \beta }\
^{\shortmid }\widetilde{\mathbf{R}}_{\alpha \beta }$ of $\ ^{\shortmid }%
\widetilde{\mathbf{D}}_{\alpha }.$ Transferring geometric data by
corresponding nonholonomic deformations and Legendre transforms from the
tangent Lorentz bundle $T\mathbf{V}$ to the contangent bundle $T^{\ast }%
\mathbf{V,}$%
\begin{equation*}
(\mathbf{g,N,D})\Leftrightarrow (\mathbf{g,N,}\widehat{\mathbf{D}}%
)\Leftrightarrow (L\mbox{ on }T\mathbf{V},\widetilde{\mathbf{g}}\mathbf{,}%
\widetilde{\mathbf{N}}\mathbf{,}\widetilde{\mathbf{D}})\longrightarrow (\
^{\shortmid }\mathbf{g,\ ^{\shortmid }N,\ ^{\shortmid }D})\Leftrightarrow (\
^{\shortmid }\mathbf{g,\ ^{\shortmid }N,}\ ^{\shortmid }\widehat{\mathbf{D}}%
)\Leftrightarrow (H\mbox{ on }T^{\ast }\mathbf{V},\ ^{\shortmid }\widetilde{%
\mathbf{g}}\mathbf{,}\ ^{\shortmid }\widetilde{\mathbf{N}}\mathbf{,}\
^{\shortmid }\widetilde{\mathbf{D}}),
\end{equation*}%
we can elaborate on quantum mechanical, QM, models determined by a
(relativistic) Hamiltonian $H$ and associated mechanical variables.

\subsubsection{Nonholonomic Einstein-Maxwell, EM, potentials and
Kaluza-Klein, KK, models}

Suppose $\pounds \rightarrow \mathbf{V}$ is the total spaces of a $U(1)$%
--bundle over $\mathbf{V},$ and $\mathbf{A=A}_{\alpha }\mathbf{e}^{\alpha },$
for a N-adapted dual basis $\mathbf{e}^{\alpha },$ is a linear connection
1-form (electromagnetic potential) on this bundle associated to $\pounds .$
The curvature of $\mathbf{A}$ is a 2-form $\mathbf{F=\{F}_{\alpha \beta }=%
\mathbf{e}_{\alpha }\mathbf{A}_{\beta }-\mathbf{e}_{\beta }\mathbf{A}%
_{\alpha }-[\mathbf{e}_{\alpha },\mathbf{e}_{\beta }]^{\gamma }\mathbf{A}%
_{\gamma }\mathbf{\},}$ for nonholonomy basis coefficients $[\mathbf{e}%
_{\alpha },\mathbf{e}_{\beta }]^{\gamma },$ representing the first Chern
class of the line bundle associated to $\pounds $ and $T\mathbf{V.}$ It is
also used the dual Hodge operator $\ast ,$ determined by $\mathbf{g},$ which
allows us to construct the dual 2-form $\ast \mathbf{F}=\{\mathbf{F}^{\alpha
\beta }\}$. The canonical energy-momentum tensor of the electromagnetic
field is defined
\begin{equation}
\mathbf{T}_{\alpha \beta }=e^{-2}[\mathbf{F}_{\alpha \gamma }\mathbf{F}%
_{\beta }^{\ \gamma }-\frac{1}{4}\mathbf{g}_{\alpha \beta }\mathbf{F}%
_{\gamma \zeta }\mathbf{F}^{\gamma \zeta \ }],  \label{emt}
\end{equation}%
where $e$ is the electromagnetic constant. It is used also the derivative $%
D_{\alpha }=\mathbf{e}_{\alpha }+e\mathbf{A}_{\alpha }.$

We shall study families of geometric and physical evolution flow models
determined by geometric data $[\mathbf{g}(\tau ),\mathbf{D}(\tau ),\mathbf{A}%
(\tau )]$ running on a positive parameter $0\leq \tau \leq \tau _{0}.$ For
simplicity, we shall use brief notations (for instance, $\mathbf{g}(\tau )=%
\mathbf{g}(\tau ,u))$ when the dependence on local coordinates is not
written if this does not result in ambiguities. In this work, it is
considered that $\beta ^{-1}=\tau $ is a temperature like parameter like in
the G. Perelman work \cite{perelman1} on the theory of Ricci flows \cite%
{hamilt1}. Here we cite also D. Friedan \cite{friedan2} who considered
equivalents of the R. Hamilton equations in physics before mathematicians
elaborated their rigorous geometric analysis and topology formalism \cite%
{monogrrf1,monogrrf2,monogrrf3}. For a Kaluza-Klein, KK, theory unifying
geometrically the gravity and electromagnetic interactions (see review \cite%
{kk}), we can introduce a 5-d dimensional metric $\underline{\mathbf{g}}=%
\mathbf{(g,A),}$ i.e. $\mathbf{g}_{\underline{\alpha }\underline{\beta }}=(%
\mathbf{g}_{\alpha \beta },\mathbf{A}_{\gamma }),$ for $\underline{\alpha },%
\underline{\beta },...=1,2...,5.$\footnote{%
The components of the 4-d metric and EM potential are taken with
corresponding dimension constants in order to get well defined dimentions
for metric fields. We also note that in this work splitting of indices of
type 2+2 and 3+1 will be considered for the base spacetime manifold $\mathbf{%
V,}$ i.e. for indices of type $\alpha ,\beta ,...$ with respect to which 5-d
constructions will be adapted.} The Einstein equations for $\underline{%
\mathbf{g}}$ (after a corresponding compactification on the fifth
coordinate) are equivalent to the system of Einstein-Maxwell, EM, equations.
We do not speculate in this work on physical implications of the KK and EM
theories but study possible generalizations and modifications for classical
and quantum GIFs. Here we note that geometric models with nonholonomic data
with $\underline{\mathbf{g}}(\tau )$ and respective adapted d-connections $%
\underline{\mathbf{D}}(\tau )$ are more convenient for encoding geometric
and physical information and elaborating on classical and quantum
information theories. For elaborating on Hamilton type canonical QM models,
we shall use different sets of equivalent (up to nonholonomic frame
transforms and deformations) geometric data $\ (\ ^{\shortmid }\underline{%
\mathbf{g}},\ ^{\shortmid }\mathbf{\underline{\mathbf{N}},\ ^{\shortmid }}%
\underline{\mathbf{D}})\Leftrightarrow (\ ^{\shortmid }\underline{\mathbf{g}}%
,\ ^{\shortmid }\mathbf{\underline{\mathbf{N}},}\ ^{\shortmid }\widehat{%
\underline{\mathbf{D}}})\Leftrightarrow (H\mbox{ on }T^{\ast }\mathbf{V,}%
\pounds ,\ ^{\shortmid }\widetilde{\underline{\mathbf{g}}},\ ^{\shortmid }%
\widetilde{\underline{\mathbf{N}}},\ ^{\shortmid }\widetilde{\underline{%
\mathbf{D}}}).$

\subsection{Geometric flow equations for nonholonomic Einstein-Maxwell
systems}

We consider geometric generalizations of the R. Hamilton equations \cite%
{hamilt1} for relativistic flows \cite{ruchin13,bubuianu19} of modified
Einstein--Maxwell and Kaluza-Klein theories using nonholonomic variables
which allow a general decoupling of geometric flow evolution and dynamical
field equations (using the canonical d-connection with "hats"). For
elaborating analogous mechanical systems and related statistical
thermodynamical and quantum mechanical models, it is convenient to work with
Lagrange--Hamilton variables with "tilde".

\subsubsection{Modified R. Hamilton equations for EM and KK systems}

A family $[\mathbf{g}(\tau ),\mathbf{D}(\tau ),\mathbf{A}(\tau )]$ defines a
nonholonomic Einstein-Maxwell geometric flow on $hTV$ if
\begin{equation}
\frac{\partial \mathbf{g}_{\alpha \beta }}{\partial \tau }=-2(\mathbf{R}%
_{\alpha \beta }-\mathbf{\Upsilon }_{\alpha \beta })\mbox{ and }\frac{%
\partial \mathbf{A}^{\alpha }}{\partial \tau }=-\mathbf{D}_{\beta }\mathbf{F}%
^{\alpha \beta },  \label{emgf}
\end{equation}%
where $\Upsilon _{ij}=\varkappa (\mathbf{T}_{ij}-\frac{1}{2}\mathbf{g}%
_{ij}T) $ is determined by the electromagnetic energy momentum tensor (\ref%
{emt}) for the gravitational constant $\varkappa $ and $T=\mathbf{T}_{ij}%
\mathbf{T}^{ij}.$ \ This system of nonlinear PDEs can be parameterized in a
form describing KK geometric flows when the nonholonomic Ricci flow
equations are written%
\begin{equation}
\frac{\partial \mathbf{g}_{\underline{\alpha }\underline{\beta }}}{\partial
\tau }=-2\mathbf{R}_{\underline{\alpha }\underline{\beta }}.  \label{kkgf}
\end{equation}%
For Riemannian metrics and respective LC-connections, the mathematical
properties and conditions of existence of solutions for systems of type (\ref%
{emgf}) were studied in \cite{streets09}. In our works, such systems were
studied for pseudo-Riemannian metrics and generalized connections (see \cite%
{vacaru18tc,bubuianu19,bubuianu18,ruchin13,vacaru09,gheorghiu16,rajpoot17}
and referecences therein), when large classes of nontrivial solutions can be
constructed for the canonical d-connection $\mathbf{D}$ with possible
nonholonomic constraints for extracting LC-configurations with $\nabla .$

\subsubsection{Canonical form of nonholonomic flow equations for EM and KK\
systems}

In explicit form, we can decouple and integrate the systems (\ref{emgf}) and
(\ref{kkgf}) (applying the AFDM \cite%
{vacaru18tc,bubuianu18,bubuianu19,vacaru19a,gheorghiu16}) if we work with
necessary type N-adapted variables, the canonical d-connection$\ \widehat{%
\mathbf{D}}$ and consider a respective normalization function $\widehat{f}.$
The equations (\ref{emgf}) are written in the form
\begin{eqnarray}
\frac{\partial \mathbf{g}_{ij}}{\partial \tau } &=&-2\left( \widehat{\mathbf{%
R}}_{ij}-\widehat{\mathbf{\Upsilon }}_{ij}\right) ;\ \frac{\partial \mathbf{g%
}_{ab}}{\partial \tau }=-2\left( \widehat{\mathbf{R}}_{ab}-\widehat{\mathbf{%
\Upsilon }}_{ab}\right) ;  \label{canhamiltevol} \\
\widehat{\mathbf{R}}_{ia} &=&\widehat{\mathbf{R}}_{ai}=0;\ \widehat{\mathbf{R%
}}_{ij}=\widehat{\mathbf{R}}_{ji};\ \widehat{\mathbf{R}}_{ab}=\widehat{%
\mathbf{R}}_{ba};  \notag \\
\partial _{\tau }\ \widehat{f} &=&-\widehat{\square }\widehat{f}+\left\vert
\ \widehat{\mathbf{D}}\ \widehat{f}\right\vert ^{2}-\ \ _{s}\widehat{R}+%
\widehat{\mathbf{\Upsilon }}_{a}^{a},  \notag
\end{eqnarray}%
where $\ \widehat{\square }(\tau )=\widehat{\mathbf{D}}^{\alpha }(\tau )\
\widehat{\mathbf{D}}_{\alpha }(\tau )$ is used for the geometric flows of
the d'Alambert operator.

Using KK\ geometric data for $\mathbf{g}_{\underline{\alpha }\underline{%
\beta }}=(\mathbf{g}_{\alpha \beta },\mathbf{A}_{\gamma })$ defined in a $%
U(1)$--bundle $\pounds \rightarrow \mathbf{V,}$ we write (\ref{canhamiltevol}%
) in a form similar to (\ref{kkgf}),
\begin{equation}
\frac{\partial \mathbf{g}_{\underline{\alpha }\underline{\beta }}}{\partial
\tau }=-2\widehat{\mathbf{R}}_{\underline{\alpha }\underline{\beta }}%
\mbox{
and }\ \partial _{\tau }\ \underline{\widehat{f}}=-\ \underline{\widehat{%
\square }}\underline{\widehat{f}}+\left\vert \underline{\widehat{\mathbf{D}}}%
\ \underline{\widehat{f}}\right\vert ^{2}-\ \ _{s}\underline{\widehat{R}},
\label{cankkhamiltevol}
\end{equation}%
where $\ \underline{\widehat{\square }}(\tau )=\widehat{\mathbf{D}}^{%
\underline{\alpha }}(\tau )\ \widehat{\mathbf{D}}_{\underline{\alpha }}(\tau
)$ is used for the geometric flows of the d'Alambert operator determined by $%
\ \underline{\widehat{\mathbf{D}}}=\ \{\widehat{\mathbf{D}}_{\underline{%
\alpha }}\}$ on the total space of $\pounds $ and the normalization function
$\underline{\widehat{f}}$ is correspondingly redefined.

We constructed and studied various examples of exact and parametric
solutions of the system (\ref{canhamiltevol}) in references \cite%
{ruchin13,vacaru09,gheorghiu16,rajpoot17}. Such (in general) generic
off-diagonal solutions describe geometric flows and nonholonomic
deformations of cosmological and/or black hole solutions.

\subsubsection{Ricci flow equations for EM systems in Lagrange mechanical
variables}

Any system of nonholonomic geometric flow evolution equations of type (\ref%
{canhamiltevol}) can be rewritten equivalently in so-called nonholonomic
Lagrange variables, see \cite{bubuianu19,vacaru19b}. We can prescribe a
family regular relativistic Lagrangians $L(\tau ,x^{i},y^{a})$ for
conventional $2+2$ splitting on $\mathbf{V.}$ Together with the family of
corresponding non-degenerated Hessians $\widetilde{g}_{ab}:=\frac{1}{2}$ $%
\frac{\partial ^{2}L}{\partial y^{a}\partial y^{b}}$ this defines a set of
geometric data
\begin{equation*}
(\widetilde{\mathbf{g}}(\tau )=\{\widetilde{\mathbf{g}}_{\mu \nu }(\tau )=[%
\widetilde{g}_{ij}(\tau ),\widetilde{g}_{ab}(\tau )]\},\widetilde{\mathbf{N}}%
(\tau )\mathbf{=\{}\widetilde{N}_{i}^{a}(\tau )\},\widetilde{\mathbf{D}}%
(\tau )).
\end{equation*}%
In such Lagrange mechanical variables, the nonlinear PDEs (\ref%
{canhamiltevol}) can be written equivalently in the form (for a
corresponding redefinition of normalizing functions and sources):
\begin{eqnarray}
\frac{\partial \widetilde{\mathbf{g}}_{ij}}{\partial \tau } &=&-2\left(
\widetilde{\mathbf{R}}_{ij}-\widetilde{\mathbf{\Upsilon }}_{ij}\right) ;\
\frac{\partial \widetilde{\mathbf{g}}_{ab}}{\partial \tau }=-2\left(
\widetilde{\mathbf{R}}_{ab}-\widetilde{\mathbf{\Upsilon }}_{ab}\right) ;
\label{hamlagr} \\
\widetilde{\mathbf{R}}_{ia} &=&\widetilde{\mathbf{R}}_{ai}=0;\ \widetilde{%
\mathbf{R}}_{ij}=\widetilde{\mathbf{R}}_{ji};\ \widetilde{\mathbf{R}}_{ab}=%
\widetilde{\mathbf{R}}_{ba};  \notag \\
\partial _{\tau }\ \widetilde{f} &=&-\widetilde{\square }\widetilde{f}%
+\left\vert \ \widetilde{\mathbf{D}}\widetilde{f}\right\vert ^{2}-\ \ _{s}%
\widetilde{R}+\widetilde{\mathbf{\Upsilon }}_{a}^{a},  \notag
\end{eqnarray}%
where $\ \widetilde{\square }(\tau )=\widetilde{\mathbf{D}}^{\alpha }(\tau
)\ \widetilde{\mathbf{D}}_{\alpha }(\tau )$ is constructed for the canonical
Lagrange d-connection.

The geometric evolution of KK systems (\ref{kkgf}) can be written in "tilde"
variables similarly to (\ref{hamlagr}), i. e. in a form similar to (\ref%
{kkgf}),
\begin{equation}
\frac{\partial \widetilde{\mathbf{g}}_{\underline{\alpha }\underline{\beta }}%
}{\partial \tau }=-2\widetilde{\mathbf{R}}_{\underline{\alpha }\underline{%
\beta }}\mbox{ and }\ \partial _{\tau }\ \underline{\widetilde{f}}=-\
\underline{\widetilde{\square }}\underline{\widetilde{f}}+\left\vert
\underline{\widetilde{\mathbf{D}}}\ \underline{\widetilde{f}}\right\vert
^{2}-\ \ _{s}\underline{\widetilde{R}},  \label{hamlagrkk}
\end{equation}%
where $\ \underline{\widetilde{\square }}(\tau )=\widetilde{\mathbf{D}}^{%
\underline{\alpha }}(\tau )\ \widetilde{\mathbf{D}}_{\underline{\alpha }%
}(\tau )$ is used for the geometric flows of the d'Alambert operator
determined by $\ \underline{\widetilde{\mathbf{D}}}=\ \{\widetilde{\mathbf{D}%
}_{\underline{\alpha }}\}$ on the total space of $\pounds $ and the
normalization function $\underline{\widetilde{f}}$ is redefined to include
distortions of d-connections and respective curvature terms.

It is not possible to find certain general decoupling properties of systems (%
\ref{hamlagrkk}) which make very difficult the task to construct exact
solutions in certain general and explicit forms. Nevertheless, such Lagrange
type variables are important for elaborating models of geometric evolution
of relativistic mechanical systems and performing certain type Lagrange type
quantization.

\subsubsection{Geometric flow equations for EM systems in Hamilton
mechanical variables}

\label{sshamvar}Let us consider a family of Hamilton fundamental functions $%
H(\tau )=H(\tau ,x,p)$ with regular Hessians (cv-metrics)
\begin{equation}
\ ^{\shortmid }\widetilde{g}^{ab}(\tau ,x,p):=\frac{1}{2}\frac{\partial
^{2}H(\tau )}{\partial p_{a}\partial p_{b}},  \label{hesshs}
\end{equation}%
when $\det |\ ^{\shortmid }\widetilde{g}^{ab}|\neq 0,$ and of constant
signature. Such functions and coefficients are defined for a conventional
2+2 splitting of on a nonholonomic $\mathbf{V}^{\ast },$ which (in this
work) is Lorentz manifold $\mathbf{V}$ but enabled with a system of "dual"
local coordinates $p_{a}$ corresponding to a nonholonomic manifold.

Lagrange and Hamilton variables are related by Legendre transforms $%
L\rightarrow H(x,p):=p_{a}y^{a}-L(x,y)$ when $y^{a}$ determining solutions
of the equations $p_{a}=\partial L(x,y)/\partial y^{a}.$ The inverse
Legendre transforms are defined $H\rightarrow L$ for $%
L(x,y):=p_{a}y^{a}-H(x,p),$ where $p_{a}$ are solutions of the equations $%
y^{a}=\partial H(x,p)/\partial p_{a}.$ We can consider families of such
Legendre transforms depending on a $\tau$-parameter.

Any $H$ defines a canonical nonlinear connection (N-connection) structure
\begin{equation}
\ ^{\shortmid }\widetilde{\mathbf{N}}:\ TT^{\ast }V=hT^{\ast }V\oplus
vT^{\ast }V  \label{nconcan}
\end{equation}%
for which the Euler-Lagrange and/or Hamilton equations are equivalent to the
spray equations, see details in \cite{bubuianu19,vacaru19b,vacaru19c}. The
geometric flows of such nonholonomic structures are determined by such
families of canonical N-connection coefficients,
\begin{equation*}
\ ^{\shortmid }\widetilde{\mathbf{N}}(\tau )=\left\{ \ ^{\shortmid }%
\widetilde{N}_{ij}(\tau ):=\frac{1}{2}\left[ \{\ \ ^{\shortmid }\widetilde{g}%
_{ij}(\tau ),H(\tau )\}-\frac{\partial ^{2}H(\tau )}{\partial p_{k}\partial
x^{i}}\ ^{\shortmid }\widetilde{g}_{jk}(\tau )-\frac{\partial ^{2}H(\tau )}{%
\partial p_{k}\partial x^{j}}\ ^{\shortmid }\widetilde{g}_{ik}(\tau )\right]
\right\}
\end{equation*}%
where $\ \ ^{\shortmid }\widetilde{g}_{ij}$ is inverse to $\ \ ^{\shortmid }%
\widetilde{g}^{ab}$ (\ref{hesshs}) for any value of the running parameter.%
\footnote{%
For a fixed value of the flow parameter, such coefficients define canonical
systems of N--adapted (co) frames,
\begin{equation*}
\ ^{\shortmid }\widetilde{\mathbf{e}}_{\alpha }=(\ ^{\shortmid }\widetilde{%
\mathbf{e}}_{i}=\frac{\partial }{\partial x^{i}}-\ ^{\shortmid }\widetilde{N}%
_{ia}(x,p)\frac{\partial }{\partial p_{a}},\ ^{\shortmid }e^{b}=\frac{%
\partial }{\partial p_{b}});\ \ ^{\shortmid }\widetilde{\mathbf{e}}^{\alpha
}=(\ ^{\shortmid }e^{i}=dx^{i},\ ^{\shortmid }\mathbf{e}_{a}=dp_{a}+\
^{\shortmid }\widetilde{N}_{ia}(x,p)dx^{i}).
\end{equation*}%
Such frames are characterized by anholonomy relations $\ [\ ^{\shortmid }%
\widetilde{\mathbf{e}}_{\alpha },\ ^{\shortmid }\widetilde{\mathbf{e}}%
_{\beta }]=\ ^{\shortmid }\widetilde{\mathbf{e}}_{\alpha }\ ^{\shortmid }%
\widetilde{\mathbf{e}}_{\beta }-\ ^{\shortmid }\widetilde{\mathbf{e}}_{\beta
}\ ^{\shortmid }\widetilde{\mathbf{e}}_{\alpha }=\ ^{\shortmid }\widetilde{W}%
_{\alpha \beta }^{\gamma }\ ^{\shortmid }\widetilde{\mathbf{e}}_{\gamma },$
with anholonomy coefficients $\widetilde{W}_{ia}^{b}=\partial _{a}\widetilde{%
N}_{i}^{b},$ $\widetilde{W}_{ji}^{a}=\widetilde{\Omega }_{ij}^{a},$ and $\
^{\shortmid }\widetilde{W}_{ib}^{a}=\partial \ ^{\shortmid }\widetilde{N}%
_{ib}/\partial p_{a}$ and $\ ^{\shortmid }\widetilde{W}_{jia}=\ \mathbf{\
^{\shortmid }}\widetilde{\Omega }_{ija}.$ Such a frame is holonomic
(integrable) if the respective anholonomy coefficients are zero.}

Using above coefficients, we can define families of canonical d-metrics,
\begin{equation}
\ ^{\shortmid }\widetilde{\mathbf{g}}(\tau )=\ ^{\shortmid }\widetilde{%
\mathbf{g}}_{\alpha \beta }(\tau ,x,p)\ ^{\shortmid }\widetilde{\mathbf{e}}%
^{\alpha }(\tau )\mathbf{\otimes \ ^{\shortmid }}\widetilde{\mathbf{e}}%
^{\beta }(\tau )=\ \ ^{\shortmid }\widetilde{g}_{ij}(\tau ,x,p)e^{i}\otimes
e^{j}+\ ^{\shortmid }\widetilde{g}^{ab}(\tau ,x,p)\ ^{\shortmid }\widetilde{%
\mathbf{e}}_{a}(\tau )\otimes \ ^{\shortmid }\widetilde{\mathbf{e}}_{b}(\tau
),  \label{cdmds}
\end{equation}%
where the canonical N-linear frames $\ ^{\shortmid }\widetilde{\mathbf{e}}%
^{\alpha }(\tau )=(e^{i},\ ^{\shortmid }\widetilde{\mathbf{e}}_{a}(\tau ))$
are canonically determined by data $(H(\tau ),\ ^{\shortmid }\widetilde{g}%
^{ab}(\tau )).$ Considering general frame (vierbein) transforms, $e_{\alpha
}(\tau)=e_{\ \alpha }^{\underline{\alpha }}(\tau,u)\partial /\partial u^{%
\underline{\alpha }}$ and $e^{\beta }(\tau)=e_{\ \underline{\beta }}^{\beta
}(\tau,u)du^{\underline{\beta }},$ any N-connection and d-metric structure
on $\mathbf{V}^{\ast }$ can be written in general form (without "tilde" on
symbols), with $\ ^{\shortmid }\mathbf{N}(\tau )=\{\ ^{\shortmid
}N_{ij}(\tau ,x,p)\}$ and
\begin{equation*}
\ ^{\shortmid }\mathbf{g}(\tau )=\ ^{\shortmid }\mathbf{g}_{\alpha \beta
}(\tau ,x,p)\ ^{\shortmid }\mathbf{e}^{\alpha }(\tau )\mathbf{\otimes \
^{\shortmid }e}^{\beta }(\tau )=\ \ ^{\shortmid }g_{ij}(\tau
,x,p)e^{i}\otimes e^{j}+\ ^{\shortmid }g^{ab}(\tau ,x,p)\ ^{\shortmid }%
\mathbf{e}_{a}(\tau )\otimes \ ^{\shortmid }\mathbf{e}_{b}(\tau ).
\end{equation*}

In Hamilton mechanical variables, the analogs of systems (\ref{canhamiltevol}%
) and (\ref{hamlagr}) are written
\begin{eqnarray}
\frac{\partial \ ^{\shortmid }\widetilde{\mathbf{g}}_{ij}}{\partial \tau }
&=&-2\left( \ ^{\shortmid }\widetilde{\mathbf{R}}_{ij}-\ ^{\shortmid }%
\widetilde{\mathbf{\Upsilon }}_{ij}\right) ;\ \frac{\partial \ ^{\shortmid }%
\widetilde{\mathbf{g}}_{ab}}{\partial \tau }=-2\left( \ ^{\shortmid }%
\widetilde{\mathbf{R}}_{ab}-\ ^{\shortmid }\widetilde{\mathbf{\Upsilon }}%
_{ab}\right) ;  \label{hamham} \\
\ ^{\shortmid }\widetilde{\mathbf{R}}_{ia} &=&\ ^{\shortmid }\widetilde{%
\mathbf{R}}_{ai}=0;\ \ ^{\shortmid }\widetilde{\mathbf{R}}_{ij}=\
^{\shortmid }\widetilde{\mathbf{R}}_{ji};\ \ ^{\shortmid }\widetilde{\mathbf{%
R}}_{ab}=\ ^{\shortmid }\widetilde{\mathbf{R}}_{ba};  \notag \\
\partial _{\tau }\ \ ^{\shortmid }\widetilde{f} &=&-\ ^{\shortmid }%
\widetilde{\square }\ ^{\shortmid }\widetilde{f}+\left\vert \ ^{\shortmid }%
\widetilde{\mathbf{D}}\ ^{\shortmid }\widetilde{f}\right\vert ^{2}-\ \
_{s}^{\shortmid }\widetilde{R}+\ ^{\shortmid }\widetilde{\mathbf{\Upsilon }}%
_{a}^{a},  \notag
\end{eqnarray}%
where $\ ^{\shortmid }\widetilde{\square }(\tau )=\ ^{\shortmid }\widetilde{%
\mathbf{D}}^{\alpha }(\tau )\ \ ^{\shortmid }\widetilde{\mathbf{D}}_{\alpha
}(\tau )$ is determined by the canonical Hamilton d-connection. This system
is most convenient for elaborating QM and QGIFs models following standard
approaches with conventional Hamiltonian variables associated to $2+2$
splitting.

The geometric flow evolution equations (\ref{hamham}) can be re-defined in
order to describe flow evolution models of KK\ systems in canonical Hamilton
variables,
\begin{equation}
\frac{\partial \ ^{\shortmid }\widetilde{\mathbf{g}}_{\underline{\alpha }%
\underline{\beta }}}{\partial \tau }=-2\ ^{\shortmid }\widetilde{\mathbf{R}}%
_{\underline{\alpha }\underline{\beta }}\mbox{ and }\ \partial _{\tau }\
^{\shortmid }\underline{\widetilde{f}}=-\ \ ^{\shortmid }\underline{%
\widetilde{\square }}\ ^{\shortmid }\underline{\widetilde{f}}+\left\vert \
^{\shortmid }\underline{\widetilde{\mathbf{D}}}\ \ ^{\shortmid }\underline{%
\widetilde{f}}\right\vert ^{2}-\ \ \ _{s}^{\shortmid }\underline{\widetilde{R%
}}),  \label{hamhamkk}
\end{equation}%
where $\ \ ^{\shortmid }\underline{\widetilde{\square }}(\tau )=\
^{\shortmid }\widetilde{\mathbf{D}}^{\underline{\alpha }}(\tau )\ \
^{\shortmid }\widetilde{\mathbf{D}}_{\underline{\alpha }}(\tau )$ is used
for the geometric flows of the d'Alambert operator determined by $\ \
^{\shortmid }\underline{\widetilde{\mathbf{D}}}=\ \{\ ^{\shortmid }%
\widetilde{\mathbf{D}}_{\underline{\alpha }}\}.$ The normalization function $%
\ ^{\shortmid }\underline{\widetilde{f}}$ is re-defined in such forms when
thare are included contributions from distortions of d-connections and
respective curvature terms.

The systems of nonlinear PDEs (\ref{cankkhamiltevol}), (\ref{hamlagrkk}) and
(\ref{hamhamkk}) for KK geometric flow models encode in different type
variables classsical EM long distance interactions. In this work, for
simplicity, we shall study how (\ref{hamhamkk}) can be derived from
nonholonomic deformations of G. Perelman functionals and respective models
of classical and quantum GIFs.

\subsubsection{KK gravity models as nonholonomic Ricci solitons}

Nonholonomic Ricci solitons are defined as self-similar geometric flow
configurations for a fixed parameter $\tau _{0}.$ In such cases we can
consider $\partial \mathbf{g}_{\alpha \beta }/\partial \tau =0$ and $%
\partial \mathbf{A}^{\alpha }/\partial \tau =0$ and transform (\ref{emgf})
into%
\begin{equation*}
\mathbf{R}_{\alpha \beta }=\mathbf{\Upsilon }_{\alpha \beta }\mbox{ and }%
\mathbf{D}_{\beta }\mathbf{F}^{\alpha \beta }=0.
\end{equation*}%
For self-similar conditions with a fixed parameter $\tau =\tau _{0}$, the
system (\ref{kkgf}) results in $\mathbf{R}_{\underline{\alpha }\underline{%
\beta }}=0.$ We can extract from such 4-d and 5-d modified Ricci soliton
equations standard EM and KK equations with LC connections imposing
additional nonholonomic constraints when $\mathbf{D}_{\mid \mathbf{Z=0}%
}=\nabla .$\footnote{%
In geometrical literature \cite{monogrrf1,monogrrf2,monogrrf3}, self-similar
equations, i.e. Ricci solitons, are described by Einstein type equations for
$\nabla $ determined by Riemannian metrics, with a cosmological constant $%
\lambda $ and a vector feld $v_{\underline{\alpha }}(u),$ $R_{\underline{%
\alpha }\underline{\beta }}[\nabla ]-\lambda g_{\underline{\alpha }%
\underline{\beta }}=\nabla _{\underline{\alpha }}v_{\underline{\beta }%
}+\nabla _{\underline{\beta }}v_{\underline{\alpha }}.$ For corresponding
normalizaton functions, one generates models with vanishing $\lambda $ and $%
v_{\underline{\alpha }}$ when $R_{\underline{\alpha }\underline{\beta }%
}[\nabla ]=0,$ see \cite{gheorghiu16} for applications of Ricci soliton
techniques in modern gravity.}

The normalizing function in (\ref{canhamiltevol}) can be chosen in such a
form when above nonholonomic Ricci soliton equations can be written in
canonical variables%
\begin{equation}
\widehat{\mathbf{R}}_{ij}=\widehat{\mathbf{\Upsilon }}_{ij};\ \widehat{%
\mathbf{R}}_{ab}=\widehat{\mathbf{\Upsilon }}_{ab};\widehat{\mathbf{R}}_{ia}=%
\widehat{\mathbf{R}}_{ai}=0;\ \widehat{\mathbf{R}}_{ij}=\widehat{\mathbf{R}}%
_{ji};\ \widehat{\mathbf{R}}_{ab}=\widehat{\mathbf{R}}_{ba}.
\label{soliteinst}
\end{equation}%
Such a system of nonlinear PDEs posses a general decoupling property and
allows use to generate various classes of exact and parametric solutions. In
this approach, the (modified) gravitational and electromagnetic interactions
consist certain examples of nonolonomic Ricci soliton equations.

Working, for instance, with Hamilton mechanical like variables, we obtain
from (\ref{hamham}) an equivalent nonholonomic Ricci soliton system with
"less decoupling" properties but which is very useful for elaborating models
of quantum gravity and QGIFs. Such a system of nonlinear PDEs is written
\begin{equation*}
\ ^{\shortmid }\widetilde{\mathbf{R}}_{ij}=\ ^{\shortmid }\widetilde{\mathbf{%
\Upsilon }}_{ij};\ ^{\shortmid }\widetilde{\mathbf{R}}_{ab}=\ ^{\shortmid }%
\widetilde{\mathbf{\Upsilon }}_{ab};\ ^{\shortmid }\widetilde{\mathbf{R}}%
_{ia}=\ ^{\shortmid }\widetilde{\mathbf{R}}_{ai}=0;\ \ ^{\shortmid }%
\widetilde{\mathbf{R}}_{ij}=\ ^{\shortmid }\widetilde{\mathbf{R}}_{ji};\ \
^{\shortmid }\widetilde{\mathbf{R}}_{ab}=\ ^{\shortmid }\widetilde{\mathbf{R}%
}_{ba}.
\end{equation*}%
This is an example of analogous gravity and electromagnetic interactions
modeled by relativistic Hamilton mechanical systems.

In this work, the Ricci solitons are generated as self-similar
configurations of geometric flow evolution systems (\ref{hamhamkk})
described by systems of nonlinear PDEs of type
\begin{equation*}
\ ^{\shortmid }\widetilde{\mathbf{R}}_{\underline{\alpha }\underline{\beta }%
}-\lambda \ ^{\shortmid }\widetilde{\mathbf{g}}_{\underline{\alpha }%
\underline{\beta }}=\ ^{\shortmid }\underline{\widetilde{\mathbf{D}}}_{%
\underline{\alpha }}\ \ ^{\shortmid }\mathbf{v}_{\underline{\beta }}+\
^{\shortmid }\underline{\widetilde{\mathbf{D}}}_{\underline{\beta }}\ \
^{\shortmid }\mathbf{v}_{\underline{\alpha }}.
\end{equation*}%
Such configurations are determined by some geometric data $(\ ^{\shortmid }%
\widetilde{\mathbf{g}}_{\underline{\alpha }\underline{\beta }},\ ^{\shortmid
}\underline{\widetilde{\mathbf{D}}}_{\underline{\beta }})$ and a
cosmological constant $\lambda $ and d-vector feld $\ ^{\shortmid }\mathbf{v}%
_{\underline{\alpha }}(\ ^{\shortmid }u).$

\section{G. Perelman functionals and thermodynamics of KK--geometric flows}

\label{s3}The Thurston conjecture (and, in particular, the Poincar\'{e}
hypothesis)\ was proven by G. Perelman \cite{perelman1}. He elaborated a
geometric approach when the R. Hamilton equations \cite{hamilt1} for
Riemannian metrics are derived for certain models of gradient flows. Such
constructions are determined by corresponding Lyapunov type F- and
W-functionals for dynamical systems, see details in \cite%
{monogrrf1,monogrrf2,monogrrf3}. For geometric evolution of
pseudo-Riemannian metrics and generalized connections, it is not clear if
and how analogs of the Thurston conjecture can be formulated and proved.
Nevertheless, various generalizations of the G. Perelman's functionals seem
to be important for characterizing physical properties of new classes of
generic off-diagonal (which can be not diagonalized by coordinate
transforms) solutions in modern gravity. Such models and, for instance,
locally anisotropic black hole and cosmological solutions \cite%
{bubuianu18,vacaru18tc,bubuianu19} are considered in GR and MGTs with
various noncommutative and/or supersymmetric, entropic modifications \cite%
{vacaru09,vacaru2012,ruchin13,gheorghiu16,rajpoot17}. Recently, G.\
Perelman's geometric thermodynamic models were developed and applied in
geometric information theory \cite{vacaru19b,vacaru19c,vacaru19d}. For
reviews on classical and quantum information theory, we cite \cite%
{preskill,witten18}) and references therein. The W-entropy was used by G.
Perelman for formulating a statistical thermodynamic models associated to
Ricci flow theories. This approach involve more general thermodynamic
constructions than the Bekenstein-Hawking entropy for black holes \cite%
{bekenstein72,bekenstein73,bardeen73,hawking75} and various
area--hypersurface generalizations for holographic gravity, entropic
gravity, conformal field theories and duality, see \cite%
{ryu06,raamsdonk10,faulkner14,swingle12,jacobson15,pastawski15,casini11}.

The goal of this section is to define nonholonomic modifications of the so
called F- and W-functionals which allow us to prove the nonholonomic
geometric flow equations (\ref{emgf}) and (\ref{kkgf}) (in respective
variables, (\ref{canhamiltevol}), (\ref{hamlagr}) and/or (\ref{hamham})).
The W-entropy will be used for constructing an associated statistical
thermodynamic model for EM and KK flows. We shall analyze how corresponding
statistical thermodynamic generating functions can be used for formulating a
geometric information flow, GIF, theory for classical EM and KK systems.

\subsection{F- and W-functionals for EM and KK geometric flows}

Generalized G. Perelman entropic like functionals can be postulated using
different types of nonholonomic variables with conventional 2+2 and 3+1
decompostion of dimensions or double fibration splitting, see details in
\cite{ruchin13,gheorghiu16,rajpoot17}.

\subsubsection{Nonholonomic 3+1 splitting adapted to 2+2 decompositions}

Let us consider that a region $U\subset $ $\mathbf{V}$ of a nonholonomic
Lorentz manifold with N-connection 2+2 splitting defined by data $(\mathbf{%
N,g}).$ We suppose that any necessary $U$ is fibered additionally into a
structure of 3-d hypersurfaces $\Xi _{t}$ parameterized by time like
coordinate $y^{4}=t.$ Locally all geometric constructions with such a 3+1
splitting can be adapted to coordinates of type $u^{\alpha }=(x^{i},y^{a})$
when the metric structure can be represented in the form
\begin{eqnarray}
\mathbf{g} &=&\mathbf{g}_{\alpha ^{\prime }\beta ^{\prime }}(\tau ,\ u)d\
\mathbf{e}^{\alpha ^{\prime }}(\tau )\otimes d\mathbf{e}^{\beta ^{\prime
}}(\tau )  \label{decomp31} \\
&=&q_{i}(\tau ,x^{k})dx^{i}\otimes dx^{i}+\mathbf{q}_{3}(\tau ,x^{k},y^{a})%
\mathbf{e}^{3}(\tau )\otimes \mathbf{e}^{3}(\tau )-[\ _{q}N(\tau
,x^{k},y^{a})]^{2}\mathbf{e}^{4}(\tau )\otimes \mathbf{e}^{4}(\tau ).  \notag
\end{eqnarray}%
In this formula, the "shift" coefficients $\mathbf{q}_{\grave{\imath}%
}=(q_{i},\mathbf{q}_{3})$ are related to the 3-d metric $\mathbf{q}%
_{ij}=diag(\mathbf{q}_{\grave{\imath}})=(q_{i},\mathbf{q}_{3})$ on a
hypersurface $\Xi _{t}$ if $\mathbf{q}_{3}=\mathbf{g}_{3}$ and $[\
_{q}N]^{2}=-\mathbf{g}_{4},$ where $\ _{q}N$ is the lapse function. We shall
use left/over labels on $N$ in order to distinguish such traditional symbols
used in GR \cite{misner} from N-connection constructions with coefficients
of type $N_{i}^{a}.$ It should be noted here that in this work the geometric
flow parameter $\tau ,0\leq \tau \leq \tau _{0},$ is a temperature like one
as in G. Perelman works for Ricci flows \cite{perelman1} and in our partner
works \cite{vacaru19b,vacaru19c,vacaru19d}.

Considering $p_{a}$ as certain duals of $y^{a}$ we can define respective
"dual" decompositions when corresponding 3-d hupersurfaces $\
^{\shortmid}\Xi _{E}$ are parameterized by a conventional "energy" parameter
$p_{4}=E$ for local coordinates on $\ ^{\shortmid }\mathbf{V}$. In result,
decompositions of type (\ref{decomp31}) can be rewritten equivalently in the
form%
\begin{eqnarray}
\ ^{\shortmid }\mathbf{g} &=&\ ^{\shortmid }\mathbf{g}_{\alpha ^{\prime
}\beta ^{\prime }}(\tau ,\ u)d\ \ ^{\shortmid }\mathbf{e}^{\alpha ^{\prime
}}(\tau )\otimes d\ ^{\shortmid }\mathbf{e}^{\beta ^{\prime }}(\tau )
\label{decomp31d} \\
&=&\ ^{\shortmid }q_{i}(\tau ,x^{k})dx^{i}\otimes dx^{i}+\ ^{\shortmid }%
\mathbf{q}^{3}(\tau ,x^{k},p_{a})\ ^{\shortmid }\mathbf{e}_{3}(\tau )\otimes
\ ^{\shortmid }\mathbf{e}_{3}(\tau )-[\ \ _{q}^{\shortmid }N(\tau
,x^{k},p_{a})]^{2}\ ^{\shortmid }\mathbf{e}_{4}(\tau )\otimes \ ^{\shortmid }%
\mathbf{e}_{4}(\tau ).  \notag
\end{eqnarray}%
Similar nonholonomic 3+1 decompositions can be considered for canonical
variables with "hats" and/or "tilde" if we elaborate on models with systems
of nonlinear PDEs with certain general decoupling and integration properties
and/or mechanical like analogous relativistic mechanical interpretation.
Such parameterizations of d-metrics can be summarized in this form:%
\begin{equation}
\begin{array}{ccc}
\mathbf{V=}(\mathbf{N}=\{N_{i}^{a}\},\mathbf{g}=\{\mathbf{g}_{\alpha \beta
}=[g_{i},\mathbf{g}_{a}]\}) & \Longleftrightarrow & \ ^{\shortmid }\mathbf{V=%
}(\ ^{\shortmid }\mathbf{N}=\{\ ^{\shortmid }N_{ia}\},\ ^{\shortmid }\mathbf{%
g}=\{\ ^{\shortmid }\mathbf{g}_{\alpha \beta }=[g_{i},\ ^{\shortmid }\mathbf{%
g}^{a}]\}) \\
\updownarrow &  & \updownarrow \\
\ \Xi _{t}:\mathbf{q}_{\grave{\imath}}=(q_{i},\mathbf{q}_{3}),_{q}N &  & \
^{\shortmid }\Xi _{E}:\ ^{\shortmid }\mathbf{q}_{\grave{\imath}}=(q_{i},\
^{\shortmid }\mathbf{q}^{3}),\ _{q}^{\shortmid }N \\
\ \widehat{\Xi }_{t}:\widehat{\mathbf{q}}_{\grave{\imath}}=(q_{i},\widehat{%
\mathbf{q}}_{3}),_{q}\widehat{N} & \updownarrow & \ ^{\shortmid }\widehat{%
\Xi }_{E}:\ ^{\shortmid }\widehat{\mathbf{q}}_{\grave{\imath}}=(q_{i},\
^{\shortmid }\widehat{\mathbf{q}}^{3}),\ _{q}^{\shortmid }\widehat{N} \\
\ \widetilde{\Xi }_{t}:\widetilde{\mathbf{q}}_{\grave{\imath}}=(q_{i},%
\widetilde{\mathbf{q}}_{3}),_{q}\widetilde{N} &  & \ ^{\shortmid }\widetilde{%
\Xi }_{E}:\ ^{\shortmid }\widetilde{\mathbf{q}}_{\grave{\imath}}=(q_{i},\
^{\shortmid }\widetilde{\mathbf{q}}^{3}),\ _{q}^{\shortmid }\widetilde{N}%
\end{array}
\label{31p}
\end{equation}

We can extend parameterizations (\ref{31p}) for extra dimension (underlined)
indices $\underline{\alpha },\underline{\beta },...=1,2...,5$ and
coordinates \underline{$u$}$=(u,u^{5})$ when, for instance,
\begin{eqnarray}
\underline{\mathbf{g}} &=&\{\mathbf{g}_{\underline{\alpha }\underline{\beta }%
}=(\mathbf{g}_{\alpha \beta },\mathbf{A}_{\gamma })\}=\{([g_{i},\mathbf{g}%
_{a}],[\mathbf{A}_{i},\mathbf{A}_{a}])\};\ ^{\shortmid }\underline{\mathbf{g}%
}=\{\ ^{\shortmid }\mathbf{g}_{\underline{\alpha }\underline{\beta }}=(\
^{\shortmid }\mathbf{g}_{\alpha \beta },\ ^{\shortmid }\mathbf{A}_{\gamma
})\}=\{([g_{i},\ ^{\shortmid }\mathbf{g}_{a}],[\ ^{\shortmid }\mathbf{A}%
_{i},\ ^{\shortmid }\mathbf{A}_{a}])\};  \notag \\
\underline{\mathbf{g}} &=&\{\mathbf{g}_{\underline{\alpha }\underline{\beta }%
}=(\mathbf{q}_{\grave{\imath}},_{q}N,\mathbf{A}_{_{\grave{\imath}}},\mathbf{A%
}_{_{4}})\}=\{[\mathbf{q}_{\grave{\imath}}=(q_{i},\mathbf{q}_{3}),_{q}N],[%
\mathbf{A}_{\grave{\imath}}=(A_{i},\mathbf{A}_{3}),\mathbf{A}_{4}]\};....;
\label{p531} \\
\ ^{\shortmid }\underline{\widetilde{\mathbf{g}}} &=&\{\ ^{\shortmid }%
\widetilde{\mathbf{g}}_{\alpha \beta }=[\widetilde{g}_{i},\ ^{\shortmid }%
\widetilde{\mathbf{g}}^{a}],[\ ^{\shortmid }\widetilde{\mathbf{A}}_{i},\
^{\shortmid }\widetilde{\mathbf{A}}^{a}]\}=\{[\ ^{\shortmid }\widetilde{%
\mathbf{q}}_{\grave{\imath}}=(\ ^{\shortmid }\widetilde{q}_{i},\ ^{\shortmid
}\widetilde{\mathbf{q}}_{3}),\ _{q}^{\shortmid }\widetilde{N}],[\
^{\shortmid }\widetilde{\mathbf{A}}_{\grave{\imath}}=(\ ^{\shortmid }%
\widetilde{A}_{i},\ ^{\shortmid }\widetilde{\mathbf{A}}^{4}),\ ^{\shortmid }%
\widetilde{\mathbf{A}}^{4}]\}.  \notag
\end{eqnarray}
in order to study geometric evolution models of KK systems.

\subsubsection{Perelman functionals for nonholonomic KK geometric flows}

In nonholonomic variables with a metric compatible d-connection $\underline{%
\mathbf{D}}=\underline{\nabla }+\underline{\mathbf{Z}}=(\ _{h}\underline{%
\mathbf{D}},\ _{v}\underline{\mathbf{D}})$ uniquely distorted from $%
\underline{\nabla }$, the relativistic versions of G. Perelman functionals
are postulated
\begin{eqnarray}
\underline{\mathcal{F}} &=&\int \left( 4\pi \tau \right) ^{-5/2}e^{-\
\underline{f}}\sqrt{|\underline{\mathbf{g}}|}d^{5}\underline{u}(\ _{s}%
\underline{R}+|\underline{\mathbf{D}}\underline{f}|^{2})\mbox{
and }  \label{ffkk} \\
\underline{\mathcal{W}} &=&\int \underline{\mu }\sqrt{|\underline{\mathbf{g}}%
|}d^{5}\underline{u}[\tau (\ _{s}\underline{R}+|\ \ _{h}\underline{\mathbf{D}%
}\ \underline{f}|+|\ \ _{v}\underline{\mathbf{D}}\ \underline{f}|)^{2}+%
\underline{f}-5].  \label{wfkk}
\end{eqnarray}%
In these formulas, we use a brief notation for the integrals on extra
dimension KK variables and the normalizing function $\ \underline{f}(\tau ,%
\underline{u})$ is subjected to the conditions $\ \int \underline{\mu }\sqrt{%
|\underline{\mathbf{g}}|}d^{5}\underline{u}=\int_{t_{1}}^{t_{2}}\int_{\Xi
_{t}}\int_{u_{1}^{5}}^{u_{2}^{5}}\underline{\mu }\sqrt{|\underline{\mathbf{g}%
}|}d^{5}\underline{u}=1$, for a classical integration measure$\ \underline{%
\mu }=\left( 4\pi \tau \right) ^{-5/2}e^{-\underline{f}}$ and the Ricci
scalar $\ _{s}\underline{R}$ is taken for the Ricci d-tensor $\underline{%
\mathbf{R}}_{\alpha \beta }$ of a d-connection $\underline{\mathbf{D}}.$

N-adapted variational calculuses on $\mathbf{g}_{\underline{\alpha }%
\underline{\beta }}$ using (\ref{ffkk}) or (\ref{wfkk}) (we omit these
cumbersome proofs, see similar details in \cite%
{ruchin13,gheorghiu16,rajpoot17} performed as nonholonomic deformations of
respective sections in \cite{monogrrf1,monogrrf2,monogrrf3}) allows us to
prove the geometric flow evolution for the KK-systems (\ref{kkgf}) and
(equivalently, for EM-systems) (\ref{emgf}). Here we note that the
functional $\underline{\mathcal{W}}$ (\ref{wfkk}) is a nonholonomic
relativistic generalizations of so-called W-entropy introduced in \cite%
{perelman1}. In our partner works \cite{vacaru19b,vacaru19c,vacaru19d},
various 4-d and 8-d versions of $\underline{\mathcal{W}}$ and associated
statistical and quantum thermodynamics values are used for elaborating
models of classical and quantum GIFs.

\subsubsection{Hamilton variables for Perelman KK-functionals}

Redefining respectively the normalization functions for respective duble
splitting (\ref{31p}) and KK d-metrics (\ref{p531}), we can write
equivalently the nonholonomic Perelman functionals in different type
nonholonomic variables. In this paper, we shall use analogous 4-d and 5-d
canonical Hamilton variables with tilde.\footnote{%
Here we note that the theory of nonholonomic geometric flows of relativistic
Hamilton mechanical systems was formulated \cite{vacaru19a} (see also
references therein on yearly works on Finsler-Lagrange geometric flows
beginning 2006) in explicit form using canonical data $(\ ^{\shortmid }%
\widetilde{\mathbf{g}}(\tau ),\ ^{\shortmid }\widetilde{\mathbf{D}}(\tau )),$
in terms of geometric objects with "tilde" values defined on 8-d cotangent
Lorentz bundles.}

The Lyapunov type functionals (\ref{ffkk}) and (\ref{wfkk}) defining
geometric flow evolution of analogous mechanical Hamilton systems for KK
systems on $\ ^{\shortmid }\mathbf{V}$ can be expressed equivalently in the
form
\begin{eqnarray}
\ ^{\shortmid }\underline{\widetilde{\mathcal{F}}} &=&\int \left( 4\pi \tau
\right) ^{-5/2}e^{-\ ^{\shortmid }\widetilde{\ \underline{f}}}\sqrt{|\
^{\shortmid }\underline{\widetilde{\mathbf{g}}}|}d^{5}\ ^{\shortmid }%
\underline{u}(\ \ _{s}^{\shortmid }\underline{\widetilde{R}}+|\ ^{\shortmid }%
\underline{\widetilde{\mathbf{D}}}\ ^{\shortmid }\underline{\widetilde{f}}%
|^{2})\mbox{
and }  \label{kkf} \\
\ ^{\shortmid }\underline{\widetilde{\mathcal{W}}} &=&\int \ ^{\shortmid }%
\underline{\widetilde{\mu }}\sqrt{|\ ^{\shortmid }\underline{\widetilde{%
\mathbf{g}}}|}d^{5}\ ^{\shortmid }\underline{u}[\tau (\ _{s}^{\shortmid }%
\underline{\widetilde{R}}+|\ \ _{h}^{\shortmid }\underline{\widetilde{%
\mathbf{D}}}\ ^{\shortmid }\underline{\widetilde{f}}|+|\ \ _{v}^{\shortmid }%
\underline{\widetilde{\mathbf{D}}}\ ^{\shortmid }\ \underline{\widetilde{f}}%
|)^{2}+\ ^{\shortmid }\underline{\widetilde{f}}-5].  \label{kkw}
\end{eqnarray}%
In these formulas, we use a brief notation for the integrals on extra
dimension KK variables and the normalizing function $\ \ ^{\shortmid }%
\underline{\widetilde{f}}(\tau ,\ ^{\shortmid }\underline{u})$ is subjected
to the conditions $\ \int \int \ ^{\shortmid }\underline{\widetilde{\mu }}%
\sqrt{|\ ^{\shortmid }\underline{\widetilde{\mathbf{g}}}|}d^{5}\ ^{\shortmid
}\underline{u}=\int_{t_{1}}^{t_{2}}\int_{\Xi _{t}}\int_{\ ^{\shortmid
}u_{1}^{5}}^{\ ^{\shortmid }u_{2}^{5}}\int \ ^{\shortmid }\underline{%
\widetilde{\mu }}\sqrt{|\ ^{\shortmid }\underline{\widetilde{\mathbf{g}}}|}%
d^{5}\ ^{\shortmid }\underline{u}=1$, for a classical integration measure$\
\ ^{\shortmid }\underline{\widetilde{\mu }}=\left( 4\pi \tau \right)
^{-5/2}e^{-\ ^{\shortmid }\widetilde{\ \underline{f}}}$ and the Ricci scalar
$\ _{s}^{\shortmid }\underline{\widetilde{R}}$ is taken for the Ricci
d-tensor $\ ^{\shortmid }\underline{\widetilde{\mathbf{R}}}_{\alpha \beta }$
of a d-connection $\ ^{\shortmid }\underline{\widetilde{\mathbf{D}}}$.

The functionals (\ref{kkf}) and (\ref{kkw}) \ are determined by families of
Hamilton fundamental functions $H(\tau ,x,p)$ and respective Hessians (\ref%
{hesshs}) and canonical d-metrics (\ref{cdmds}). Performing respective
canonical N-adapted variational calculus for $\ ^{\shortmid }\underline{%
\widetilde{\mathcal{F}}}$ and/or $\ ^{\shortmid }\underline{\widetilde{%
\mathcal{W}}},$ we prove the geometric flow evolution equations for
analogous Hamilton mechanical systems modeling KK-flows (see (\ref{hamham})
and (\ref{hamhamkk})). Such proofs follow from abstract symbolic calculus
(when geometric objects with "tilde" are changed into respective ones with
"hats" or other type ones on $\underline{\mathbf{V}}$ and/or $\ ^{\shortmid }%
\underline{\mathbf{V}})$. In result, we can similarly define values of type $%
\ \underline{\widetilde{\mathcal{F}}}$ and/or $\underline{\widetilde{%
\mathcal{W}}}$ (allowing proofs via respective N-adapted calculus of (\ref%
{hamlagrkk}) and (\ref{hamlagr})); and $\ \underline{\widehat{\mathcal{F}}}$
and/or $\ \underline{\widehat{\mathcal{W}}},$ with respective proofs of (\ref%
{cankkhamiltevol}) and (\ref{canhamiltevol}).

Considering LC-configurations with $\ ^{\shortmid }\widetilde{\underline{%
\mathbf{D}}}_{\mid \ ^{\shortmid }\widetilde{\underline{\mathbf{T}}}=0}=\
^{\shortmid }$\underline{$\nabla $} and/or$\ ^{\shortmid }\widehat{%
\underline{\mathbf{D}}}_{\mid \ ^{\shortmid }\widehat{\underline{\mathbf{T}}}%
=0}=\ ^{\shortmid }\underline{\nabla },$ the values (\ref{ffkk}) or (\ref%
{wfkk}) transform respectively into analogous KK and/or EM versions of
nonholonomically generalized Perelman's F-entropy and W-entropy. It should
be noted that $\ ^{\shortmid }\widetilde{\underline{\mathcal{W}}}$ (\ref%
{wfkk}) and/or $\ ^{\shortmid }\widehat{\underline{\mathcal{W}}}$ \ (\ref%
{kkw}) do not have a character of entropy for pseudo--Riemannian metrics but
can be treated as respective values characterizing relativistic nonholonomic
geometric hydrodynamic type spacetimes and extra dimension flows.

\subsection{Statistical thermodynamic models for KK flows}

We can characterize MGTs and GR by analogous analogous thermodynamic models
\cite{bubuianu19,bubuianu18,ruchin13,vacaru09,gheorghiu16,rajpoot17}
generalizizng G. Perelman's constructions for geometric flows of Riemannian
metrics \cite{perelman1}.

\subsubsection{Basic concepts of statistical KK thermodynamics}

\label{ssbasicth}We shall underline geometric and thermodynamical values for
geometric flows of KK systems in order to distinguish the construtions from
similar ones elaborated for GIF and QGIF theories in partner works \cite%
{bubuianu19,vacaru19b,vacaru19c}. We consider the partition function $%
\underline{Z}=\int \exp (-\beta \underline{E})d\underline{\omega }(%
\underline{E})$ for a canonical ensemble at temperature $\beta ^{-1}=T.$ The
measure as the density of states includes $\omega (E)$ for certain models in
GR and extensions to 5-d spaceteims with $\underline{\omega }(\underline{E}%
). $ We compute in standard form%
\begin{eqnarray*}
\mbox{average  flow energy:  }\underline{\mathcal{E}} = \ \left\langle
\underline{E}\right\rangle &:=& -\partial \log \underline{Z}/\partial \beta ,
\\
\mbox{ flow entropy: }\underline{\mathcal{S}} &:=&\beta \left\langle
\underline{E}\right\rangle +\log \underline{Z}, \\
\mbox{ flow fluctuation: }\underline{\eta } &:=&\left\langle \left(
\underline{E}-\left\langle \underline{E}\right\rangle \right)
^{2}\right\rangle =\partial ^{2}\log \underline{Z}/\partial \beta ^{2}.
\end{eqnarray*}

A value $\underline{Z}$ allows us to define a conventional \textit{state
density} (for quantum models, a \textit{density matrix})%
\begin{equation}
\underline{\sigma }(\beta ,\underline{E})=\underline{Z}^{-1}e^{-\beta
\underline{E}}.  \label{statedens}
\end{equation}%
The \textit{relative entropy} between any state density $\underline{\rho }$
and $\underline{\sigma }$ is defined/computed%
\begin{equation*}
\underline{\mathcal{S}}(\underline{\rho }\shortparallel \underline{\sigma }%
):=-\underline{\mathcal{S}}(\underline{\rho })+\int (\beta \underline{%
\mathcal{E}}\mathcal{+}\log \underline{Z})\underline{\rho }d\underline{%
\omega }(\underline{E})=\beta \lbrack \underline{\mathcal{E}}(\underline{%
\rho })-T\underline{\mathcal{S}}(\underline{\rho })]+\log \underline{Z},
\end{equation*}%
where the \textit{average energy} is computed for the density matrix $%
\underline{\rho },$ $\underline{\mathcal{E}}(\underline{\rho })=\int
\underline{\mathcal{E}}\underline{\rho }d\underline{\omega }(\underline{E}),$
and the formula $\log \underline{\sigma }=-\beta \underline{\mathcal{E}}%
\mathcal{-}\log \underline{Z}$ is used.

The \textit{free energy} is introduced by formula $\ \underline{\mathcal{F}}(%
\underline{\rho }):=\underline{\mathcal{E}}(\underline{\rho })-T\underline{%
\mathcal{S}}(\underline{\rho }).$ If $\log \underline{Z}$ is independent on $%
\underline{\rho },$ we get $\underline{\mathcal{S}}(\underline{\sigma }%
\shortparallel \underline{\sigma })=0$ and%
\begin{equation}
\underline{\mathcal{S}}(\underline{\rho }\shortparallel \underline{\sigma }%
)=\beta \lbrack \underline{\mathcal{F}}(\underline{\rho })-\underline{%
\mathcal{F}}(\underline{\sigma })].  \label{relentr}
\end{equation}%
In our models of geometric flow evolution and analogous thermodynamics
systems, we consider that under evolution it is preserved the thermal
equilibrium at temperature $\beta $ with maps of density states $\underline{%
\rho }\rightarrow \underline{\rho }^{\prime }$ keeping the same density
state $\underline{\sigma }.$ Such systems are characterized by inequalities
\begin{equation}
\underline{\mathcal{S}}(\underline{\rho }\shortparallel \underline{\sigma }%
)\geq \underline{\mathcal{S}}(\underline{\rho }^{\prime }\shortparallel
\underline{\sigma }),\mbox{ i.e. }\underline{\mathcal{F}}(\underline{\rho }%
)\geq \underline{\mathcal{F}}(\underline{\rho }^{\prime }).
\label{secondthlaw}
\end{equation}%
Above presented formulas allow us to connect KK, EM and mechanical flow
models to the second low of thermodynamics.

\subsubsection{Thermodynamic values for KK and Hamilton mechanical flows}

\label{ssthmodels}We associate to functionals (\ref{ffkk}), (\ref{wfkk}) and
(\ref{kkf}), (\ref{kkw}) respective thermodynamic generating functions%
\begin{eqnarray}
\underline{\mathcal{Z}}[\underline{\mathbf{g}}(\tau )] &=&\int \left( 4\pi
\tau \right) ^{-5/2}e^{-\ \underline{f}}\sqrt{|\underline{\mathbf{g}}|}d^{5}%
\underline{u}(-\underline{f}+5/2),\mbox{ for }\underline{\mathbf{V}}\mathbf{;%
}  \label{genfkk} \\
\ ^{\shortmid }\widetilde{\underline{\mathcal{Z}}}[\ ^{\shortmid }\widetilde{%
\underline{\mathbf{g}}}(\tau )] &=&\int \left( 4\pi \tau \right)
^{-5/2}e^{-\ ^{\shortmid }\widetilde{\ \underline{f}}}\sqrt{|\ ^{\shortmid }%
\underline{\widetilde{\mathbf{g}}}|}d^{5}\ ^{\shortmid }\underline{u}(-\
^{\shortmid }\underline{\widetilde{f}}+5/2),\mbox{ for }\ ^{\shortmid }%
\underline{\mathbf{V}}\mathbf{,}  \label{genfkkh}
\end{eqnarray}%
Such values are with functional dependence on $\underline{\mathbf{g}}(\tau )$
and $\ ^{\shortmid }\widetilde{\underline{\mathbf{g}}}(\tau )$ (we shall
omit to write this in explicit forms if that will not result in
ambiguities). A density state is a functional $\underline{\sigma }[%
\underline{\mathbf{g}}(\tau )]$ when the geometric evolution involve
densities $\underline{\rho }[\ _{1}\underline{\mathbf{g}}]$ and $\underline{%
\rho }^{\prime }[\ _{1}\underline{\mathbf{g}}],$ where the left label 1 is
used in order to distinguish two KK d-metrics $\underline{\mathbf{g}}$ and $%
\ _{1}\underline{\mathbf{g}}.$ Similar values can be defined in canonical
Hamilton variables with tilde and duality labels, for instance, $\underline{%
\widetilde{\sigma }}[\ ^{\shortmid }\widetilde{\underline{\mathbf{g}}}(\tau
)].$

Using (\ref{genfkk}) and respective 3+1 parameterizations of d-metrics (see
formulas (\ref{31p}) and (\ref{p531})), we define and compute analogous
thermodynamic values for geometric evolution flows of KK systems,%
\begin{eqnarray}
\underline{\mathcal{E}}\ &=&-\tau ^{2}\int (4\pi \tau )^{-5/2}e^{-\
\underline{f}}\sqrt{|q_{1}q_{2}\mathbf{q}_{3}(_{q}N)q_{5}|}\delta ^{5}%
\underline{u}(\ _{s}\underline{R}+|\underline{\mathbf{D}}\underline{f}|^{2}%
\mathbf{\ }-\frac{5}{2\tau }),  \label{thvalkk} \\
\underline{\mathcal{S}}\ &=&-\int (4\pi \tau )^{-5/2}e^{-\ \underline{f}}%
\sqrt{|q_{1}q_{2}\mathbf{q}_{3}(_{q}N)q_{5}|}\delta ^{5}\underline{u}\left[
\tau \left( \ _{s}\underline{R}+|\underline{\mathbf{D}}\underline{f}%
|^{2}\right) +\underline{f}-5\right] ,  \notag \\
\underline{\eta }\ &=&2\tau ^{4}\int (4\pi \tau )^{-5/2}e^{-\ \underline{f}}%
\sqrt{|q_{1}q_{2}\mathbf{q}_{3}(_{q}N)q_{5}|}\delta ^{5}\underline{u}[|\
\underline{\mathbf{R}}_{\alpha \beta }+\underline{\mathbf{D}}_{\alpha }\
\underline{\mathbf{D}}_{\beta }\underline{f}-\frac{1}{2\tau }\underline{%
\mathbf{g}}_{\alpha \beta }|^{2}],  \notag
\end{eqnarray}%
where $\delta ^{5}\underline{u}$ contains N-elongated differentials in order
to compute such integrals in N-adapted forms and $q_{5}=1$ can be considered
for compactifications on the 5th coordinate. Using such values, we can
compute the respective free energy and relative entropy (\ref{relentr}),%
\begin{equation*}
\underline{\mathcal{F}}\ (\ _{1}\underline{\mathbf{g}})=\underline{\mathcal{E%
}}(\ _{1}\underline{\mathbf{g}})-\beta ^{-1}\underline{\mathcal{S}}(\ _{1}%
\underline{\mathbf{g}})\mbox{ and }\underline{\mathcal{S}}(\ _{1}\underline{%
\mathbf{g}}\shortparallel \underline{\mathbf{g}})=\beta \lbrack \underline{%
\mathcal{F}}(\ _{1}\underline{\mathbf{g}})-\underline{\mathcal{F}}(%
\underline{\mathbf{g}})],\mbox{ where }
\end{equation*}%
\begin{eqnarray*}
\underline{\mathcal{E}}(\ _{1}\underline{\mathbf{g}}) &=&-\tau ^{2}\int
(4\pi \tau )^{-5/2}e^{-\ \underline{f}}\sqrt{|q_{1}q_{2}\mathbf{q}%
_{3}(_{q}N)q_{5}|}\delta ^{5}\underline{u}[\ _{s}\underline{R}(\ _{1}%
\underline{\mathbf{g}})+|\underline{\mathbf{D}}(\ _{1}\underline{\mathbf{g}})%
\underline{f}(\tau ,\underline{u})|^{2}\mathbf{\ }-\frac{5}{2\tau }], \\
\underline{\mathcal{S}}(\ _{1}\underline{\mathbf{g}}) &=&-\int (4\pi \tau
)^{-5/2}e^{-\ \underline{f}}\sqrt{|q_{1}q_{2}\mathbf{q}_{3}(_{q}N)q_{5}|}%
\delta ^{5}\underline{u}\left[ \tau \left( \ _{s}\underline{R}(\ _{1}%
\underline{\mathbf{g}})+|\underline{\mathbf{D}}(\ _{1}\underline{\mathbf{g}})%
\underline{f}(\tau ,\underline{u})|^{2}\right) +\underline{f}(\tau ,%
\underline{u})-5\right] .
\end{eqnarray*}

For geometric evolution flows described in nonholonomic Hamilton mechanical
variables on $\ ^{\shortmid }\mathbf{V}$ and generating function (\ref%
{genfkkh}), the thermodynamic values are computed
\begin{eqnarray}
\ ^{\shortmid }\widetilde{\underline{\mathcal{E}}}\ &=&-\tau ^{2}\ \int
(4\pi \tau )^{-5/2}e^{-\ ^{\shortmid }\widetilde{\ \underline{f}}}\sqrt{|\
^{\shortmid }\widetilde{q}_{1}\ ^{\shortmid }\widetilde{q}_{2}\ ^{\shortmid }%
\widetilde{\mathbf{q}}_{3}\ _{q}^{\shortmid }\widetilde{N}\ ^{\shortmid }%
\widetilde{q}_{5}\ |}d^{5}\ ^{\shortmid }\underline{u}(\ \ _{s}^{\shortmid }%
\widetilde{\underline{R}}+|\ ^{\shortmid }\widetilde{\underline{\mathbf{D}}}%
\ ^{\shortmid }\widetilde{\underline{f}}|^{2}\mathbf{\ }-\frac{5}{\tau }),
\label{thvalkkh} \\
\ ^{\shortmid }\widetilde{\underline{\mathcal{S}}}\ &=&-\int (4\pi \tau
)^{-5/2}e^{-\ ^{\shortmid }\widetilde{\ \underline{f}}}\sqrt{|\ ^{\shortmid }%
\widetilde{q}_{1}\ ^{\shortmid }\widetilde{q}_{2}\ ^{\shortmid }\widetilde{%
\mathbf{q}}_{3}\ _{q}^{\shortmid }\widetilde{N}\ ^{\shortmid }\widetilde{q}%
_{5}\ |}d^{5}\ ^{\shortmid }\underline{u}\left[ \tau \left( \
_{s}^{\shortmid }\widetilde{\underline{R}}+|\ ^{\shortmid }\widetilde{%
\underline{\mathbf{D}}}\ ^{\shortmid }\widetilde{\underline{f}}|^{2}\right)
+\ ^{\shortmid }\widetilde{\underline{f}}-5\right] ,  \notag \\
\ ^{\shortmid }\widetilde{\underline{\eta }}\ &=&-2\tau ^{4}\ ^{\shortmid
}\int (4\pi \tau )^{-5/2}e^{-\ ^{\shortmid }\widetilde{f}}\sqrt{%
|q_{1}q_{2}q_{3}\breve{N}\ ^{\shortmid }q_{5}\ ^{\shortmid }q_{6}\
^{\shortmid }q_{7}\ ^{\shortmid }\check{N}|}\delta ^{5}\ ^{\shortmid }u[|\ \
^{\shortmid }\widetilde{\underline{\mathbf{R}}}_{\alpha \beta }+\
^{\shortmid }\widetilde{\underline{\mathbf{D}}}_{\alpha }\ \ ^{\shortmid }%
\widetilde{\underline{\mathbf{D}}}_{\beta }\ \ ^{\shortmid }\widetilde{%
\underline{f}}-\frac{1}{2\tau }\ ^{\shortmid }\widetilde{\underline{\mathbf{g%
}}}_{\alpha \beta }|^{2}].  \notag
\end{eqnarray}%
Such formulas are important for elaborating quantum information models with
geometric flows. Geometric evolution of EM systesm are encoded in such
formulas using with compactication on 5th coordinate. Similar analogous
relativistic thermodynamic models were studied in our partner papers \cite%
{bubuianu19,vacaru19b,vacaru19c} for (co) tangent Lorentz bundles of total
space dimension 8-d. In this article, we work with nonholonomic Lorentz 5-d
manifolds with compactification to 4-d.

Finally we note that generating functions (\ref{genfkk}) and (\ref{genfkkh})
and respective thermodynamical values (\ref{thvalkk}) and (\ref{thvalkkh})
can be written equivalently in terms of the canonical d--connections $%
\widehat{\underline{\mathbf{D}}}$ and $\widetilde{\underline{\mathbf{D}}}$
if we consider nonholonomic deformations to certain systems of nonlinear
PDEs with general decoupling and related models in canonical Lagrange
variables on $\underline{\mathbf{V}}.$

\section{Classical and quantum geometric information flows of KK systems}

\label{s4} We consider basic aspects of (quantum) geometric information
flows (respectively, GIFs and QGIFs) of KK systems.

\subsection{Geometric information flow theory of classical KK systems}

In this subsection, we follow classical information theory with fundamental
concepts of Shannon, conditional and relative entropies and applications in
modern physics \cite%
{preskill,witten18,nielsen,cover,wilde,weedbrook11,hayashi17,watrous18}. To
elaborate on GIFs of KK systems there are used W-entropy functionals (\ref%
{wfkk}) and (\ref{kkw}) and associated thermodynamical models elaborated in
section \ref{ssthmodels} and \cite{vacaru19b,vacaru19c}.

\subsubsection{Shannon entropy and GIF entropy}

Let us consider a random variable $B$ taking certain values $b_{\underline{1}%
},b_{\underline{2}},...,b_{\underline{k}}$ (for instance, a long message of
symbols $\underline{N}\gg 1$ containing different $k$ letters) when the
respective probabilities to observe such values are $p_{\underline{1}},p_{%
\underline{2}},...,p_{\underline{k}}.$ By definition, the Shannon entropy $%
S_{B}:=-\sum\limits_{\underline{j}=1}^{\underline{k}}p_{\underline{j}}\log
p_{\underline{j}}\geq 0$ for $\sum\limits_{\underline{j}=1}^{\underline{k}%
}p_{\underline{j}}=1.$ The value $\underline{N}S_{B}$ is the number of bits
of information which can be extracted from a message consisting from $%
\underline{N}$ symbols. Real messages contain correlations between letters
(grammar and syntax) for a more complex random process. Ignoring
correlations (the ideal gaze limit), we approximate the entropy of a long
message to be $\underline{N}S$ with $S$ being the entropy of a message
consisting of only one letter. In a statistical thermodynamical model, we
can consider a classical Hamiltonian $H$ determining the probability of a $i$%
-th symbol $b_{i}$ via formula $p_{\underline{i}}=2^{-H(b_{\underline{i}})}.$

For GIFs of classical KK systems, the thermodynamic values are determined by
data $\left[ \underline{\mathcal{W}};\underline{\mathcal{Z}},\underline{%
\mathcal{E}},\underline{\mathcal{S}},\underline{\eta }\right] $ (\ref%
{thvalkk}) and/or $\left[ \ ^{\shortmid }\underline{\widetilde{\mathcal{W}}};%
\underline{\ ^{\shortmid }\widetilde{\mathcal{Z}}},\ ^{\shortmid }\underline{%
\widetilde{\mathcal{E}}},\ ^{\shortmid }\underline{\widetilde{\mathcal{S}}}%
,\ ^{\shortmid }\underline{\widetilde{\eta }}\right] $ (\ref{thvalkkh})
(tilde are used for analogous mechanical variables determined by a
conventional Hamilton density $H$ and respective Hessian $\ ^{\shortmid }%
\widetilde{g}^{ab}$ (\ref{hesshs})). We can introduce probabilities on a
discrete network with random variables, for instance, $\ ^{\shortmid }%
\widetilde{p}_{\underline{n}}=2^{-\ H(b_{\underline{n}})},$ or, \ for
statistical ansambles, $\ ^{\shortmid }\widetilde{p}_{\underline{n}}=2^{-\
^{\shortmid }\widetilde{\mathcal{E}}(b_{\underline{n}})}.$

We can elaborate on continuous GIF models encoding geometric evolution of KK
systems in general covariant and conventional analogous mechanical veriables
using the thermodynamic entropies $\underline{\mathcal{S}}[\underline{%
\mathbf{g}}(\tau )]$ and $\ ^{\shortmid }\widetilde{\underline{\mathcal{S}}}%
[\ ^{\shortmid }\widetilde{\underline{\mathbf{g}}}(\tau )].$ In such an
approach, we can not involve in the constructions probability distributions
which appear for discrete random variables. We can study GIF KK systems
using only W-entropies $\widetilde{\mathcal{W}}[\widetilde{\mathbf{g}}(\tau
)]$ and $\ ^{\shortmid }\widetilde{\mathcal{W}}[\ ^{\shortmid }\widetilde{%
\mathbf{g}}(\tau )])$ for certain constructions wihout statistical
thermodynamics values. KK systems under geometric evolution flows are
denoted in general form as $\underline{B}=\underline{B}[\underline{\mathbf{g}%
}(\tau )]$ and $\ ^{\shortmid }\widetilde{\underline{B}}=\ ^{\shortmid }%
\widetilde{\underline{B}}[\ ^{\shortmid }\widetilde{\underline{\mathbf{g}}}%
(\tau )]$\ determined by corresponding canonical d-metrics on nonholonomic
Lorentz spacetimes.

\subsubsection{Conditional entropy for GIFs}

Let us consider sending a message with many letters (any letter is a random
variable $X$ taking possible values $x_{\underline{1}},...,x_{\underline{k}%
}).$ A receiver see a random variable $Y$ consisting from letters $y_{%
\underline{1}},...,y_{\underline{r}}$. In the classical information theory,
the goal is to compute how many bits of information does such a receiver get
form a message with $N$ letters when the random variables are denoted $X,Y,Z$
etc. In the simplest case, we can consider one variable when the probability
to observe $X=x_{\underline{i}}$ is denoted $P_{X}(x_{\underline{i}})$ for $%
\sum_{\underline{i}}$ $P_{X}(x_{\underline{i}})=1.$ The communication
between a sender and receiver is a random process of two variables defined
by a joint distribution $P_{X,Y}(x_{\underline{i}},y_{\underline{j}})$ as
the probability that to send $X=x_{\underline{i}}$ and hear $Y=y_{\underline{%
j}}.$ The value $\ P_{Y}(y_{\underline{j}})=\sum_{\underline{i}}P_{X,Y}(x_{%
\underline{i}},y_{\underline{j}})$ is the probability to receive $Y=y_{%
\underline{j}}$ (summation is over all choices that could be send). By
definition, the \textit{conditional probability} $P_{X|Y}(x_{\underline{i}%
}|y_{\underline{j}}):=\frac{P_{X,Y}(x_{\underline{i}},y_{\underline{j}})}{%
P_{Y}(y_{\underline{j}})}$ is a value characterizing receiving $Y=y_{%
\underline{j}},$ one can estimate the probability that it was sent $x_{i}.$
We can write for receiver's messages $P_{X}(x_{\underline{i}})=\sum_{%
\underline{j}}$ $P_{X,Y}(x_{\underline{i}},y_{\underline{j}})$ or consider $%
P_{X}(x_{\underline{i}})$ as an independent probability density.

In classical information theory, there are defined such important values:
\begin{eqnarray*}
\mbox{Shanon entropy of the conditional probability:  } &&S_{X|Y=y_{j}}:=-%
\sum_{\underline{i}}P_{X|Y}(x_{\underline{i}}|y_{\underline{j}})\log
P_{X|Y}(x_{\underline{i}}|y_{\underline{j}}); \\
\mbox{ entropy of joint distribution: } &&S_{XY}:=-\sum_{\underline{i},%
\underline{j}}P_{X,Y}(x_{\underline{i}},y_{\underline{j}})\log P_{X,Y}(x_{%
\underline{i}},y_{\underline{j}}); \\
\mbox{ total received information content: } &&S_{Y}:=-\sum_{\underline{i},%
\underline{j}}P_{X,Y}(x_{\underline{i}},y_{\underline{j}})\log P_{Y}(y_{%
\underline{j}}); \\
\mbox{ total sent information content: } &&S_{X}:=-\sum_{\underline{i},%
\underline{j}}P_{X,Y}(x_{\underline{i}},y_{\underline{j}})\log P_{X}(x_{%
\underline{i}}).
\end{eqnarray*}%
Using such formulas, one prove that for the \textit{conditional entropy}
\begin{equation}
S_{X|Y}:=\sum_{\underline{j}}P_{Y}(y_{\underline{j}%
})S_{X|Y=y_{j}}=S(X|Y)=S_{XY}-S_{Y}\geq 0  \label{condentr}
\end{equation}%
and the \textit{mutual information} between $X$ and $Y$ (a measure of how
much we learn about $X$ observing $Y)$%
\begin{equation}
I(X;Y):=S_{X}-S_{XY}+S_{Y}\geq 0.  \label{mutualinf}
\end{equation}

Now, let us consider how basic concepts from classical information theory
can be generalized for models of information thermodynamics determined by
geometric flows of KK systems. Conventionally, there are considered two such
KK GIFs, $\underline{A}=\underline{A}[\underline{\mathbf{g}}(\tau )],$ or $\
^{\shortmid }\widetilde{\underline{A}}=\ ^{\shortmid }\widetilde{\underline{A%
}}[\ ^{\shortmid }\widetilde{\underline{\mathbf{g}}}(\tau )],$ and $%
\underline{B}=\underline{B}[\ _{1}\underline{\mathbf{g}}(\tau )],$ or $\
^{\shortmid }\widetilde{\underline{B}}=\ ^{\shortmid }\widetilde{\underline{B%
}}[\ _{1}^{\shortmid }\widetilde{\underline{\mathbf{g}}}(\tau )]. $
Hereafter, for simplicity, we shall omit tilde formulas considering that
they can be always introduced if certain Hamiton mechanical variables are
important for study certain physical and/or information processes and when
probability densities can be introduced in any point and along causal lines
on a KK spacetime manifold $\underline{\mathbf{V}}\mathbf{.}$

In a general covariant form, we shall work with a thermodynamic generating
function $\underline{\mathcal{Z}}[\underline{\mathbf{g}}(\tau )]$ and
respective thermodynamic model $\left[ \underline{\mathcal{W}};\underline{%
\mathcal{Z}},\underline{\mathcal{E}},\underline{\mathcal{S}},\underline{\eta
}\right] $ (\ref{thvalkk}) for GIFs on $\underline{\mathbf{V}}\mathbf{.}$ To
study conditional GIFs we shall use geometric flow models on $\underline{%
\mathbf{V}}\otimes \underline{\mathbf{V}}$ when the local coordinates are $(%
\underline{u},\ _{1}\underline{u})$ and the normalizing functions are of
type $\ _{AB}\underline{f}(\underline{u},\ _{1}\underline{u}).$ A d-metric
structure on such tensor products of nonholonomic Lorentz manifolds is of
type $\ _{AB}\underline{\mathbf{g}}=\{\underline{\mathbf{g}}=[q_{1},q_{2},%
\mathbf{q}_{3},_{q}N,q_{5}],\ _{1}\underline{\mathbf{g}}=[\ _{1}q_{1},\
_{1}q_{2},\ _{1}\mathbf{q}_{3},_{1q}N,\ _{1}q_{5}]\}.$ Respectively, we
introduce a metric compatible d--connection $\ _{AB}\underline{\mathbf{D}}=\
_{A}\underline{\mathbf{D}}+\ _{B}\underline{\mathbf{D}}$ and corresponding
scalar curvature $\ _{sAB}\underline{R}=\ _{s}\underline{R}+\ _{s1}%
\underline{R}.$

The thermodynamic GIF entropies for respective systems are $\underline{%
\mathcal{S}}[\underline{A}]$ and $\ \underline{\mathcal{S}}[\underline{B}]$
defined by $\underline{\mathbf{g}}(\tau )$ and $\ _{1}\underline{\mathbf{g}}%
(\tau )$ as in (\ref{thvalkk}). They can be considered as analogs of $S_{X}$
and $S_{Y}$ used in formulas (\ref{condentr}) and (\ref{mutualinf}). As an
analog of $S_{XY}$ for GIFs, we introduce the thermodynamic generating
function (as a generalization of (\ref{genfkk}))
\begin{equation*}
\ _{AB}\underline{\mathcal{Z}}[\ \underline{\mathbf{g}}(\tau ),\ _{1}%
\underline{\mathbf{g}}(\tau )]=\ \int \ \ _{1}\int (4\pi \tau )^{-5}e^{-\
_{AB}\underline{f}}\sqrt{|\ \underline{\mathbf{g}}|}\sqrt{|\ _{1}\underline{%
\mathbf{g}}|}d^{5}\underline{u}\ d^{5}\ _{1}\underline{u}(-\ _{AB}\underline{%
f}+10),\mbox{ for
}\underline{\mathbf{V}}\mathbf{\otimes \ \underline{\mathbf{\mathbf{V}}}.}
\end{equation*}%
This results in a GIF thermodynamic entropy function
\begin{eqnarray*}
\ _{AB}\underline{\mathcal{S}}=\underline{\mathcal{S}}\ [\underline{A},%
\underline{B}] &=&-\ \int \ _{1}\int (4\pi \tau )^{-5}e^{-\ _{AB}\
\underline{f}}\sqrt{|q_{1}q_{2}\mathbf{q}_{3}(_{q}N)q_{5}|}\sqrt{|\
_{1}q_{1}\ _{1}q_{2}\ \ _{1}\mathbf{q}_{3}(_{1q}N)\ _{1}q_{5}|}\delta ^{5}%
\underline{u}\ \delta ^{5}\ _{1}\underline{u} \\
&&\ \left[ \tau \left( \ _{s}\underline{R}+\ _{s1}\underline{R}+|\
\underline{\mathbf{D}}\ _{AB}\underline{f}+\ _{1}\underline{\mathbf{D}}\
_{AB}\underline{f}|^{2}\right) +\ _{AB}\underline{f}-10\right] .
\end{eqnarray*}%
Using such formulas, we claim (this can be proven in any point of respective
causal curves on Lorentz manifolds) that for GIFs the formulas for the
conditional entropy (\ref{condentr}) and mutual information (\ref{mutualinf}%
) are respectively generalized
\begin{equation*}
\ \underline{\mathcal{S}}\ [\underline{A}|\underline{B}]:=\ _{AB}\underline{%
\mathcal{S}}-\ \underline{\mathcal{S}}[\underline{B}]\geq 0\mbox{
and }\ \underline{\mathcal{J}}\ [\underline{A};\underline{B}]:=\ \underline{%
\mathcal{S}}[\underline{A}]-\ _{AB}\underline{\mathcal{S}}+\ \underline{%
\mathcal{S}}[\underline{B}]\geq 0.
\end{equation*}

Similar claims can be formulated if we use the W-entropy (for diversity, we
use the variant with Hamilton mechanical variables but such formulas can be
proven in general form without tilde)$\ ^{\shortmid }\widetilde{\underline{%
\mathcal{W}}}$ (\ref{kkw}):
\begin{equation*}
\ ^{\shortmid }\widetilde{\underline{\mathcal{W}}}\ [\widetilde{\underline{A}%
}|\widetilde{\underline{B}}]:=\ _{AB}^{\shortmid }\widetilde{\underline{%
\mathcal{W}}}-\ ^{\shortmid }\widetilde{\underline{\mathcal{W}}}[\widetilde{%
\underline{B}}]\geq 0\mbox{ and }\ ^{\shortmid }\widetilde{\underline{%
\mathcal{J}}}\ [\widetilde{\underline{A}};\widetilde{\underline{B}}]:=\
^{\shortmid }\widetilde{\underline{\mathcal{W}}}[\widetilde{\underline{A}}%
]-\ _{AB}^{\shortmid }\widetilde{\underline{\mathcal{W}}}+\ ^{\shortmid }%
\widetilde{\underline{\mathcal{W}}}[\widetilde{\underline{B}}]\geq 0.
\end{equation*}%
These formulas are computed respectively for the W--entropy instead of the
S-entropy in the standard probability theory. For information flows of KK
systems, such formulas can be applied without additional assumptions on
formulating associated statistical thermodynamic models.

Finally, we note that above formulas can be defined and proven respectively,
and in similar forms, on $\mathbf{\underline{\mathbf{V}},\underline{\mathbf{V%
}}\otimes \underline{\mathbf{V}}},$ and other tensor products involving
different types of d-metrics, d-connections and generating functions. For
instance,
\begin{eqnarray*}
\underline{\mathcal{S}}\ [\underline{A}|\underline{B}] &:=&\ _{AB}\underline{%
\mathcal{S}}-\ \underline{\mathcal{S}}[\underline{B}]\geq 0\mbox{ and }\
\underline{\mathcal{J}}\ [\underline{A};\underline{B}]:=\underline{\mathcal{S%
}}[\underline{A}]-\ _{AB}\underline{\mathcal{S}}+\ \underline{\mathcal{S}}[%
\underline{B}]\geq 0; \\
\underline{\mathcal{W}}\ [\underline{A}|\underline{B}] &:=&\ _{AB}\underline{%
\mathcal{W}}-\ \underline{\mathcal{W}}[\underline{B}]\geq 0\mbox{
and }\underline{\mathcal{J}}_{\ \mathcal{W}}\ [\underline{A};\underline{B}]:=%
\underline{\mathcal{W}}[\underline{A}]-\ _{AB}\underline{\mathcal{W}}+\
\underline{\mathcal{W}}[\underline{B}]\geq 0.
\end{eqnarray*}
The models with dual conventional momentum variables are important for
elaborating QM theories of GIFs with Hamilton generating functions. In their
turn, the QGIF models on "pure" tangent bundles are important for encoding
quantum field theories following the Lagrange formalism.

\subsubsection{Relative KK GIF entropy and monotonicity}

The relative entropy is introduced for two probability distributions $P_{X}$
and $Q_{X}.$ For $X=x_{\underline{i}},$ with \underline{$i$}$=\{1,2,...s\},$
one states $p_{\underline{i}}=P_{X}(x_{\underline{i}})$ and $q_{\underline{i}%
}=Q_{X}(x_{\underline{i}}),$ for some long messages with $\underline{N}$
letters. The main problem is to decide which distribution describes a random
process more realistically. One defines the relative entropy per observation
$S(P_{X}||Q_{X}):=\sum_{\underline{i}}p_{\underline{i}}(\log p_{\underline{i}%
}-\log q_{\underline{i}})\geq 1$ under assumption that $\underline{N}%
S(P_{X}||Q_{X})\gg 1.$ This value is asymmetric on $P_{X}$ and $Q_{X}.$ It
measures the difference between these two probability distributions when $%
P_{X}$ is for the correct answer and $Q_{X}$ is taken as an initial
hypothesis.\footnote{%
Let us remember some basic formulas which are necessary for our further
considerations. We consider a pair of random variables $X$ and $Y$ and
respective two probability distributions. The fist one is a possible
correlated joint distribution $P_{X,Y}(x_{\underline{i}},y_{\underline{j}})$
and $P_{X}(x_{\underline{i}}):=\sum_{\underline{j}}P_{X,Y}(x_{\underline{i}%
},y_{\underline{j}}),P_{Y}(y_{\underline{j}}):=\sum_{\underline{i}%
}P_{X,Y}(x_{\underline{i}},y_{\underline{j}}).$ We also consider a second
probability distribution $Q_{X,Y}(x_{\underline{i}},y_{\underline{j}%
})=P_{X}(x_{\underline{i}})$ $P_{Y}(y_{\underline{j}})$ which is defined in
a form ignoring correlations between $X$ and $Y.$ In a general context, $%
Q_{X,Y}(x_{\underline{i}},y_{\underline{j}})$ can be with correlations of
type $Q_{X}(x_{\underline{i}}):=\sum_{\underline{j}}Q_{X,Y}(x_{\underline{i}%
},y_{\underline{j}}).$ We can introduce three random variables $X,Y,Z$
described by a joint probability distribution and related values, $%
P_{X,Y,Z}(x_{\underline{i}},y_{\underline{j}},z_{\underline{k}})$ and $%
P_{X}(x_{\underline{i}}):=\sum_{\underline{j},\underline{k}}P_{X,Y,Z}(x_{%
\underline{i}},y_{\underline{j}},z_{\underline{k}}),P_{Y,Z}(y_{\underline{j}%
},z_{\underline{k}}):=\sum_{\underline{i}}P_{X,Y,Z}(x_{\underline{i}},y_{%
\underline{j}},z_{\underline{k}})$. If we ignore the correlations between $X$
and $YZ,$ we define $Q_{X,Y,Z}(x_{\underline{i}},y_{\underline{j}},z_{%
\underline{k}}):=P_{X}(x_{\underline{i}})P_{Y,Z}(y_{\underline{j}},z_{%
\underline{k}}).$ Other type values can be defined if we observe the
subsystem $XY,$ when $P_{X,Y}(x_{\underline{i}},y_{\underline{j}}):=\sum_{%
\underline{k}}P_{X,Y,Z}(x_{\underline{i}},y_{\underline{j}},z_{\underline{k}%
}),Q_{X,Y}(x_{\underline{i}},y_{\underline{j}}):=\sum_{\underline{k}%
}Q_{X,Y,Z}(x_{\underline{i}},y_{\underline{j}},z_{\underline{k}})=P_{X}(x_{%
\underline{i}})P_{Y}(y_{\underline{j}})$.}

We can define and calculate the relative entropy $S$ and mutual information $%
I$ between two distributions
\begin{eqnarray*}
&&S(P_{X}||Q_{X}):=\sum_{i,j}P_{X,Y}(x_{\underline{i}},y_{\underline{j}%
})[\log P_{X,Y}(x_{\underline{i}},y_{\underline{j}})-\log (P_{X}(x_{%
\underline{i}})P_{Y}(y_{\underline{j}}))]=S_{X}-S_{XY}+S_{Y}=I(X;Y); \\
&&S(P_{X,Y}||Q_{X,Y}):=S_{X}-S_{XY}+S_{Y}=I(X;Y);S(P_{X,Y,Z}||Q_{X,Y,Z}):=S_{XY}-S_{XYZ}-S_{YZ}=I(X;YZ).
\end{eqnarray*}%
There are important inequalities
\begin{eqnarray*}
&&I(X;Y):=S_{X}+S_{Y}-S_{XY}\geq 0,\mbox{ subadditivity of entropy }; \\
&&S(P_{X,Y}||Q_{X,Y})\geq S(P_{X}||Q_{X}),S(P_{X,Y,Z}||Q_{X,Y,Z})\geq
S(P_{X,Y}||Q_{X,Y}),\mbox{ monotonicity of relative entropy}.
\end{eqnarray*}%
For three random variables, there is also the condition of strong
subadditivity
\begin{equation*}
S_{X}-S_{XYZ}-S_{YZ}\geq S_{X}-S_{XY}+S_{Y},\mbox{ or }S_{XY}+S_{YZ}\geq
S_{Y}+S_{XYZ},
\end{equation*}%
which is equivalent for the condition of monotonicity of mutual information $%
I(X;YZ)\geq I(X;Y).$

The formulas for $S$ and $I$ can be generalized respectively for the
relative entropy and mutual information of geometric flows of KK systems
(proofs can be provided for causal lines and nonholonomic variables
generated by certain relativistic Hamilton generating functions $H(x,p)$).
For KK GIV, we can consider thermodynamic generating functions$\ _{A}%
\underline{\mathcal{Z}}:=\underline{\mathcal{Z}}[\ \mathbf{g}(\tau )]$ and $%
\ _{B}\underline{\mathcal{Z}}:=\ _{1}\underline{\mathcal{Z}}[\ _{1}%
\underline{\mathbf{g}}(\tau )],$ see (\ref{genfkk}), as analogs of $%
p_{i}=P_{X}(x_{i})$ and $q_{i}=Q_{X}(x_{i}).$ We can consider GIFs of three
KK systems $\underline{A},\underline{B},\underline{C}.$ We claim (for
certain configurations, we can prove using standard methods in any point of
causal curves and applying explicit integral N-adapted calculations on $%
\mathbf{\underline{\mathbf{V}}\otimes \underline{\mathbf{V}}\otimes
\underline{\mathbf{V}}}$ such properties{\small
\begin{eqnarray*}
&&\underline{\mathcal{J}}\ [\underline{A};\underline{B}] :=\underline{%
\mathcal{S}}[\underline{A}]-\ _{AB}\underline{\mathcal{S}}+\ \underline{%
\mathcal{S}}[\underline{B}]\geq 0,\mbox{ subadditivity of entropy}; \\
&&\underline{\mathcal{S}}\ [\ _{AB}\underline{\mathcal{Z}}||\ _{AB}%
\underline{\mathcal{Z}}] \geq \underline{\mathcal{S}}\ [\ _{A}\underline{%
\mathcal{Z}}||\ _{A}\underline{\mathcal{Z}}],\underline{\mathcal{S}}\ [\
_{ABC}\underline{\mathcal{Z}}||\ _{ABC}\underline{\mathcal{Z}}]\geq
\underline{\mathcal{S}}\ [\ _{AB}\underline{\mathcal{Z}}||\ _{AB}\underline{%
\mathcal{Z}}],\mbox{ monotonicity of relative entropy}.
\end{eqnarray*}%
} The conditions of strong subadditivity for GIF entropies are claimed
\begin{equation*}
\ _{A}\underline{\mathcal{S}}-\ _{ABC}\underline{\mathcal{S}}-\ _{BC}%
\underline{\mathcal{S}}\geq \ _{A}\underline{\mathcal{S}}-\ _{AB}\underline{%
\mathcal{S}}+\ _{B}\underline{\mathcal{S}},\mbox{ or }\ _{AB}\underline{%
\mathcal{S}}+\ _{BC}\underline{\mathcal{S}}\geq \ _{B}\underline{\mathcal{S}}%
+\ _{ABC}\underline{\mathcal{S}}.
\end{equation*}%
In equivalent form, these formulas can be written as the condition of
monotonicity of KK GIFs mutual information, $\ \underline{\mathcal{J}}\ [%
\underline{A};\underline{B}\underline{C}]\geq \underline{\mathcal{J}}\ [%
\underline{A};\underline{B}].$

Above inequalities can be proven for any point along causal curves on $%
\mathbf{\underline{\mathbf{V}}.}$ They involve the thermodynamic generating
function (as a generalization of (\ref{genfkk})),
\begin{eqnarray}
\ _{ABC}\mathcal{Z}[\underline{\mathbf{g}}(\tau ),\ _{1}\underline{\mathbf{g}%
}(\tau ),\ _{2}\underline{\mathbf{g}}(\tau )] &=&\int \ \ _{1}\int \
_{2}\int (4\pi \tau )^{-15/2}e^{-\ _{ABC}\underline{f}}\sqrt{|\ \underline{%
\mathbf{g}}|}\sqrt{|\ _{1}\underline{\mathbf{g}}|}\sqrt{|\ _{2}\underline{%
\mathbf{g}}|}d^{5}\underline{u}\ d^{5}\ _{1}\underline{u}\ d^{5}\ _{2}%
\underline{u}  \notag \\
&&(-\ _{ABC}\underline{f}+15),\mbox{ for }\mathbf{\underline{\mathbf{V}}%
\otimes \underline{\mathbf{V}}\otimes \ \underline{\mathbf{V}},}
\label{3relz}
\end{eqnarray}%
with a normalizing function $\ _{ABC}\underline{f}(\ \underline{u},\ _{1}%
\underline{u},\ _{2}\underline{u}).$ On such tensor products of KK manifolds
is of type%
\begin{equation*}
\ _{ABC}\underline{\mathbf{g}}=\{\underline{\mathbf{g}}=[q_{1},q_{2},\mathbf{%
q}_{3},_{q}N,q_{5}],\ _{1}\underline{\mathbf{g}}=[\ _{1}q_{1},\ _{1}q_{2},\
_{1}\mathbf{q}_{3},_{1q}N,\ _{1}q_{5}],\ _{2}\underline{\mathbf{g}}=[\
_{2}q_{1},\ _{2}q_{2},\ _{2}\mathbf{q}_{3},_{2q}N,\ _{2}q_{5}]\}.
\end{equation*}%
We can consider a canonical d--connection $\ _{ABC}\underline{\mathbf{D}}=\
\underline{\mathbf{D}}+\ _{B}\ \underline{\mathbf{D}}$ $+\ _{C}\ \underline{%
\mathbf{D}}$ and respective scalar curvature $\ _{sABC}\underline{R}=\ _{s}%
\underline{R}+\ _{s1}\underline{R}+\ _{s2}\underline{R}.$ The resulting
entropy function
\begin{eqnarray}
\ _{ABC}\underline{\mathcal{S}}=\ \underline{\mathcal{S}}[\underline{A},%
\underline{B},\underline{C}] &=&-\int \ \ _{1}\int \ _{2}\int (4\pi \tau
)^{-15/2}e^{-\ _{ABC}\underline{f}}\sqrt{|\ \underline{\mathbf{g}}|}\sqrt{|\
_{1}\underline{\mathbf{g}}|}\sqrt{|\ _{2}\underline{\mathbf{g}}|}d^{5}%
\underline{u}\ d^{5}\ _{1}\underline{u}\ d^{5}\ _{2}\underline{u}  \notag \\
&&\left[ \tau \left( \ _{s}\underline{R}+\ _{s1}\underline{R}+\ _{s2}%
\underline{R}+|\underline{\mathbf{D}}\ _{ABC}\underline{f}|^{2}\right) +\
_{ABC}\underline{f}-15\right] .  \label{3rels}
\end{eqnarray}%
Similar formulas can be derived for W-entropies and for canonical (with
hats) Lagrange-Hamilton GIFs on respective tensor products of KK spaces.

We conclude this introduction to the GIF theory of canonical classical
mechanical systems with three remarks: First, such constructions can be
generalized for different types of d-connections for stochastic maps and
nonholonomic flow evolution and kinetic processes of Lagrange-Hamilton
systems as we studied in \cite{vacaru2000,vacaru2012}. In this work, QGIF
analogs with quantum relative entropy is monotonic including those
associated to the evolution of KK and associated Hamiltonian QM systems.
Relativistic mechanical QGIFs are studied in details in our partner works
\cite{vacaru19b,vacaru19c}.

\subsection{Density matrix and entropies of KK quantum geometric flows, QGIFs%
}

We develop the density matrix formalism for elaborating models of KK GIFs
and QGIFs. Similar constructions for Hamilton mechanical systems are
provided in sections 4 and 5 in \cite{vacaru19b} and, for entangled QG
systems, in \cite{vacaru19c}.

\subsubsection{Entanglement and density matric for KK systems and QGIFs}

QGIF models for KK systems can be elaborated using canonical mechanical
variables $(\widetilde{H},\ ^{\shortmid }\widetilde{\underline{g}}^{ab})$
and generating thermodynamical functions $\ ^{\shortmid }\widetilde{%
\underline{\mathcal{Z}}}[\ ^{\shortmid }\widetilde{\underline{\mathbf{g}}}%
(\tau )]$ (\ref{genfkkh}). Certain special QM systems can be described by
pure states. In a more general context, quantum theories with probabilities
involve not only pure quantum states but also density matrices. The goal of
this subsection is to study how GIFs of KK systems can be generalized using
basic concepts of QM and information theory. We shall elaborate on QGIFs
described in terms of density matrices defined as quantum analogs of state
densities of type $\ ^{\shortmid }\underline{\sigma }$ (\ref{statedens}).

Let us consider a thermodynamical model $\ ^{\shortmid }\underline{%
\widetilde{\mathcal{A}}}=\left[ \ ^{\shortmid }\underline{\widetilde{%
\mathcal{Z}}},\ ^{\shortmid }\underline{\widetilde{\mathcal{E}}},\
^{\shortmid }\underline{\widetilde{\mathcal{S}}},\ ^{\shortmid }\underline{%
\widetilde{\eta }}\right] $ (\ref{thvalkkh}) for GIFs defined in Hamilton
mechanical variables on $\ ^{\shortmid }\underline{\mathbf{V}}\mathbf{.}$ In
any point $\ ^{\shortmid }\underline{u}\in \ ^{\shortmid }\underline{\mathbf{%
V}}$ (we can consider causal curves covering open regions with such points),
we associate a typical Hilbert space $\ ^{\shortmid }\underline{\widetilde{%
\mathcal{H}}}_{\mathcal{A}}.$ A state vector $\psi _{\mathcal{A}}\in $ $\
^{\shortmid }\underline{\widetilde{\mathcal{H}}}_{\mathcal{A}}$ is an
infinite dimensional complex vector function which in quantum information
theory is approximated to vectors in complex spaces of finite dimensions.
Such a $\psi _{\mathcal{A}}$ is a solution of the Schr\"{o}dinger equation
with as a well-defined quantum version of a canonical Hamiltonian $%
\widetilde{H},$ see details in \cite{preskill,witten18}. For nonholonomic
structures, such constructions are considered in \cite{vacaru19b,vacaru19c}.

The combined Hilbert space for KK QGIFs is defined as a tensor product, $\
^{\shortmid }\underline{\widetilde{\mathcal{H}}}_{\mathcal{AB}}=$ $\
^{\shortmid }\underline{\widetilde{\mathcal{H}}}_{\mathcal{A}}\otimes \
^{\shortmid }\underline{\widetilde{\mathcal{H}}}_{\mathcal{B}}$, with an
associate Hilbert space $\ ^{\shortmid }\underline{\widetilde{\mathcal{H}}}_{%
\mathcal{A}}$ considered for a complementary system $\ ^{\shortmid }%
\underline{\widetilde{\mathcal{A}}}.$ Here we note that symbols $\
^{\shortmid }\underline{\widetilde{\mathcal{A}}},\ ^{\shortmid }\underline{%
\widetilde{\mathcal{B}}},\ ^{\shortmid }\underline{\widetilde{\mathcal{C}}}$
etc. are used as labels for certain KK systems under geometric flow
evolution. The state vectors for a combined QGIF system are written $\psi _{%
\mathcal{AB}}=\psi _{\mathcal{A}}\otimes \psi _{\mathcal{B}}\in \
^{\shortmid }\underline{\widetilde{\mathcal{H}}}_{\mathcal{AB}}$ for $\psi _{%
\mathcal{A}}=1_{\mathcal{A}}$ taken as the unity state vector. Quantum
systems subjected only to quantum evolution and not to geometric flows are
denoted $A,B,C,...$ (we do not underline such symbols if they do not involve
KK structures).

A pure state $\psi _{\mathcal{AB}}\in \ ^{\shortmid }\underline{\widetilde{%
\mathcal{H}}}_{\mathcal{AB}}$ may be not only a tensor product of complex
vectors. A quantum system can be also \textit{entangled} and represented by
a matrix of dimension $\underline{N}\times \underline{M}$ if $\dim \
^{\shortmid }\underline{\widetilde{\mathcal{H}}}_{\mathcal{A}}=\underline{N}$
and $\dim \ ^{\shortmid }\underline{\widetilde{\mathcal{H}}}_{\mathcal{B}}=%
\underline{M}.$ We underline symbols for dimensions (one should not be
confused with underlining for KK structures) in order to avoid ambiguities
with the N-connection symbol $\mathbf{N.}$ For any pure state, we can
perform a Schmidt decomposition $\ $%
\begin{equation}
\psi _{\mathcal{AB}}=\sum_{\underline{i}}\sqrt{p_{\underline{i}}}\psi _{%
\mathcal{A}}^{\underline{i}}\otimes \psi _{\mathcal{B}}^{\underline{i}},
\label{schmidt1}
\end{equation}
for any index $\underline{i}=1,2,....$(up to a finite value). A state vector
$\psi _{\mathcal{A}}^{\underline{i}}$ is orthonormal if $<\psi _{\mathcal{A}%
}^{\underline{i}},\psi _{\mathcal{A}}^{\underline{j}}>=<\psi _{\mathcal{B}}^{%
\underline{i}},\psi _{\mathcal{B}}^{\underline{j}}>=\delta ^{\underline{i}%
\underline{j}},$ where $\delta ^{\underline{i}\underline{j}}$ is the
Kronecker symbol. Taking $p_{\underline{i}}>0$ and $\sum_{\underline{i}}p_{%
\underline{i}}=1,$ we can treat $p_{\underline{i}}$ as probabilities. We
note that, in general, such $\psi _{\mathcal{A}}^{\underline{i}}$ and/or $%
\psi _{\mathcal{B}}^{\underline{i}}$ do not define bases of $\ ^{\shortmid }%
\underline{\widetilde{\mathcal{H}}}_{\mathcal{A}}$ and/or $\ ^{\shortmid }%
\underline{\widetilde{\mathcal{H}}}_{\mathcal{B}}.$

The quantum density matrix for a KK QGIF system $\ ^{\shortmid }\underline{%
\widetilde{\mathcal{A}}}$ is defined $\ ^{\shortmid }\underline{\rho }_{%
\mathcal{A}}:=\sum_{\underline{a}}p_{\underline{a}}|\psi _{\mathcal{A}}^{%
\underline{a}}><\otimes \psi _{\mathcal{A}}^{\underline{a}}|$ as a Hermitian
and positive semi-definite operator with trace $Tr_{\mathcal{H}_{\mathcal{A}%
}}\rho _{\mathcal{A}}=1.$ This allows us to compute the \textit{expectation}
value of any operator $\ ^{\shortmid }\underline{\mathcal{O}}_{\mathcal{A}}$
characterizing additionally such a system,%
\begin{eqnarray}
<\ ^{\shortmid }\underline{\mathcal{O}}>_{\mathcal{AB}} &=&<\psi _{\mathcal{%
AB}}|\ ^{\shortmid }\underline{\mathcal{O}}_{\mathcal{A}}\otimes 1_{\mathcal{%
B}}|\psi _{\mathcal{AB}}>=\sum_{\underline{i}}p_{\underline{i}}<\psi _{%
\mathcal{A}}^{\underline{i}}|\ ^{\shortmid }\underline{\mathcal{O}}_{%
\mathcal{A}}|\psi _{\mathcal{A}}^{\underline{i}}><\psi _{\mathcal{B}}^{%
\underline{i}}|1_{\mathcal{B}}|\psi _{\mathcal{B}}^{\underline{i}}>=  \notag
\\
<\ ^{\shortmid }\underline{\mathcal{O}}>_{\mathcal{A}} &=&\sum_{\underline{i}%
}p_{\underline{i}}<\psi _{\mathcal{A}}^{\underline{i}}|\ ^{\shortmid }%
\underline{\mathcal{O}}_{\mathcal{A}}|\psi _{\mathcal{A}}^{\underline{i}%
}>=Tr_{\mathcal{H}_{\mathcal{A}}}\ ^{\shortmid }\underline{\rho }_{\mathcal{A%
}}\ ^{\shortmid }\underline{\mathcal{O}}_{\mathcal{A}}.  \label{expectvalues}
\end{eqnarray}%
In general covariant form, we can write for KK-systems that $<\underline{%
\mathcal{O}}>_{\mathcal{A}}=Tr_{\mathcal{H}_{\mathcal{A}}}\underline{\rho }_{%
\mathcal{A}}\underline{\mathcal{O}}_{\mathcal{A}}.$

We can elaborate on models encoding both quantum information and geometric
flow evolution of bipartite systems of type $\ ^{\shortmid }\underline{%
\widetilde{\mathcal{A}}},\ ^{\shortmid }\underline{\widetilde{\mathcal{B}}},$
and $\ ^{\shortmid }\underline{\widetilde{\mathcal{A}}}\ ^{\shortmid }%
\underline{\widetilde{\mathcal{B}}}$ with both quantum and geometric
entanglement defined by density matrices. In general form, bipartite KK
QGIFs systems are described in general form by quantum density matrices of
type $\underline{\rho }_{\mathcal{AB}}$ or (in canonical Hamilton variables)
$\ ^{\shortmid }\underline{\rho }_{\widetilde{\mathcal{A}}\widetilde{%
\mathcal{B}}}.$ In classical theory of pobability \cite{preskill,witten18},
a bipartite system $XY$ by a \textit{joint probability} distribution $%
P_{X,Y}(x_{\underline{i}},y_{\underline{j}}),$ where $P_{X}(x_{\underline{i}%
}):=\sum_{\underline{j}}P_{X,Y}(x_{\underline{i}},y_{\underline{j}}).$
Considering for KK systems $\ ^{\shortmid }\underline{\widetilde{\mathcal{A}}%
}\ ^{\shortmid }\underline{\widetilde{\mathcal{B}}}$ as a bipartite quantum
system with Hilbert space $\ ^{\shortmid }\underline{\widetilde{\mathcal{H}}}%
_{\mathcal{AB}},$ we can define and parameterize a KK QGIF density matrix: $%
\ ^{\shortmid }\underline{\rho }_{\mathcal{AB}}=\sum_{\underline{a},%
\underline{a}^{\prime },\underline{b},\underline{b}^{\prime }}\ ^{\shortmid }%
\underline{\rho }_{\underline{a}\underline{a}^{\prime }\underline{b}%
\underline{b}^{\prime }}|\underline{a}>_{\mathcal{A}}\otimes |\underline{b}%
>_{\mathcal{B}}\ _{\mathcal{A}}<\underline{a}^{\prime }|\otimes \ _{\mathcal{%
B}}<\underline{b}^{\prime }|,$ where $|\underline{a}>_{A},$ $\underline{a}%
=1,2,...,$\underline{$n$} is an orthonormal basis of $\mathcal{H}_{\mathcal{A%
}}$ and $|\underline{b}>_{\mathcal{B}},$ $\underline{b}=1,2,...,$\underline{$%
m$} as an orthonormal basis of $\ ^{\shortmid }\underline{\widetilde{%
\mathcal{H}}}_{\mathcal{B}}.$

A \textit{measurement} of the KK-system $\ ^{\shortmid }\underline{%
\widetilde{\mathcal{A}}}$ is characterized by a \textit{reduced density
matrix}
\begin{equation*}
\ ^{\shortmid }\underline{\rho }_{\mathcal{A}}=Tr_{\mathcal{H}_{\mathcal{B}%
}}\ ^{\shortmid }\underline{\rho }_{\mathcal{AB}}=\sum_{\underline{a},%
\underline{a}^{\prime },\underline{b},\underline{b}}\ ^{\shortmid }%
\underline{\rho }_{\underline{a}\underline{a}^{\prime }\underline{b}%
\underline{b}}|\underline{a}>_{\mathcal{A}}\ _{\mathcal{A}}<\underline{a}%
^{\prime }|,\mbox{ for }|\underline{b}>_{\mathcal{B}}\ _{\mathcal{B}}<%
\underline{b}|=1.
\end{equation*}%
In a similar form, we can define and compute $\underline{\rho }_{\mathcal{B}%
}=Tr_{\mathcal{H}_{\mathcal{A}}}\underline{\rho }_{\mathcal{AB}}.$ Using
above introduced concepts and formulas, we can elaborate on KK QGIF models
formulated in Hamilton variables or in a general covariant form.

\subsubsection{Quantum density matrix and von Neumann entropy for KK QGIFs}

For such systems, the quantum density matrix $\ \underline{\sigma }_{%
\mathcal{AB}}$ for a state density $\ \underline{\sigma }$ of type (\ref%
{statedens}) can be defined and computed using formulas (\ref{expectvalues}%
),
\begin{eqnarray}
\ \underline{\sigma }_{\mathcal{AB}} &=&<\underline{\sigma }>_{\mathcal{AB}%
}=<\psi _{\mathcal{AB}}|\underline{\sigma }\otimes 1_{\mathcal{B}}|\psi _{%
\mathcal{AB}}>=\sum_{\underline{i}}p_{\underline{i}}<\psi _{\mathcal{A}}^{%
\underline{i}}|\underline{\sigma }|\psi _{\mathcal{A}}^{\underline{i}}><\psi
_{\mathcal{B}}^{\underline{i}}|1_{\mathcal{B}}|\psi _{\mathcal{B}}^{%
\underline{i}}>=  \notag \\
\underline{\sigma }_{\mathcal{A}} &=&<\underline{\sigma }>_{\mathcal{A}%
}=\sum_{\underline{i}}p_{\underline{i}}<\psi _{\mathcal{A}}^{\underline{i}}|%
\underline{\sigma }|\psi _{\mathcal{A}}^{\underline{i}}>=Tr_{\mathcal{H}_{%
\mathcal{A}}}\underline{\rho }_{\mathcal{A}}\underline{\sigma },
\label{aux01}
\end{eqnarray}%
where the density matrix $\ \underline{\rho }_{\mathcal{A}}$ is taken for
computing the KK QGIF density matrix$\ \underline{\sigma }_{\mathcal{A}}.$ A
matrix (\ref{aux01}) is determined by a state density of the thermodynamical
model for KK GIFs of a classical system $\underline{\sigma }.$ In explicit
form, we can work with quantum density matrices $\ \underline{\sigma }_{%
\mathcal{AB}},$ $\underline{\sigma }_{\mathcal{A}}=Tr_{\mathcal{H}_{\mathcal{%
B}}}\ \underline{\sigma }_{\mathcal{AB}}$ and $\underline{\sigma }_{\mathcal{%
B}}=Tr_{\mathcal{H}_{\mathcal{A}}}\underline{\sigma }_{\mathcal{AB}}.$ Such
formulas can be written in respective coefficient forms
\begin{equation*}
\ \underline{\sigma }_{\mathcal{AB}}=\sum_{\underline{a},\underline{a}%
^{\prime },\underline{b},\underline{b}^{\prime }}\underline{\sigma }_{%
\underline{a}\underline{a}^{\prime }\underline{b}\underline{b}^{\prime }}|%
\underline{a}>_{\mathcal{A}}\otimes |\underline{b}>_{\mathcal{B}}\ _{%
\mathcal{A}}<\underline{a}^{\prime }|\otimes \ _{\mathcal{B}}<\underline{b}%
^{\prime }|\mbox{ and }\underline{\sigma }_{\mathcal{A}}=\sum_{\underline{a},%
\underline{a}^{\prime },\underline{b},\underline{b}}\underline{\sigma }_{%
\underline{a}\underline{a}^{\prime }\underline{b}\underline{b}}|\underline{a}%
>_{\mathcal{A}}\ _{\mathcal{A}}<\underline{a}^{\prime }|.
\end{equation*}

QGIFs can be characterized by quantum analogs of entropy values used for
classical geometric flows. We can consider both an associated thermodynamics
entropy and a W-entropy in classical variants and then quantize such systems
using a Hamiltonian for a fixed system of nonholonomic variables and which
allows a self-consistent QM formulation. Such values can be computed in
explicit form using formulas of type (\ref{aux01}) for classical conditional
and mutual entropy considered for GIFs and in information theory \cite%
{preskill,witten18,vacaru19b}. For instance, this allow to compute define
and compute%
\begin{eqnarray*}
\ _{q}\underline{\mathcal{W}}_{\mathcal{AB}} &=&Tr_{\mathcal{H}_{\mathcal{AB}%
}}[(\underline{\sigma }_{\mathcal{AB}})(_{\mathcal{AB}}\underline{\mathcal{W}%
})]\mbox{ and }\ _{q}\underline{\mathcal{W}}_{\mathcal{A}}=Tr_{\mathcal{H}_{%
\mathcal{A}}}[(\underline{\sigma }_{\mathcal{A}})(\ _{\mathcal{A}}\underline{%
\mathcal{W}})],\ _{q}\underline{\mathcal{W}}_{\mathcal{B}}=Tr_{\mathcal{H}_{%
\mathcal{B}}}[(\underline{\sigma }_{\mathcal{B}})(\ _{\mathcal{B}}\underline{%
\mathcal{W}})]; \\
\ _{q}\underline{\mathcal{S}}_{\mathcal{AB}} &=&Tr_{\mathcal{H}_{\mathcal{AB}%
}}[(\underline{\sigma }_{\mathcal{AB}})(\ _{\mathcal{AB}}\underline{\mathcal{%
S}})]\mbox{ and }\ _{q}\underline{\mathcal{S}}_{\mathcal{A}}=Tr_{\mathcal{H}%
_{\mathcal{A}}}[(\underline{\sigma }_{\mathcal{A}})(\ _{\mathcal{A}}%
\underline{\mathcal{S}})],\ _{q}\underline{\mathcal{S}}_{\mathcal{B}}=Tr_{%
\mathcal{H}_{\mathcal{B}}}[(\underline{\sigma }_{\mathcal{B}})(\ _{\mathcal{B%
}}\underline{\mathcal{S}})].
\end{eqnarray*}%
Such values describe additional entropic properties of quantum KK systems
with rich geometric structure under QGIFs.

\subsubsection{Quantum generalizations of the W- and thermodynamic entropy
of KK GIFs}

We describe KK QGIFs in standard QM form for the von Neumann entropy
determined by $\ \underline{\sigma }_{\mathcal{A}}$ (\ref{aux01}) \ as a
probability distribution,%
\begin{equation}
\ \ _{q}\underline{\mathcal{S}}(\ \underline{\sigma }_{\mathcal{A}}):=Tr\
\underline{\sigma }_{\mathcal{A}}\log \ \underline{\sigma }_{\mathcal{A}}.
\label{neumgfentr}
\end{equation}%
Using $\ _{q}\underline{\mathcal{S}}(\ \underline{\sigma }_{\mathcal{A}})$ (%
\ref{neumgfentr}), we can consider generalizations for $\underline{\mathcal{A%
}}\underline{\mathcal{B}}$ and $\underline{\mathcal{A}}$ systems,
respectively,
\begin{equation*}
\ _{q}\underline{\mathcal{S}}(\ \underline{\sigma }_{\mathcal{AB}}):=Tr\
\underline{\sigma }_{\mathcal{AB}}\log \underline{\sigma }_{\mathcal{AB}}%
\mbox{ and }\ \ _{q}\underline{\mathcal{S}}(\ \underline{\sigma }_{\mathcal{A%
}}):=Tr\ \ \underline{\sigma }_{\mathcal{A}}\log \ \underline{\sigma }_{%
\mathcal{A}},\ \ _{q}\underline{\mathcal{S}}(\ \underline{\sigma }_{\mathcal{%
B}}):=Tr\ \ \underline{\sigma }_{\mathcal{B}}\log \ \underline{\sigma }_{%
\mathcal{B}}.
\end{equation*}

The von Neumann entropy for KK QGIFs, $\ \ _{q}\underline{\mathcal{S}}(\
\underline{\sigma }_{\mathcal{A}}),$ has a purifying property not existing
for classical analogs. For a bipartite system $\psi _{\mathcal{AB}}=\sum_{%
\underline{i}}\sqrt{p_{\underline{i}}}\psi _{\mathcal{A}}^{\underline{i}%
}\otimes \psi _{\mathcal{B}}^{\underline{i}}$ and $\ \underline{\sigma }_{%
\mathcal{A}}:=\sum_{\underline{i}}p_{\underline{i}}|\psi _{\mathcal{A}}^{%
\underline{i}}>\otimes <\psi _{\mathcal{A}}^{\underline{i}}|,$ we compute
\begin{equation}
\ \underline{\sigma }_{\mathcal{A}}:=\sum_{\underline{a},\underline{a}%
^{\prime },\underline{b},\underline{b}}\ \sum_{\underline{k}}^{{}}\
\underline{\sigma }_{\underline{a}\underline{a}^{\prime }\underline{b}%
\underline{b}}p_{\underline{k}}\ _{\mathcal{A}}<\underline{a}^{\prime
}||\psi _{\mathcal{A}}^{\underline{k}}><\otimes \psi _{\mathcal{A}}^{%
\underline{k}}||\underline{a}>_{\mathcal{A}},\ \underline{\sigma }_{\mathcal{%
B}}:=\sum_{\underline{a},\underline{a}^{\prime },\underline{b},\underline{b}%
}\ \sum_{\underline{k}}^{{}}\ \underline{\sigma }_{\underline{a}\underline{a}%
^{\prime }\underline{b}\underline{b}}p_{\underline{k}}\ _{\mathcal{B}}<%
\underline{a}^{\prime }||\psi _{\mathcal{B}}^{\underline{k}}><\otimes \psi _{%
\mathcal{B}}^{\underline{k}}||\underline{b}>_{\mathcal{B}}.  \label{aux03}
\end{equation}

We can consider both an associated thermodynamics entropy and a W-entropy in
classical variants and then quantize KK systems using a respective
Hamiltonian which allows a self-consistent QM formulation. Using respective
formulas (\ref{aux01}), (\ref{aux03}) for classical conditional and mutual
entropy considered for KK GIFs and in information theory, there are defined
and computed respectively%
\begin{eqnarray*}
\ _{q}\underline{\mathcal{W}}_{\mathcal{AB}} &=&Tr_{\mathcal{H}_{\mathcal{AB}%
}}[(\underline{\sigma }_{\mathcal{AB}})(\ _{\mathcal{AB}}\underline{\mathcal{%
W}})]\mbox{ and }\ _{q}\underline{\mathcal{W}}_{\mathcal{A}}=Tr_{\mathcal{H}%
_{\mathcal{A}}}[(\underline{\sigma }_{\mathcal{A}})(\ _{\mathcal{A}}%
\underline{\mathcal{W}})],\ _{q}\underline{\mathcal{W}}_{\mathcal{B}}=Tr_{%
\mathcal{H}_{\mathcal{B}}}[(\underline{\sigma }_{\mathcal{B}})(\ _{\mathcal{B%
}}\underline{\mathcal{W}})]; \\
\ _{q}\underline{\mathcal{S}}_{\mathcal{AB}} &=&Tr_{\mathcal{H}_{\mathcal{AB}%
}}[(\underline{\sigma }_{\mathcal{AB}})(\ _{\mathcal{AB}}\underline{\mathcal{%
S}})]\mbox{ and }\ _{q}\underline{\mathcal{S}}_{\mathcal{A}}=Tr_{\mathcal{H}%
_{\mathcal{A}}}[(\underline{\sigma }_{\mathcal{A}})(\ _{\mathcal{A}}%
\underline{\mathcal{S}})],\ _{q}\underline{\mathcal{S}}_{\mathcal{B}}=Tr_{%
\mathcal{H}_{\mathcal{B}}}[(\underline{\sigma }_{\mathcal{B}})(\ _{\mathcal{B%
}}\underline{\mathcal{S}})].
\end{eqnarray*}%
Such values describe complimentary entropic properties of quantum KK systems
with rich geometric structure under quantum GIF evolution. Additionally, the
quantum probabilistic values are described by the von Neumann entropy $\ \
_{q}\underline{\mathcal{S}}(\ \underline{\sigma }_{\mathcal{A}})$ (\ref%
{neumgfentr}).

\section{Entanglement and QGIFs of KK systems}

\label{s5} We generalize and study the concept of entanglement for QGIFs of
KK systems. The notion of bipartite entanglement was introduced for pure
states and density matrices in description of finite-dimensional QM systems
\cite{preskill,witten18,aolita14,nishioka18}. Thermodynamic and QM analogs
of GIFs are characterized by a series of different types of entropies (G.
Perelman's W-entropy and geometric thermodynamic entropy, and the
entanglement entropy in the von Neumann sense etc.). It will be shown how
each of such entropic values characterise classical and quantum correlations
determined by KK\ QGIFs and quantifies the amount of quantum entanglement. A
set of inequalities involving Pereleman and entanglement entropies playing a
crucial role in definition and description of such systems will be provided.

\subsection{Geometric KK-flows with entanglement}

\subsubsection{Bipartite entanglement for KK QGIFs}

For the KK theory, we can consider various (relativistic) mechanic,
continuous or lattice models of quantum field theory, thermofield theory,
QGIF models etc. A QM model can be characterized by a pure ground state $|%
\underline{\Psi }>$ for a total Hilbert space $\ _{t}\underline{\mathcal{H}}%
\mathcal{\ }$(we underline symbols for KK theories with compactification to
EM models for any fixed value of an geometric flow parameter). The density
matrix
\begin{equation}
\ _{t}\underline{\rho }=|\underline{\Psi }><\underline{\Psi }|
\label{pureground}
\end{equation}%
can be normalized following the conditions $<\underline{\Psi }|$ $\underline{%
\Psi }>=1$ and total trace $\ _{t}tr(\ _{t}^{\shortmid }\underline{\rho }%
)=1. $ We suppose that such a conventional total quantum system is divided
into a two subsystems $\underline{\mathcal{A}}$ and $\underline{\mathcal{B}}$
with respective Hamilton mechanical variables and analogous GIF
thermodynamic models $\ ^{\shortmid }\underline{\widetilde{\mathcal{A}}}$
and $\ ^{\shortmid }\underline{\widetilde{\mathcal{B}}}.$ In this section,
we consider that, for instance $\ ^{\shortmid }\underline{\widetilde{%
\mathcal{A}}}=\left[ \ ^{\shortmid }\underline{\widetilde{\mathcal{Z}}},\
^{\shortmid }\underline{\widetilde{\mathcal{E}}},\ ^{\shortmid }\underline{%
\widetilde{\mathcal{S}}},\ ^{\shortmid }\underline{\widetilde{\eta }}\right]
$ (\ref{thvalkkh}) is a typical KK GIF system which in a general covariant
form on $\ \underline{\mathbf{V}}$ can be parameterized $\ \underline{%
\mathcal{A}}=\left[ \underline{\mathcal{Z}},\ \underline{\mathcal{E}},%
\underline{\mathcal{S}},\underline{\eta }\right] $ (\ref{thvalkk}). Similar
models (they can be with a different associated relativistic Hamiltonian and
d-metric $\ _{1}\mathbf{g}$) are considered for $\ ^{\shortmid }\underline{%
\widetilde{\mathcal{B}}}$ and $\underline{\mathcal{B}}.$ Such subsystems $%
\underline{\mathcal{A}}$ and $\underline{\mathcal{B}}=\overline{\underline{%
\mathcal{A}}} $ are complimentary to each other if there is a common
boundary $\partial \underline{\mathcal{A}}=\partial \underline{\mathcal{B}}$
of codimension 2, where the non-singular flow evolution $\underline{\mathcal{%
A}}$ transforms into a necessary analytic class of flows on $\overline{%
\underline{\mathcal{A}}}.$ We can consider that for bipartite KK QGIFs $\
_{t}\underline{\mathcal{H}}\mathcal{=}\ \underline{\mathcal{H}}_{\mathcal{AB}%
}=$ $\underline{\mathcal{H}}_{\mathcal{A}}\otimes \underline{\mathcal{H}}_{%
\mathcal{B}}.$\footnote{%
Such an assumption can not be correct if there are considered theories with
gauge symmetries (in special for non Abbelian gauge model), see discussion
and references in footnote 3 of \cite{nishioka18}. In this work, we do not
enter in details of entanglement of EM systems as quantum gauge field
theories but consider only constructions with a classical KK gravity theory
and related geometric flow evolution models with associated thermodynamic
and QM values.}

The measure of entanglement of a KK QGIF subsystem $\underline{\mathcal{A}}$
is just the von Neumann entropy $\ _{q}\underline{\mathcal{S}}$ (\ref%
{neumgfentr}) but defined for the reduced density matrix $\underline{\rho }_{%
\mathcal{A}}=Tr_{\mathcal{H}_{\mathcal{B}}}(\ _{t}\underline{\rho }).$ We
define and compute the \textit{entanglement entropy} of $\underline{\mathcal{%
A}}$ as

\begin{equation}
\ _{q}\underline{\mathcal{S}}(\ \underline{\rho }_{\mathcal{A}}):=Tr(\
\underline{\rho }_{\mathcal{A}}\ \log \underline{\rho }_{\mathcal{A}}),
\label{entangentr}
\end{equation}%
when $\ \underline{\rho }_{\mathcal{A}}$ is associated to a state density $\
\underline{\rho }(\beta ,\ \underline{\mathcal{E}}\ ,\underline{\mathbf{g}})$
of type (\ref{statedens}). We note that the total entropy $\ _{t}^{\shortmid
}\underline{\mathcal{S}}=0$ for a pure grand state (\ref{pureground})
associated to $\underline{\mathbf{V}}.$ \ More rich nonholonomic KK QGIF
models encoding Hamilton mechanical variables are defined by
\begin{equation}
\ _{q}^{\shortmid }\underline{\widetilde{\mathcal{S}}}(\ ^{\shortmid }\rho _{%
\mathcal{A}}):=Tr(\ ^{\shortmid }\widetilde{\underline{\rho }}_{\mathcal{A}%
}\ \log \ ^{\shortmid }\widetilde{\underline{\rho }}_{\mathcal{A}}),
\label{entangentrh}
\end{equation}%
when effective Hamilton functions can be defined in any point belonging to a
causal region in $\ ^{\shortmid }\underline{\mathbf{V}}.$ As QM models, the
systems characterized by (\ref{entangentr}) encode less informaton than
those described in conventional mechanical variables (\ref{entangentrh}).

\subsubsection{Separable and entangled KK QGIFs}

Such concepts were introduced for QGIFs in our partner works \cite%
{vacaru19b,vacaru19c} extending to analogous thermodynamic models standard
constructions in quantum information theory \cite%
{preskill,witten18,aolita14,nishioka18}. Let us show how those formulas can
be generalized for KK systems. We consider $\{|\underline{a}>_{\mathcal{A}};%
\underline{a}=1,2,...k_{a}\}\in \mathcal{H}_{\mathcal{A}}$ and $\{|%
\underline{b}>_{\mathcal{B}};\underline{b}=1,2,...k_{b}\}\in \underline{%
\mathcal{H}}_{\mathcal{B}}$ as orthonormal bases when a pure total ground
state is parameterized in the form%
\begin{equation}
|\underline{\Psi }>=\sum_{\underline{a}\underline{b}}C_{\underline{a}%
\underline{b}}|\underline{a}>_{\mathcal{A}}\otimes |\underline{b}>_{\mathcal{%
B}},  \label{groundstate}
\end{equation}%
where $C_{\underline{a}\underline{b}}$ is a complex matrix of dimension $%
\dim \underline{\mathcal{H}}_{\mathcal{A}}\times \dim \underline{\mathcal{H}}%
_{\mathcal{B}}.$ If such coefficients factorize, $C_{\underline{a}\underline{%
b}}=C_{\underline{a}}C_{\underline{b}},$ there are defined separable ground
states (equivalently, pure product states), when
\begin{equation*}
|\underline{\Psi }>=|\underline{\Psi }_{\mathcal{A}}>\otimes |\underline{%
\Psi }_{\mathcal{B}}>,\mbox{ for }|\underline{\Psi }_{\mathcal{A}}>=\sum_{%
\underline{a}}C_{\underline{a}}|\underline{a}>_{\mathcal{A}}\mbox{ and }|%
\underline{\Psi }_{\mathcal{B}}>=\sum_{\underline{b}}C_{\underline{b}}|%
\underline{b}>_{\mathcal{B}}.
\end{equation*}%
The entanglement entropy vanishes, $\ _{q}\underline{\mathcal{S}}(\underline{%
\rho }_{\mathcal{A}})=0,$ if and only if the pure ground state is separable.
Here we note that such a system with nonzero $\ _{q}^{\shortmid }\underline{%
\widetilde{\mathcal{S}}}(\ ^{\shortmid }\underline{\widetilde{\rho }}_{%
\mathcal{A}})$ reflects a more rich nonholonomic structure induced by the
conventional mechanical variables (it is natural that more information is
contained in such cases). For KK QGIFs, such definitions are motivated
because all sub-systems are described by corresponding effective
relativistic Hamilton functions, $\widetilde{\underline{H}}_{\mathcal{A}}$
and $\widetilde{\underline{H}}_{\mathcal{B}},$ and/or effective
thermodynamics energies, $\ _{\mathcal{A}}^{\shortmid }\underline{\mathcal{E}%
}$ and $\ _{\mathcal{B}}^{\shortmid }\underline{\mathcal{E}}.$ Similar
values can be defined and computed for W-entropies.

A ground state $|\underline{\Psi }>$ (\ref{groundstate}) is \textit{%
entangled (inseparable )} if $C_{\underline{a}\underline{b}}\neq C_{%
\underline{a}}C_{\underline{b}}$. For such a state, the entanglement entropy
is positive, $\ _{q}\underline{\mathcal{S}}(\underline{\rho }_{\mathcal{A}%
})>0.$ Using quantum Schmidt decompositions (\ref{schmidt1}), we prove for
any point along a causal curve on $\underline{\mathbf{V}}$ that
\begin{equation}
\ _{q}\underline{\mathcal{S}} = \mathcal{-}\sum_{\underline{a}}^{\min (%
\underline{a},\underline{b})}p_{\underline{a}}\log p_{\underline{a}}%
\mbox{
and } \ _{q}\underline{\mathcal{S}}_{|\max } = \log \min (\underline{a},%
\underline{b})\mbox{ for }\sum_{\underline{a}}p_{\underline{a}}=1\mbox{ and }%
p_{\underline{a}}=1/\min (\underline{a},\underline{b}),\forall a.
\label{aux04}
\end{equation}

An entangled state of KK QGIFs is a superposition of several quantum states
associated to respective GIFs. An observer having access only to a subsystem
$\underline{\mathcal{A}}$ will find him/ herself in a mixed state when the
total ground state $|\underline{\Psi }>$ is entangled following such
conditions:%
\begin{equation*}
|\underline{\Psi }>:\mbox{ separable }\longleftrightarrow \underline{\ \rho }%
_{\mathcal{A}}:\mbox{ pure state},\mbox{ or }|\underline{\Psi }>:%
\mbox{
entangled }\longleftrightarrow \underline{\rho }_{\mathcal{A}}:%
\mbox{ mixed
state}.
\end{equation*}

We note that the von Neumann entanglement entropies$\ _{q}\underline{%
\mathcal{S}}$ (and $\ _{q}^{\shortmid }\underline{\widetilde{\mathcal{S}}}%
\mathcal{)}$ encodes three (four) types of information data: 1) how the
geometric evolution is quantum flow correlated; 2) how much a given QGIF
state differs from a separable QM state; 3) how KK gravity models are
subjected to quantum flow evolution; (and in which forms such KK\ QGIFs are
modelled in nonholonomic Hamilton mechanical like variables). A maximum
value of quantum correlations is reached when a given KK QGIF state is a
superposition of all possible quantum states with an equal weight. But there
are also additional KK GIF properties which are characterized by W-entropies
$\ \underline{\mathcal{W}}$ (\ref{wfkk}) and $\ ^{\shortmid }\underline{%
\widetilde{\mathcal{W}}}$ (\ref{kkw}) and thermodynamic entropies, $\
\underline{\mathcal{S}}$ (\ref{thvalkk}) and $\ ^{\shortmid }\underline{%
\widetilde{\mathcal{S}}}$ (\ref{thvalkkh}), which can be computed in certain
quasi-classical QM limits for a 3+1 splitting and respective dualization of
variables, for instance, along a time like curve.

Let us consider a KK QGIF model with entanglement elaborated for different
associated relativistic Hamiltonians and respective d-metrics $\underline{%
\mathbf{g}}$ and $\ _{1}\underline{\mathbf{g}}.$ We suppose that the
conventional Hilbert spaces are spanned by two orthonormal basic states in
the form $\{|\underline{a}>_{\mathcal{A}};\underline{a}=1,2\}\in \underline{%
\mathcal{H}}_{\mathcal{A}}$ and $\{|\underline{b}>_{\mathcal{B}};\underline{b%
}=1,2\}\in \underline{\mathcal{H}}_{\mathcal{B}},$ when $_{\mathcal{A},%
\mathcal{B}}<\underline{a}|\underline{b}>_{\mathcal{A},\mathcal{B}}=\delta _{%
\underline{a}\underline{b}}.$ The total Hilbert space $\underline{\mathcal{H}%
}_{\mathcal{AB}}=\underline{\mathcal{H}}_{\mathcal{A}}\otimes \underline{%
\mathcal{H}}_{\mathcal{B}}$ has a 4-dim orthonormal basis $\underline{%
\mathcal{H}}_{\mathcal{AB}}=\{|11>,|12>,|21>,|22>\},$ where $|\underline{a}%
\underline{b}>=|\underline{a}>_{\mathcal{A}}\otimes |\underline{b}>_{%
\mathcal{B}}$ are tensor product states.

As a general state for such a bipartite KK QGIF system, we can consider
\begin{equation}
|\underline{\Psi }>=\cos \theta |12>-\sin \theta |21>,  \label{2flow}
\end{equation}%
where $0\leq \theta \leq \pi /2.$ The corresponding entanglement entropy (%
\ref{entangentr}) is computed%
\begin{equation*}
\ _{q}\underline{\mathcal{S}}(\underline{\rho }_{\mathcal{A}})=-\cos
^{2}\theta \log (\cos ^{2}\theta )-\sin ^{2}\theta \log (\sin ^{2}\theta ).
\end{equation*}%
These formulas show that for $\theta =0,\pi /2$ we obtain pure product
states with zero entanglement entropy. For a quantum system $|\underline{%
\Psi }>=\frac{1}{\sqrt{2}}(|12>-|21>),$ when the density matrix
\begin{equation*}
\ \underline{\rho }_{\mathcal{A}}=\frac{1}{2}(|1>_{\mathcal{A}}\ _{\mathcal{A%
}}<1|+|2>_{\mathcal{A}}\ _{\mathcal{A}}<2|)=\frac{1}{2}diag(1,1)
\end{equation*}%
results in $\ _{q}\underline{\mathcal{S}}(\underline{\rho }_{\mathcal{A}%
})=-tr_{\mathcal{A}}(\underline{\rho }_{\mathcal{A}}\log \underline{\rho }_{%
\mathcal{A}})=\log 2.$ For such quantum systems, the maximal entanglement is
for $\theta =\pi /4.$ If the KK GIF structure is "ignored" for such a
quantum system (see formula (\ref{2flow})), we can associate a conventional
QM system which is similar to spin ones, for instance, with up-spin $|1>$
and down-spin $|2>.$

\subsubsection{Entanglement with thermofield double KK QGIF states and
W-entropy}

In this series of works \cite{bubuianu19,vacaru19b,vacaru19c,vacaru19d} the
evolution parameter $\beta =T^{-1}$ is treated as a temperature one like
similarly to the standard G. Perelman's approach \cite{perelman1}. This
allows us to elaborate on GIF theories as certain classical and/or quantum
thermofield models. For KK GIFs, such a nontrivial example with entanglement
and a thermofield double state is defined by a ground state (\ref%
{groundstate}) parameterized in the form
\begin{equation}
|\underline{\Psi }>=\underline{Z}^{-1/2}\sum\limits_{\underline{k}}e^{-\beta
E_{\underline{k}}/2}|\underline{k}>_{\mathcal{A}}\otimes |\underline{k}>_{%
\mathcal{B}}.  \label{dtfst}
\end{equation}%
In this formula, the normalization of the states is taken for the partition
function $\underline{Z}=\sum\limits_{\underline{k}}e^{-\beta E_{\underline{k}%
}/2}$. Such values are associated to the thermodynamic generating function$\
\underline{\mathcal{Z}}\ [\underline{\mathbf{g}}(\tau )]$ (\ref{genfkk}) and
state density matrix $\underline{\sigma }(\beta ,\underline{\mathcal{E}},%
\underline{\mathbf{g}})$ (\ref{statedens}) the energy $\ \underline{\mathcal{%
E}}_{\mathcal{A}}=\{E_{\underline{k}}\}\ $\ is considered quantized with a
discrete spectrum for a KK QGIF system $\underline{\mathcal{A}}=\left[
\underline{\mathcal{Z}},\ \underline{\mathcal{E}},\underline{\mathcal{S}},%
\underline{\eta }\right] $ (\ref{thvalkk}). We compute the density matrix
for this subsystem determining a Gibbs state, \
\begin{equation*}
\ \underline{\rho }_{\mathcal{A}}=\underline{Z}^{-1}\sum\limits_{\underline{k%
}}e^{-\beta E_{\underline{k}}/2}|\underline{k}>_{\mathcal{A}}\otimes \ _{%
\mathcal{A}}<\underline{k}|=\underline{Z}^{-1}e^{-\beta \ ^{\shortmid }%
\mathcal{E}_{\mathcal{A}}}.
\end{equation*}%
In above formulas, we consider $\ \ \underline{\mathcal{E}}$ as \ a
(modular) Hamiltonian $\underline{\mathcal{E}}_{\mathcal{A}}$ such that $\ \
\underline{\mathcal{E}}_{\mathcal{A}}|\underline{k}>_{\mathcal{A}}=E_{%
\underline{k}}|\underline{k}>_{\mathcal{A}}.$

Thermofield double states are certain entanglement purifications of thermal
states with Boltzmann weight $p_{k}=\underline{Z}^{-1}\sum\limits_{%
\underline{k}}e^{-\beta E_{\underline{k}}}.$ Transferring state vectors $\{|%
\underline{k}>_{\mathcal{B}}\}$ from $\underline{\mathcal{H}}_{\mathcal{A}}$
to $\underline{\mathcal{H}}_{\mathcal{B}},$ we can purify $\underline{%
\mathcal{A}}$ in the extended Hilbert space $\underline{\mathcal{H}}_{%
\mathcal{A}}$ $\otimes $ $\underline{\mathcal{H}}_{\mathcal{B}}.$ So, every
expectation of local operators in $\underline{\mathcal{A}}$ can be
represented using the thermofield double state $|\underline{\Psi }>$ (\ref%
{dtfst}) of the total system $\underline{\mathcal{A}}\mathcal{\cup }%
\underline{\mathcal{B}}.$ \ The entanglement entropy can be treated as a
measure of the thermal entropy of the subsystem $\underline{\mathcal{A}},$
\begin{equation*}
\underline{\mathcal{S}}(\underline{\rho }_{\mathcal{A}})=-tr_{\mathcal{A}}[%
\underline{\rho }_{\mathcal{A}}(-\beta \underline{\mathcal{E}}_{\mathcal{A}%
}-\log \underline{Z})]=\beta (<\underline{\mathcal{E}}_{\mathcal{A}}>-\
\underline{\mathcal{F}}_{\mathcal{A}}),
\end{equation*}%
where $\ ^{\shortmid }\mathcal{F}_{\mathcal{A}}=-\log Z$ the thermal free
energy. For thermofield values, we omit the label "q" considered, for
instance, for $\ _{q}\underline{\mathcal{S}}$ (\ref{entangentr}).

Finally, we note that thermofield KK GIF configurations are also
characterized by W-entropy $\ \underline{\mathcal{W}}$ (\ref{wfkk}), see
examples in \cite{ruchin13}.

\subsubsection{Bell like KK QGIF states}

For two KK QGIF systems, a state (\ref{2flow}) is maximally entangled for $%
\theta =\pi /4$. We can defined analogs of Bell state, i.e.
Einstein-Podolsky-Rosen pairs, in quantum geometric flow theory are defined
{\small
\begin{equation}
|\underline{\Psi }_{\mathcal{B}}^{1}>=\frac{1}{\sqrt{2}}(|11>+|22>),|%
\underline{\Psi }_{\mathcal{B}}^{2}>=\frac{1}{\sqrt{2}}(|11>-|22>), |%
\underline{\Psi }_{\mathcal{B}}^{3}>=\frac{1}{\sqrt{2}}(|12>+|21>),|%
\underline{\Psi }_{\mathcal{B}}^{4}>=\frac{1}{\sqrt{2}}(|12>-|21>).
\label{bellqgif}
\end{equation}%
} Such states violate the Bell's inequalities and encode also the
information characterized by W-entropy.

The constructions of type (\ref{bellqgif}) can be extended for systems of $k$
qubits. For instance, the Greenberger-Horne-Zelinger, GHZ, states \cite%
{nishioka18} are $\ |\underline{\Psi }_{\mathcal{B}}^{GHZ}>=\frac{1}{\sqrt{2}%
}(|1>^{\otimes k}+|2>^{\otimes k}).$ In quantum information theory, there
are considered also (W states), $|\underline{\Psi }_{\mathcal{B}}^{W}>=\frac{%
1}{\sqrt{2}}(|21...11>+|121...1>+...+|11...12>).$ A state $|\underline{\Psi }%
_{\mathcal{B}}^{GHZ}>$ is fully separable but not $|\underline{\Psi }_{%
\mathcal{B}}^{W}>$ as we prove below:

For $k=3,$ a tripartite KK\ QGIF with subsystems $\underline{\mathcal{A}},%
\underline{\mathcal{B}}$ and $\underline{\mathcal{C}},$ we write $|%
\underline{\Psi }_{\mathcal{B}}^{GHZ}>=\frac{1}{\sqrt{2}}(|111>+|222>)$ and $%
|\underline{\Psi }_{\mathcal{B}}^{W}>=\frac{1}{\sqrt{2}}(|112>+|121>+|211>).$
The reduced density matrices for the system $\underline{\mathcal{A}}\mathcal{%
\cup }\underline{\mathcal{B}}$ $\ $is defined using $Tr_{\mathcal{C}},\
\underline{\rho }_{\mathcal{A\cup B}}^{GHZ}=\frac{1}{2}(|11><11|+|22><22|)$
and$\ \underline{\rho }_{\mathcal{A\cup B}}^{W}=\frac{2}{3}|\underline{\Psi }%
_{\mathcal{B}}^{3}><\underline{\Psi }_{\mathcal{B}}^{3}|+\frac{1}{3}|11><11|.
$ This way, there are described two different KK QGIF states. The first
state is fully separable and can be represented in the form $\ \underline{%
\rho }_{\mathcal{A\cup B}}^{GHZ}=\sum\limits_{\underline{k}=1}^{2}p_{%
\underline{k}}\ \underline{\rho }_{\mathcal{A}}^{\underline{k}}\otimes
\underline{\rho }_{\mathcal{B}}^{\underline{k}},$ for $p_{\underline{k}}=1/2$
and $\ \underline{\rho }_{\mathcal{A},\mathcal{B}}^{\underline{1}}=|11><11|$
and $\ \underline{\rho }_{\mathcal{A},\mathcal{B}}^{\underline{2}}=|22><22|.$
Containing a Bell state $|\underline{\Psi }_{\mathcal{B}}^{3}>$ (\ref%
{bellqgif}), $\ $the $\ \underline{\rho }_{\mathcal{A\cup B}}^{W}$ can not
be written in a separable form. So, the state $|\Psi _{\mathcal{B}}^{W}>$ is
still entangled even we have taken $Tr_{\mathcal{C}}.$

\subsection{Entanglement inequalities for entropies of KK QGIFs}

We study certain important inequalities and properties of the entanglement
entropy (\ref{entangentr}) for KK QGIFs using the density matrix $\underline{%
\rho }_{\mathcal{A}}=Tr_{\mathcal{H}_{\mathcal{B}}}(\ _{t}\underline{\rho }%
). $ We omit technical proofs which are similar to those presented in \cite%
{nielsen10}.\footnote{%
Rigorous mathematical proofs involve a geometric analysis technique \cite%
{perelman1,monogrrf1,monogrrf2,monogrrf3} developed for applications in
modern in modern gravity and particle physics theories in \cite%
{ruchin13,gheorghiu16,bubuianu19}. For any classes of solutions (re-defining
normalizing functions), we can always compute Perelman's like entropy
functionals at least in the quasi-classical limit with respective measures
and related to $\ _{q}\underline{\mathcal{S}}$ (\ref{entangentr}) for a KK
QGIF or a thermofield KK GIF model.} For any $\underline{\rho }_{\mathcal{A}%
}\ $ associated to a state density $\underline{\rho }(\beta ,\underline{%
\mathcal{E}}\ ,\underline{\mathbf{g}})$ (\ref{statedens}), we can compute
the respective W-entropy and geometric thermodynamic entropy taking measures
determined by $\underline{\mathbf{g}}$ and/or respective Hamilton variables.

\subsubsection{(Strong) subadditivity for entangled KK systems}

Let us analyze three important properties of KK QGIFs resulting in a strong
subadditivity property of entanglement and Perelman's entropies.

\begin{enumerate}
\item \textit{Entanglement entropy for complementary KK QGIF subsystems: \ }%
If $\underline{\mathcal{B}}=\overline{\underline{\mathcal{A}}},$ $\ _{q}%
\underline{\mathcal{S}}_{\mathcal{A}}=\ _{q}\underline{\mathcal{S}}_{%
\overline{\mathcal{A}}}.$ This follows from formulas (\ref{aux04}) for a
pure ground state wave function. We can prove similar equalities for the
W-entropy $\ \underline{\mathcal{W}}$ (\ref{wfkk}) and/or thermodynamic
entropy $\ \underline{\mathcal{S}}$ (\ref{thvalkk}) if we use the same
d-metric $\underline{\mathbf{g}}$ and respective normalization on $%
\underline{\mathcal{A}}$ and $\overline{\underline{\mathcal{A}}}.$ For
quantum models, $\ _{q}\underline{\mathcal{S}}_{\mathcal{A}}\neq \ _{q}%
\underline{\mathcal{S}}_{\mathcal{B}}$ if $\underline{\mathcal{A}}\mathcal{%
\cup }\underline{\mathcal{B}}$ is a mixed state, for instance, at a finite
temperature. So, in general, $\ _{q}\underline{\mathcal{S}}_{\mathcal{A}%
}\neq \ _{q}\underline{\mathcal{S}}_{\mathcal{B}}$ and$\ \ _{q}\underline{%
\mathcal{W}}_{\mathcal{A}}\neq \ _{q}\underline{\mathcal{W}}_{\mathcal{B}}$%
and such inequalities hold true in any quasi-classical limit. In principle,
for some special subclasses of nonholonomic deformations such conditions may
transform in equalities. Similar formulas hold in Hamilton variables with
"tilde". \

\item \textit{Subadditivity: }Such conditions are satisfied for \ disjoint
subsystems $\underline{\mathcal{A}}$ and $\underline{\mathcal{B}},$
\begin{equation}
\ _{q}\underline{\mathcal{S}}_{\mathcal{A\cup B}}\leq \ \ _{q}\underline{%
\mathcal{S}}_{\mathcal{A}}+\ \ _{q}\underline{\mathcal{S}}_{\mathcal{B}}%
\mbox{
and }|\ \ _{q}\underline{\mathcal{S}}_{\mathcal{A}}-\ \ _{q}\underline{%
\mathcal{S}}_{\mathcal{B}}|\leq \ \ _{q}\underline{\mathcal{S}}_{\mathcal{%
A\cup B}}.  \label{subaditcond}
\end{equation}%
The second equation is the triangle inequality \cite{araki70} which is
satisfied also in the quasi-classical limit for $\ \underline{\mathcal{S}}$ (%
\ref{thvalkk}). We claim that similar conditions hold for the W-entropy $\
\underline{\mathcal{W}}$ (\ref{wfkk}) and respective quantum versions. They
can be computed (and proved in any point of causal curves) as quantum
perturbations in a QM model associated to a bipartite KK QGIF model $\ _{q}%
\underline{\mathcal{W}}_{\mathcal{A\cup B}}\leq \ _{q}\underline{\mathcal{W}}%
_{\mathcal{A}}+\ _{q}\underline{\mathcal{W}}_{\mathcal{B}}$ and$|\ \ _{q}%
\underline{\mathcal{W}}_{\mathcal{A}}-\ _{q}\underline{\mathcal{W}}_{%
\mathcal{B}}|\leq \ \ _{q}\underline{\mathcal{W}}_{\mathcal{A\cup B}}.$ Such
KK flow evolution and QM scenarios are elaborated for mixed geometric and
quantum probabilistic information flows.

\item \textit{Strong subadditivity} is considered for three disjointed KK
QGIF subsystems $\underline{\mathcal{A}},\underline{\mathcal{B}}$ and $%
\underline{\mathcal{C}}$ and certain conditions of convexity of a function
built from respective density matrices and unitarity of systems \cite%
{lieb73,narnhofer85,witten18,nishioka18}. In any point of causal curves,
there are proved the following inequalities:%
\begin{equation*}
\ _{q}\underline{\mathcal{S}}_{\mathcal{A\cup B\cup C}}+\ _{q}\underline{%
\mathcal{S}}_{\mathcal{B}}\leq \ \ _{q}\underline{\mathcal{S}}_{\mathcal{%
A\cup B}}+\ _{q}\underline{\mathcal{S}}_{B\mathcal{\cup C}}\mbox{ and }\ \
_{q}\underline{\mathcal{S}}_{\mathcal{A}}+\ _{q}\underline{\mathcal{S}}_{%
\mathcal{C}}\leq \ \ _{q}\underline{\mathcal{S}}_{\mathcal{A\cup B}}+\ _{q}%
\underline{\mathcal{S}}_{B\mathcal{\cup C}}.
\end{equation*}%
Using these formulas, the conditions of subadditivity (\ref{subaditcond})
can be derived as particular cases. Along causal curves on respective
cotangent Lorentz manifolds, we can prove similar formulas for the W-entropy
and small quantum perturbations $\ _{q}\underline{\mathcal{W}}_{\mathcal{%
A\cup B\cup C}}+\ _{q}\underline{\mathcal{W}}_{\mathcal{B}}\leq \ \ _{q}%
\underline{\mathcal{W}}_{\mathcal{A\cup B}}+\ _{q}\underline{\mathcal{W}}_{B%
\mathcal{\cup C}}$ and$\ \ _{q}\underline{\mathcal{W}}_{\mathcal{A}}+\ _{q}%
\underline{\mathcal{W}}_{\mathcal{C}}\leq \ _{q}\underline{\mathcal{W}}_{%
\mathcal{A\cup B}}+\ _{q}\underline{\mathcal{W}}_{B\mathcal{\cup C}}.$
\end{enumerate}

Such properties are claimed for KK QGIFs.

\subsubsection{Relative entropy of KK systems with QGIF entanglement}

The concept of\textit{\ relative entropy} in geometric information theories.
\begin{equation}
\ \underline{\mathcal{S}}(\underline{\rho }_{\mathcal{A}}\shortparallel
\underline{\sigma }_{\mathcal{A}})=Tr_{\mathcal{H}_{\mathcal{B}}}[\
\underline{\rho }_{\mathcal{A}}(\log \underline{\rho }_{\mathcal{A}}-\log
\underline{\sigma }_{\mathcal{A}})],  \label{relativentr}
\end{equation}%
where $\ \ \underline{\mathcal{S}}(\underline{\rho }_{\mathcal{A}%
}\shortparallel \underline{\rho }_{\mathcal{A}})=0.$ This way we introduce a
measure of "distance" between two KK QGIFs with a norm $||\ \underline{\rho }%
_{\mathcal{A}}||=tr(\sqrt{(\underline{\rho }_{\mathcal{A}})(\underline{\rho }%
_{\mathcal{A}}^{\dag })}),$ see reviews \cite{preskill,witten18,nishioka18}.
In straightforward form, we can check that there are satisfied certain
important properties and inequalities.

Two KK QGIF systems are characterized by formulas and conditions for
relative entropy:

\begin{enumerate}
\item for tensor products of density matrices, $\ \ \underline{\mathcal{S}}%
(\ _{1}\underline{\rho }_{\mathcal{A}}\otimes \ _{2}\underline{\rho }_{%
\mathcal{A}}\shortparallel \ _{1}\underline{\sigma }_{\mathcal{A}}\otimes \
_{2}\underline{\sigma }_{\mathcal{A}})=\ \underline{\mathcal{S}}(\ _{1}%
\underline{\rho }_{\mathcal{A}}\shortparallel \ _{1}\underline{\sigma }_{%
\mathcal{A}})+\ \underline{\mathcal{S}}(\ _{2}\underline{\rho }_{\mathcal{A}%
}\shortparallel \ _{2}\underline{\sigma }_{\mathcal{A}});$

\item positivity: $\ \underline{\mathcal{S}}(\ \underline{\rho }_{\mathcal{A}%
}\shortparallel \underline{\sigma }_{\mathcal{A}})\geq \frac{1}{2}||\ \
\underline{\rho }_{\mathcal{A}}-\underline{\sigma }_{\mathcal{A}}||^{2},$
i.e.$\ \underline{\mathcal{S}}(\ \ \underline{\rho }_{\mathcal{A}%
}\shortparallel \underline{\sigma }_{\mathcal{A}})\geq 0;$

\item monotonicity: $\underline{\ \mathcal{S}}(\ \underline{\rho }_{\mathcal{%
A}}\shortparallel \underline{\sigma }_{\mathcal{A}})\geq \ \underline{%
\mathcal{S}}(tr_{s}\ \ \underline{\rho }_{\mathcal{A}}|tr_{s}\ \underline{%
\sigma }_{\mathcal{A}}),$ where $tr_{s}$ denotes the trase for a subsystem
of $\underline{\mathcal{A}}.$
\end{enumerate}

Above positivity formula and the Schwarz inequality $||X||\geq tr(XY)/||X||$
result in $\ \ 2\underline{\mathcal{S}}(\ \underline{\rho }_{\mathcal{A}%
}\shortparallel \underline{\sigma }_{\mathcal{A}})\geq (\langle \mathcal{O}%
\rangle _{\rho }-\langle \mathcal{O}\rangle _{\sigma })^{2}/||\mathcal{O}%
||^{2},$ for any expectation value $\langle \mathcal{O}\rangle _{\rho }$ of
an operator $\mathcal{O}$ computed with the density matrix $\underline{\rho }%
_{\mathcal{A}},$ see formulas (\ref{expectvalues}). The relative entropy $\
\underline{\mathcal{S}}(\ \underline{\rho }_{\mathcal{A}}\shortparallel
\underline{\sigma }_{\mathcal{A}})$ (\ref{relativentr}) can be related to
the entanglement entropy $\ _{q}\underline{\mathcal{S}}(\ \underline{\rho }_{%
\mathcal{A}})$ (\ref{entangentr}). We use formula $\ \ \underline{\mathcal{S}%
}(\ \underline{\rho }_{\mathcal{A}}\shortparallel 1_{\mathcal{A}}/k_{%
\mathcal{A}})=\log k_{\mathcal{A}}-\ _{q}\underline{\mathcal{S}}(\
\underline{\rho }_{\mathcal{A}}),$ where $1_{\mathcal{A}}$ is the $k_{%
\mathcal{A}}\times k_{\mathcal{A}}$ unit matrix for a $k_{\mathcal{A}}$%
-dimensional Hilbert space associated to the region $\underline{\mathcal{A}}%
. $

For three KK QGIF systems, we denote by $\ \ \underline{\rho }_{\mathcal{%
A\cup B\cup C}}$ the density matrix of $\underline{\mathcal{A}}\mathcal{\cup
\underline{\mathcal{B}}\cup }\underline{\mathcal{C}}.$ When, for instance, $%
\ \underline{\rho }_{\mathcal{A\cup B}}$ is written for its restriction on $%
\underline{\mathcal{A}}\mathcal{\cup }\underline{\mathcal{B}}$ and $%
\underline{\rho }_{\mathcal{B}}$ is stated for its restriction on $%
\underline{\mathcal{B}}.$ Using the formula
\begin{equation*}
tr_{\mathcal{A\cup B\cup C}}[\ \underline{\rho }_{\mathcal{A\cup B\cup C}}(%
\mathcal{O}_{\mathcal{A\cup B}}\otimes 1_{\mathcal{C}}/k_{\mathcal{C}})]=tr_{%
\mathcal{A\cup B}}(\ \underline{\rho }_{\mathcal{A\cup B}}\mathcal{O}_{%
\mathcal{A\cup B}}),
\end{equation*}%
we prove such identities
\begin{eqnarray*}
\ \ \underline{\mathcal{S}}(\underline{\rho }_{\mathcal{A\cup B\cup C}%
}\shortparallel 1_{\mathcal{A\cup B\cup C}}/k_{\mathcal{A\cup B\cup C}})
&=&\ \underline{\mathcal{S}}(\underline{\rho }_{\mathcal{A\cup B}%
}\shortparallel 1_{\mathcal{A\cup B}}/k_{\mathcal{A\cup B}})+\ \underline{%
\mathcal{S}}(\underline{\rho }_{\mathcal{A\cup B\cup C}}\shortparallel
\underline{\rho }_{\mathcal{A\cup B}}\otimes 1_{\mathcal{C}}/k_{\mathcal{C}%
}), \\
\ \underline{\mathcal{S}}(\underline{\rho }_{\mathcal{B\cup C}%
}\shortparallel 1_{\mathcal{B\cup C}}/k_{\mathcal{B\cup C}}) &=&\ \underline{%
\mathcal{S}}(\underline{\rho }_{\mathcal{B}}\shortparallel 1_{\mathcal{B}%
}/k_{\mathcal{B}})+\ \underline{\mathcal{S}}(\underline{\rho }_{\mathcal{%
B\cup C}}\shortparallel \underline{\rho }_{\mathcal{B}}\otimes 1_{\mathcal{C}%
}/k_{\mathcal{C}});
\end{eqnarray*}
{\small
\begin{eqnarray*}
\mbox{ and inequalities } &&\ \underline{\mathcal{S}}(\underline{\rho }_{%
\mathcal{A\cup B\cup C}}\shortparallel \underline{\rho }_{\mathcal{A\cup B}%
}\otimes 1_{\mathcal{C}}/k_{\mathcal{C}})\geq \ \underline{\mathcal{S}}(%
\underline{\rho }_{\mathcal{B\cup C}}\shortparallel \underline{\rho }_{%
\mathcal{B}}\otimes 1_{\mathcal{C}}/k_{\mathcal{C}}), \\
\ \underline{\mathcal{S}}(\underline{\rho }_{\mathcal{A\cup B\cup C}}
&\shortparallel &1_{\mathcal{A\cup B\cup C}}/k_{\mathcal{A\cup B\cup C}})+\
\underline{\mathcal{S}}(\underline{\rho }_{\mathcal{B}}\shortparallel 1_{%
\mathcal{B}}/k_{\mathcal{B}})\geq \ \underline{\mathcal{S}}(\underline{\rho }%
_{\mathcal{A\cup B}}\shortparallel 1_{\mathcal{A\cup B}}/k_{\mathcal{A\cup B}%
})+\ \underline{\mathcal{S}}(\underline{\rho }_{\mathcal{B\cup C}%
}\shortparallel 1_{\mathcal{B\cup C}}/k_{\mathcal{B\cup C}}).
\end{eqnarray*}%
} These formulas can be re-written for the entanglement entropies $\ _{q}%
\underline{\mathcal{S}}$ and Hamilton mechanical variables with "tilde".

\subsubsection{Mutual information for KK QGIFs}

We can characterize the correlation between two KK QGIF systems $\underline{%
\mathcal{A}}$ and $\underline{\mathcal{B}}$ (it can be involved also a third
system $\underline{\mathcal{C}}$) by \textit{mutual information}
\begin{equation*}
\underline{\mathcal{J}}(\underline{\mathcal{A}},\underline{\mathcal{B}}):=\
\underline{\mathcal{S}}_{\mathcal{A}}+\underline{\mathcal{S}}_{\mathcal{B}}-%
\underline{\mathcal{S}}_{\mathcal{A\cup B}}\geq 0\mbox{ and }\ \underline{%
\mathcal{J}}(\underline{\mathcal{A}},\underline{\mathcal{B}}\mathcal{\cup }%
\underline{\mathcal{C}})\leq \underline{\mathcal{J}}(\underline{\mathcal{A}},%
\underline{\mathcal{B}}).
\end{equation*}%
Using formula $\underline{\mathcal{J}}(\underline{\mathcal{A}},\underline{%
\mathcal{B}})=\underline{\mathcal{S}}(\underline{\rho }_{\mathcal{A\cup B}%
}\shortparallel \underline{\rho }_{\mathcal{A}}\otimes \underline{\rho }_{%
\mathcal{B}}),$ we can introduce similar concepts and inequalities for the
entanglement of KK QGIF systems,%
\begin{equation*}
\ _{q}\underline{\mathcal{J}}(\underline{\mathcal{A}},\underline{\mathcal{B}}%
):=\ _{q}\underline{\mathcal{S}}_{\mathcal{A}}+\ _{q}\underline{\mathcal{S}}%
_{\mathcal{B}}-\ _{q}\underline{\mathcal{S}}_{\mathcal{A\cup B}}\geq 0%
\mbox{
and }\ _{q}\underline{\mathcal{J}}(\underline{\mathcal{A}},\underline{%
\mathcal{B}}\mathcal{\cup }\underline{\mathcal{C}})\leq \ _{q}\underline{%
\mathcal{J}}(\underline{\mathcal{A}},\underline{\mathcal{B}}),\mbox{ for }\
_{q}\underline{\mathcal{J}}(\underline{\mathcal{A}},\underline{\mathcal{B}}%
)=\ _{q}\underline{\mathcal{S}}(\underline{\rho }_{\mathcal{A\cup B}%
}\shortparallel \underline{\rho }_{\mathcal{A}}\otimes \underline{\rho }_{%
\mathcal{B}}).
\end{equation*}

We can write similar formulas for classical KK GIFs and associated
thermodynamic models with statistical density $\underline{\rho }(\beta ,%
\underline{\mathcal{E}}\ ,\underline{\mathbf{g}})$ (\ref{statedens}) and/or
for constructions using the W-entropy. With tilde values, this can be proven
for causal configurations in nonholonomic Hamilton variables \cite{vacaru19b}%
.

The mutual information between two KK QGIFs is a measure how much the
density matrix $\underline{\rho }_{\mathcal{A\cup B}}$ differs from a
separable state $\ \underline{\rho }_{\mathcal{A}}\otimes \underline{\rho }_{%
\mathcal{B}}.$ Quantum correlations entangle even spacetime disconnected
regions of the phase spacetime under geometric flow evolution. Under
geometric information KK flow evolution in respective regions, $\ 2%
\underline{\mathcal{J}}(\underline{\mathcal{A}},\underline{\mathcal{B}})\geq
(\langle \mathcal{O}_{\mathcal{A}}\mathcal{O}_{\mathcal{B}}\rangle -\langle
\mathcal{O}_{\mathcal{A}}\rangle \langle \mathcal{O}_{\mathcal{B}}\rangle
)^{2}/||\mathcal{O}_{\mathcal{A}}||^{2}||\mathcal{O}_{\mathcal{B}}||^{2},$
for bounded operators $\mathcal{O}_{\mathcal{A}}$ and $\mathcal{O}_{\mathcal{%
B}}$.

\subsubsection{The R\'{e}nyi entropy for KK QGIFs}

The R\'{e}nyi entropy \cite{renyi61} is important for computing the
entanglement entropy of QFTs using the replica method (see section IV of
\cite{nishioka18} and \cite{bao19} for a holographic dual of R\'{e}nyi
relative entropy). Such constructions are possible in KK QGIF theory because
the thermodynamic generating function $\underline{\mathcal{Z}}\ [\underline{%
\mathbf{g}}(\tau )]$ (\ref{genfkk}) and related statistical density $%
\underline{\rho }(\beta ,\underline{\mathcal{E}}\ ,\underline{\mathbf{g}})$ (%
\ref{statedens}) can be used for defining $\ \underline{\sigma }_{\mathcal{A}%
}$ (\ref{aux01}) \ as a probability distribution. \

Let us begin with the extension of replica method to G. Perelman's
thermodynamical model and related classical and quantum information
theories. We consider an integer $r$ (replica parameter) and introduce the R%
\'{e}nyi entropy%
\begin{equation}
\ _{r}\ \underline{\mathcal{S}}(\underline{\mathcal{A}}):=\frac{1}{1-r}\log
[tr_{\mathcal{A}}(\underline{\rho }_{\mathcal{A}})^{r}]  \label{renentr}
\end{equation}%
for a KK QGIF\ system determined by a density matrix $\underline{\rho }_{%
\mathcal{A}}.$ A computational formalism is elaborated for an analytic
continuation of $r$ to a real number with a well defined limit $\ _{q}%
\underline{\mathcal{S}}(\underline{\rho }_{\mathcal{A}})=\lim_{r\rightarrow
1}\ \ _{r}\ \underline{\mathcal{S}}(\underline{\mathcal{A}})$ and
normalization $tr_{\mathcal{A}}(\underline{\rho }_{\mathcal{A}})$ for $%
r\rightarrow 1.$ For such limits, the R\'{e}nyi entropy (\ref{renentr})
reduces to the entanglement entropy (\ref{entangentr}).

Using similar formulas proven in \cite{zycz03}, there are introduced such
important inequalities for derivatives on replica parameter, $\partial _{r},$
\begin{equation}
\partial _{r}(\ _{r}\ \underline{\mathcal{S}}\mathcal{)}\leq 0,\ \partial
_{r}\left( \frac{r-1}{r}\ _{r}\ \underline{\mathcal{S}}\right) \geq 0,\
\partial _{r}[(r-1)\ _{r}\ \underline{\mathcal{S}}\mathcal{]}\geq 0,\
\partial _{rr}^{2}[(r-1)](\ _{r}\ \underline{\mathcal{S}}\mathcal{)}\leq 0.
\label{aux07}
\end{equation}%
A usual thermodynamical interpretation of such formulas follows for a system
with a modular Hamiltonian $\underline{H}_{\mathcal{A}}$ and effective
statistical density $\underline{\rho }_{\mathcal{A}}:=e^{-2\pi \underline{H}%
_{\mathcal{A}}}.$ The value $\beta _{r}=2\pi r$ is considered as the inverse
temperature and the effective "thermal" statistical generation (partition)
function is defined
\begin{equation*}
\ _{r}\underline{\mathcal{Z}}(\beta _{r}):=tr_{\mathcal{A}}(\underline{\rho }%
_{\mathcal{A}})^{r}=tr_{\mathcal{A}}(e^{-\beta _{r}\underline{H}_{\mathcal{A}%
}}).
\end{equation*}%
similarly to $\underline{\mathcal{Z}}\ [\underline{\mathbf{g}}(\tau )]$ (\ref%
{genfkk}). We compute using canonical relations such statistical mechanics
values%
\begin{eqnarray*}
\mbox{ for  the modular energy}:&& \ _{r}\underline{\mathcal{E}}(\beta
_{r}):=-\partial _{\beta _{r}}\log [\ _{r}\underline{\mathcal{Z}}(\beta
_{r})]\geq 0; \\
\mbox{ for the modular entropy}: && \ _{r}^{\shortmid }\mathcal{\breve{S}}%
(\beta _{r}):=\left( 1-\beta _{r}\partial _{\beta _{r}}\right) \log [\
_{r}^{\shortmid }\mathcal{Z}(\beta _{r})]\geq 0; \\
\mbox{ for the modular capacity}: && \ _{r}^{\shortmid }\mathcal{C}(\beta
_{r}):=\beta _{r}^{2}\partial _{\beta _{r}}^{2}\log [\ _{r}^{\shortmid }%
\mathcal{Z}(\beta _{r})]\geq 0.
\end{eqnarray*}%
These inequalities are equivalent to the second line in (\ref{aux07}) and
characterize the stability if KK GIFs as a thermal system with replica
parameter regarded as the inverse temperature for a respective modular
Hamiltonian. Such replica criteria of stability define a new direction for
the theory of geometric flows and applications in modern physics \cite%
{vacaru09,rajpoot17,
ruchin13,gheorghiu16,vacaru19,vacaru19a,vacaru19b,bubuianu18}.

We note that the constructions with the modular entropy can be transformed
into models derived with the R\'{e}nyi entropy and inversely. Such
transforms can be performed using formulas%
\begin{equation*}
\ _{r}\underline{\mathcal{\breve{S}}}:=r^{2}\partial _{r}\left( \frac{r-1}{r}%
\ _{r}\underline{\mathcal{S}}\right) \mbox{ and, inversely, }\ _{r}%
\underline{\mathcal{S}}\mathcal{=}\frac{r}{r-1}\int_{1}^{r}dr^{\prime }\frac{%
\ _{r^{\prime }}\underline{\mathcal{\breve{S}}}}{(r^{\prime })^{2}}.
\end{equation*}%
The implications of the inequalities for the R\'{e}nyi entropy were analyzed
for the gravitational systems with holographic description, see reviews \cite%
{solodukhin11,preskill,nishioka18}.

The concept of relative entropy $\underline{\mathcal{S}}(\ \ \underline{\rho
}_{\mathcal{A}}\shortparallel \ \underline{\sigma }_{\mathcal{A}})$ (\ref%
{relativentr}) can be extended to that of relative R\'{e}nyi entropy (for a
review, see section II.E.3b in \cite{nishioka18}). For a system QGIFs with
two density matrices $\ \underline{\rho }_{\mathcal{A}}$ and $\underline{%
\sigma }_{\mathcal{A}},$ we introduce
\begin{eqnarray}
\ _{r}\underline{\mathcal{S}}(\ \underline{\rho }_{\mathcal{A}%
}\shortparallel \ \underline{\sigma }_{\mathcal{A}}) &=&\frac{1}{r-1}\log %
\left[ tr\left( (\ \underline{\sigma }_{\mathcal{A}})^{(1-r)/2r}\ \underline{%
\rho }_{\mathcal{A}}(\ \underline{\sigma }_{\mathcal{A}})^{(1-r)/2r}\right)
^{r}\right] ,\mbox{ for }r\in (0,1)\cup (1,\infty );  \label{relatrenyi} \\
\mbox{ or }\ _{1}\underline{\mathcal{S}}(\ \underline{\rho }_{\mathcal{A}%
}\shortparallel \ \underline{\sigma }_{\mathcal{A}}) &=&\ \underline{%
\mathcal{S}}(\ \underline{\rho }_{\mathcal{A}}\shortparallel \ \underline{%
\sigma }_{\mathcal{A}})\mbox{ and }\ _{\infty }\underline{\mathcal{S}}(\
\underline{\rho }_{\mathcal{A}}\shortparallel \ \underline{\sigma }_{%
\mathcal{A}})=\log ||(\ \underline{\sigma }_{\mathcal{A}})^{-1/2}\
\underline{\rho }_{\mathcal{A}}(\ \underline{\sigma }_{\mathcal{A}%
})^{-1/2}||_{\infty }.  \notag
\end{eqnarray}%
In any point of causal curves, we prove certain monotonic properties, $\ _{r}%
\underline{\mathcal{S}}(\ \underline{\rho }_{\mathcal{A}}\shortparallel \
\underline{\sigma }_{\mathcal{A}})\geq \ _{r}\underline{\mathcal{S}}(tr_{s}\
\underline{\rho }_{\mathcal{A}}|tr_{s}\ \ \underline{\sigma }_{\mathcal{A}})$
and $\partial _{r}[\ _{r}\underline{\mathcal{S}}(\ \underline{\rho }_{%
\mathcal{A}}\shortparallel \ \underline{\sigma }_{\mathcal{A}})]\geq 0,$ and
to reduce the relative R\'{e}nyi entropy to the R\'{e}nyi entropy using \
formula $\ _{r}\underline{\mathcal{S}}(\underline{\rho }_{\mathcal{A}%
}\shortparallel 1_{\mathcal{A}}/k_{\mathcal{A}})=\log k_{\mathcal{A}}-\ \
_{r}\underline{\mathcal{S}}(\underline{\mathcal{A}}).$

The values (\ref{relatrenyi}) do not allow a naive generalization of the
concept of mutual information and interpretation as an entanglement measure
of quantum information. There are possible negative values of relative R\'{e}%
nyi entropy for $r\neq 1$ \cite{adesso12}. This problem can be solved if it
is introduced the concept of the $r$-R\'{e}nyi mutual information \cite%
{beigi13},
\begin{equation*}
\ _{r}\underline{\mathcal{J}}(\underline{\mathcal{A}},\underline{\mathcal{B}}%
)\ :=\min_{\ ^{\shortmid }\sigma _{\mathcal{B}}}\ _{r}\underline{\mathcal{S}}%
(\ \underline{\rho }_{\mathcal{A\cup B}}\shortparallel \ \underline{\rho }_{%
\mathcal{A}}\otimes \underline{\sigma }_{\mathcal{B}})\geq 0
\end{equation*}%
for the minimum is taken over all $\underline{\sigma }_{\mathcal{B}}.$ We
obtain the standard definition of mutual information for $r=1.$ In result,
we can elaborate a self--consistent geometric--information thermodynamic
theory for KK QGIFs.

\section{Decoupling and integrability of KK cosmological flow equations}

\label{s6a} In this section, we prove that the system of nonlinear PDEs for
geometric evolution of KK gravity (\ref{canhamiltevol}) (and self-similar
configurations defining nonholonomic EM equations as Ricci soliton equations
(\ref{soliteinst})) can be decoupled and integrated in general forms. Such
solutions are defined by generic off-diagonal metrics $\mathbf{g}(\tau )$ (%
\ref{decomp31}) (for canonical d-connections, $\widehat{\mathbf{D}}(\tau ),$
and, in particular, for LC-configurations, $\nabla (\tau )$). The
coefficients of respective geometric objects depend on all spacetime
coordinates via generating and integration functions and (effective) matter
sources $\ ^{tot}\Upsilon _{\alpha \beta }(\tau )$ (\ref{totsourc}) via
nonholonomic entropic deformations of the energy-momentum tensor (\ref{emt}%
). We apply and develop the AFDM \cite%
{vacaru18tc,bubuianu18,bubuianu19,vacaru19a,gheorghiu16}.

\subsection{Quasiperiodic spacetime and quasicrystal like KK configurations}

We study two examples of space and time quasiperiodic structures defined in
a curved spacetime following our works on quasicrystal, QC, models in modern
cosmology \cite{vacaru18tc,bubuianu18,amaral17,aschheim18}. Here we cite
other type QC models studied in \cite{ghosh}. Our approach to locally
anisotropic cosmology and dark energy and dark matter physics with spacetime
quasicristal, STQC, structures is based on F. Wilczek and co-authors ideas
originally elaborated in condensed matter physics \cite%
{shapere12,wilczek12,wilczek13a,shapere17}.

\begin{itemize}
\item 1-d relativistic time QC structures are deformed by a scalar field $%
\varsigma (x^{i},y^{a})$ and respective Lagrange density on a space-time $(%
\mathbf{V,g,N),}$
\begin{equation}
\acute{L}(\varsigma )=\frac{1}{48}(\mathbf{g}^{\alpha \beta }(\mathbf{e}%
_{\alpha }\varsigma )(\mathbf{e}_{\beta }\varsigma ))^{2}-\frac{1}{4}\mathbf{%
g}^{\alpha \beta }(\mathbf{e}_{\alpha }\varsigma )(\mathbf{e}_{\beta
}\varsigma )-\acute{V}(\varsigma ),  \label{1tcqc}
\end{equation}%
resulting in N-adapted variational motion equations $\ [\frac{1}{2}\mathbf{g}%
^{\alpha \beta }(\mathbf{e}_{\alpha }\varsigma )(\mathbf{e}_{\beta
}\varsigma )-1](\mathbf{D}^{\gamma }\mathbf{D}_{\gamma }\varsigma )=2\frac{%
\partial \acute{V}}{\partial \varsigma },$ where $\mathbf{e}_{\alpha }$ are
N-adapted partial derivatives. \ In these formulas, $\acute{V}(\varsigma )$
is a nonlinear potential when the field $\varsigma $ defines a 1-d time QC
structure, 1-TQC, as a solution of motion equations.\footnote{%
There are used brief notations for partial derivatives when, for instance, $%
\partial q/\partial x^{i}=\partial _{i}q,$ $\partial q/\partial
y^{3}=\partial _{3}q=q^{\ast },$ and $\partial q/\partial y^{4}=\partial
_{4}q=\partial _{t}q=q^{\star },$ for a function $q(x^{i},y^{3},t).$ For
non-relativistic limits, we can consider $g_{\alpha \beta }=[1,1,1,-1]$ and $%
\varsigma \rightarrow \varsigma (t),$ $\acute{L}\rightarrow \frac{1}{12}%
(\varsigma ^{\bullet })^{4}-\frac{1}{2}(\varsigma ^{\bullet })^{2}-\acute{V}%
(\varsigma ),$ which relates (\ref{1tcqc}) to an effective energy $E=\frac{1%
}{4}[(\varsigma ^{\bullet })^{2}-1]^{2}+\acute{V}(\varsigma )-\frac{1}{4}$
derived for the motion equations $[(\varsigma ^{\bullet })^{2}-1]\varsigma
^{\bullet \bullet }=-\frac{\partial \acute{V}}{\partial \varsigma }$ \
introduced in \cite{shapere17}.}

\item 3-d QC structures on curved spaces are constructed as analogous
dynamic phase field crystal models with flow evolution on real parameter ${%
\tau}$. Such a QC structure is determined by a generating function $\
\overline{q}=\overline{q}(x^{i},y^{3},\tau )$ as a solution of an evolution
equation with conserved dynamics,
\begin{equation}
\frac{\partial \overline{b}}{\partial \tau }=\widehat{\Delta }\left[ \frac{%
\delta \overline{F}}{\delta \overline{b}}\right] =-\widehat{\Delta }(\Theta
\overline{b}+Q\overline{b}^{2}-\overline{b}^{3}).  \label{evoleq}
\end{equation}%
Such evolution equations can be considered on any 3-d spacelike hypersurface
$\Xi _{t}$ with a canonically nonholonomically deformed hypersurface Laplace
operator $\ \widehat{\Delta }:=(\widehat{\mathbf{D}})^{2}=q^{\grave{\imath}%
\grave{j}}\ \widehat{\mathbf{D}}_{\grave{\imath}}\ \widehat{\mathbf{D}}_{%
\grave{j}},$ for $\grave{\imath},\grave{j},...1,2,3.$ This operator is a
distortion of $\ \Delta :=(\nabla )^{2}$ constructed in 3-d Riemannian
geometry.\footnote{%
In \cite{bubuianu18,vacaru18tc}, \ we explain all details on such
quasiperiodic structures. Here we also note that the functional $\overline{F}
$ in (\ref{evoleq}) is characterized by an effective free energy $\
\overline{F}[\overline{q}]=\int \left[ -\frac{1}{2}\overline{b}\Theta
\overline{b}-\frac{Q}{3}\overline{b}^{3}+\frac{1}{4}\overline{b}^{4}\right]
\sqrt{q}dx^{1}dx^{2}\delta y^{3},$ where $q=\det |q_{\grave{\imath}\grave{j}%
}|,\delta y^{3}=\mathbf{e}^{3}$ and the operators $\Theta $ and $Q$ are
chosed in certain forms when nonlinear interactions are stabilized by the
cubic term with $Q$ and the second order resonant interactions. An average
value $<\overline{b}>$ is conserved for any fixed time variable $t$ and/or
evolution parameter $\tau _{0}$ (we can fix $<\overline{b}>_{|\tau =\tau
_{0}}=0$ by redefining $\Theta $ and $Q).$}
\end{itemize}

\subsection{KK flows with entropic (elastic) configurations}

\label{sselastic}Various models of modified emergent/entropic gravity can be
derived from nonholonomic modifications of G. Perelman functionals \cite%
{vacaru19,vacaru19a}. In this section, we re-define the constructions for KK
GIFs and nonholonomic Ricci soliton equations. Using the canonical
d--connection $\widehat{\mathbf{D}},$ we introduce three "hat" values:
\begin{eqnarray*}
\widehat{\varepsilon }_{\alpha \beta } &=&\widehat{\mathbf{D}}_{\alpha }%
\mathbf{u}_{\beta }-\widehat{\mathbf{D}}_{\beta }\mathbf{u}_{\alpha }%
\mbox{\
- the elastic strain tensor };\phi =u/\sqrt{|\Lambda |}%
\mbox{ - a
dimensionless scalar }; \\
\widehat{\chi } &=&\alpha (\widehat{\mathbf{D}}_{\mu }\mathbf{u}^{\mu })(%
\widehat{\mathbf{D}}_{\nu }\mathbf{u}^{\nu })+\beta (\widehat{\mathbf{D}}%
_{\mu }\mathbf{u}_{\nu })(\widehat{\mathbf{D}}^{\mu }\mathbf{u}^{\nu
})+\gamma (\widehat{\mathbf{D}}_{\mu }\mathbf{u}_{\nu })(\widehat{\mathbf{D}}%
^{\nu }\mathbf{u}^{\mu })\mbox{ - a general kinetic term for }\mathbf{u}%
^{\mu }.
\end{eqnarray*}%
In these formulas, we consider a conventional displacement vector field $%
\mathbf{u}^{\alpha },$ cosmological constant $\Lambda $ and some constants $%
\alpha ,\beta ,\gamma ;$ there are used short hand notations: $u:=\sqrt{|%
\mathbf{u}_{\alpha }\mathbf{u}^{\alpha }|},\widehat{\varepsilon }=\widehat{%
\varepsilon }_{\beta }^{\beta },$ and $\mathbf{n}^{\alpha }:=\mathbf{u}%
^{\alpha }/u.$

The Lagrange densities for gravitational and electromagnetic fileds for
models of emergent KK\ gravity with nonholonomic distributions determined by
quasiperiodic geometric flow evolution (or self-similar cofigurations) on $%
\mathbf{V}$ are chosen in the form
\begin{equation}
\ ^{tot}\mathcal{L}=\ ^{g}\mathcal{L+}\ ^{m}\mathcal{L+}\ ^{int}\mathcal{L+}%
\ ^{\chi }\mathcal{L},\mbox{ for }\ ^{g}\mathcal{L}=M_{P}^{2}\ ^{s}\widehat{R%
},\ ^{int}\mathcal{L}=-\sqrt{|\Lambda |}\ \widehat{\mathbf{T}}_{\mu \nu }%
\mathbf{u}^{\mu }\mathbf{u}^{\nu }/u,\ ^{\chi }\mathcal{L=}M_{P}^{2}|\Lambda
|(\chi ^{3/2}+|\Lambda ||u[\varsigma ,\overline{b}]|^{2z}),  \label{lagrs}
\end{equation}%
when the Plank gravitational mass is denoted $M_{P}$ and the gravitational
Lagrangian $\ ^{g}\mathcal{L}$ is determined by the Ricci scalar $\ ^{s}%
\widehat{R}$ of $\widehat{\mathbf{D}}$ and $\widehat{\mathbf{T}}_{\mu \nu }$
is the electromagnetic energy momentum tensor (\ref{emt}) determined by $\
^{m}\mathcal{L}(\mathbf{g},\widehat{\mathbf{D}},\mathbf{A}),$ with $\widehat{%
\mathbf{D}}$ used instead of $\nabla .$\footnote{%
On $\mathbf{V}$, the curve space covariant derivative is computed using both
$\widehat{\mathbf{D}}$ and $\mathbf{A.}$ We can construct KK GIF models with
MGTs as in \cite{nojiri17,capozziello,hossain15,bubuianu18} \ when the
gravitational Lagrangian is a functional $f$ of $\ ^{s}R,$ $\ ^{g}\mathcal{L}%
=M_{P}^{2}f(\ ^{s}R),$ and a more general energy momentum tensor (with
scalar field interactions ) \ $\ \ ^{m}\widehat{\mathbf{T}}_{\mu \nu }$
instead of $\ \widehat{\mathbf{T}}_{\mu \nu }.$ We can fix $z=1$ if we
search for compatibility with relativistic entropic gravity models \cite%
{hossenfelder17}, or $z=2$ if we search for a limit to the standard de
Sitter space solution \cite{dai17a,dai17b}, see details in \cite%
{vacaru19,vacaru19a}. We can model by GIFs certain STQC structures, for
instance as in entropic gravity if we consider that the displacement vector
field $\mathbf{u}^{\alpha }[\varsigma ,\overline{b}]$ as a functional of
functions $\varsigma ,\overline{b}$ subjected to certain conditions of type (%
\ref{1tcqc}) and/or (\ref{evoleq}). More general KK GIF cosmological
scenarious can be elaborated if we consider functionals for pattern forming,
nonlinear wave soliton structures, fractional and diffusion processes etc.,
see details in \cite{vacaru18tc,bubuianu18,amaral17,aschheim18}.}

For the full system of Lagrangians in (\ref{lagrs}), the effective
energy-momentum tensor is computed $\ \ ^{tot}\widehat{\mathbf{T}}_{\mu \nu
}=\widehat{\mathbf{T}}_{\mu \nu }+\ ^{int}\widehat{\mathbf{T}}_{\mu \nu }+\
^{\chi }\widehat{\mathbf{T}}_{\mu \nu }.$ We can model "pure" elastic
spacetime modifications of the Einstein gravity if we consider $\mathbf{%
D=\nabla ,}$ which results in respective formulas (10)-(13) in \cite{dai17a}
$\ $(for$\ ^{int}\mathbf{T}_{\mu \nu }$ and $\ ^{\chi }\mathbf{T}_{\mu \nu
}).$ \ In this section, we work with a generalized (effective) source
splitting into respective three components,%
\begin{equation}
\ ^{tot}\widehat{\mathbf{\Upsilon }}_{\mu \nu }:=\varkappa \left( \ ^{tot}%
\widehat{\mathbf{T}}_{\mu \nu }-\frac{1}{2}\mathbf{g}_{\mu \nu }\ ^{tot}%
\widehat{\mathbf{T}}\right) =\ \widehat{\mathbf{\Upsilon }}_{\mu \nu }+\
^{int}\widehat{\mathbf{\Upsilon }}_{\mu \nu }+\ ^{\chi }\widehat{\mathbf{%
\Upsilon }}_{\mu \nu },  \label{totsourc}
\end{equation}%
where $\varkappa $ is determined in standard form by the Newton
gravitational constant $G.$

\subsection{Parametric modified Einstein equations encoding GIFs}

The geometric flow equations can be written in nonholonomic canonical
variable (\ref{canhamiltevol}) as modified Einstein equations with
dependence of geometric objects and sources on a temperature like parameter $%
\tau .$ Such systems of nonlinear PDEs can be decoupled and integrated in
general forms for various classes of effective sources, see proofs and
examples in \cite{vacaru18tc,bubuianu18,bubuianu19,vacaru19a,gheorghiu16}.

In this work, we study KK GIFs determined by two effective generating
sources $\ _{h}\widehat{Y}(\tau ,{x}^{k})$ and $\widehat{Y}(\tau
,x^{k},y^{c})$ prescribing total sources of type (\ref{totsourc}) up to
frame transforms with respective tetradic fields $\mathbf{e}_{\ \mu ^{\prime
}}^{\mu }(\tau )=\mathbf{e}_{\ \mu ^{\prime }}^{\mu }(\tau ,u^{\gamma })$
and their dual $\mathbf{e}_{\nu }^{\ \nu ^{\prime }}(\tau )=\mathbf{e}_{\nu
}^{\ \nu ^{\prime }}(\tau ,u^{\gamma }).$ The effective sources are
parameterized in such a general N--adapted form,
\begin{equation}
\ ^{e}\widehat{Y}_{\mu \nu }(\tau )=\mathbf{e}_{\ \mu }^{\mu ^{\prime
}}(\tau )\mathbf{e}_{\nu }^{\ \nu ^{\prime }}(\tau )[~\ ^{tot}\widehat{%
\mathbf{\Upsilon }}_{\mu ^{\prime }\nu ^{\prime }}(\tau )+\frac{1}{2}%
~\partial _{\tau }\mathbf{g}_{\mu ^{\prime }\nu ^{\prime }}(\tau )]=[~\ _{h}%
\widehat{Y}(\tau ,{x}^{k})\delta _{j}^{i},\widehat{Y}(\tau
,x^{k},y^{c})\delta _{b}^{a}].  \label{effsourc}
\end{equation}%
In result, the system of nonholonomic R. Hamilton equations for KK GIFs with
encoding entropic modifications (see, for instance, (\ref{cankkhamiltevol})
and their equivalent form (\ref{canhamiltevol})) can be written in a
generalized Einstein form with an effective source (\ref{effsourc}),%
\begin{equation}
\widehat{\mathbf{R}}_{\alpha \beta }(\tau )=\ ^{e}\widehat{Y}_{\alpha \beta
}(\tau ).  \label{entropfloweq}
\end{equation}
Such systems of nonlinear PDEs equations are for an undetermined
normalization function $\widehat{f}(\tau )=\ \widehat{f}(\tau ,u^{\gamma})$
which can be defined explicitly for respective classes of exact or
parametric solutions with prescribed generating functions and (effective)
sources. The generating functions $\ _{h}\widehat{Y}(\tau )$ and $\widehat{Y}%
(\tau )$ for $\ ^{e}\widehat{Y}_{\alpha \beta }(\tau )$ (\ref{effsourc}) can
be considered as certain prescribed data for (effective) matter sources.
Such values impose certain nonholonomic frame constraints on geometric
evolution and self-similar configurations. This also prescribe the type of
GIF entropic corrections and quasiperiodic configuration.

For locally anisotropic cosmological configurations with coordinates $%
(x^{k},y^{4}=t),$ the $\tau $-evolution of d--metric $\mathbf{g}(\tau )$ (%
\ref{decomp31}) can be parameterized by N-adapted coefficients,
\begin{equation}
g_{i}(\tau )=e^{\psi {(\tau ,x^{k})}},\,\,\,\,g_{a}(\tau )=\omega ({\tau ,}%
x^{k},y^{b})\overline{h}_{a}({\tau ,}x^{k},t),\ N_{i}^{3}(\tau )=\overline{n}%
_{i}({\tau ,}x^{k},t),\,\,\,\,N_{i}^{4}(\tau )=\overline{w}_{i}({\tau ,}%
x^{k},t),\mbox{ for }\omega =1.  \label{cosmf}
\end{equation}%
For simplicity, we shall construct cosmological evolution models determined
by nonholonomic configurations possessing at least one Killing symmetry, for
instance, on $\partial _{3}=\partial _{\varphi }.$ Here we note that
different types of ansatz and parameterizations of d-metrics and effective
sources (\ref{effsourc}) are considered, for instance, for generating
stationary solutions, see \cite{bubuianu18,bubuianu19,vacaru19a,gheorghiu16}.

\subsection{Decoupling of quasiperiodic KK cosmological flow equations}

\subsubsection{d-metric ansatz and sources with decoupling}

Tables similar to Table 1 (see below) are provided in \cite%
{bubuianu18,vacaru19a} for certain general ansatz which allow constructing
exact and parametric solutions in MGTs and entropic gravity. In this
subsection, we introduce locally anisotropic cosmological ansatz for generic
off-diagonal metrics and effective sources when is possible to prove general
decoupling and integration properties of (modified) Einstein equations (\ref%
{entropfloweq}) with effective sources (\ref{effsourc}).

We model nonholonomic deformations of d-metrics using $\eta $-polarization
functions, $\mathbf{\mathring{g}\rightarrow g}(\tau ),$ of a 'prime' metric,
$\mathbf{\mathring{g}}$, into a family of 'target' d-metrics $\mathbf{g}%
(\tau )$ (\ref{decomp31}) for parameterizations
\begin{equation}
\mathbf{g}(\tau )=\eta _{i}(\tau ,x^{k})\mathring{g}_{i}dx^{i}\otimes
dx^{i}+\eta _{a}(\tau ,x^{k},y^{b})\mathring{h}_{a}\mathbf{e}^{a}[\eta
]\otimes \mathbf{e}^{a}[\eta ].  \label{dme}
\end{equation}%
In this and similar formulas, we do not consider summation on repeating
indices if they are not written as contraction of "up-low" ones. The target
N-elongated basis in (\ref{dme}) is determined by $N_{i}^{a}(\tau ,u)=\eta
_{i}^{a}(\tau ,x^{k},y^{b})\mathring{N}_{i}^{a}(\tau ,x^{k},y^{b}),$ when $%
\mathbf{e}^{\alpha }[\eta ]=(dx^{i},\mathbf{e}^{a}=dy^{a}+\eta _{i}^{a}%
\mathring{N}_{i}^{a}dx^{i})$ and $\eta _{i}(\tau )=\eta _{i}(\tau
,x^{k}),\eta _{a}(\tau )=\eta _{a}(\tau,x^{k},y^{b})$ and $\eta
_{i}^{a}(\tau )=\eta _{i}^{a}(\tau ,x^{k},y^{b})$ are called respectively
gravitational polarization functions, or $\eta $-polarizations. We consider
than any $\mathbf{g}(\tau )$ defines a solution of (\ref{entropfloweq}) even
a general prime metric $\mathbf{\mathring{g}}$ may be not a physically
important metric. For certain cosmological models, we can consider $\mathbf{%
\mathring{g}}$ as standard isotropic or anisotropic metric. Using a general
coordinate parametrization, a prime metric $\mathbf{\mathring{g}}=\mathring{g%
} _{\alpha \beta }(x^{i},y^{a})du^{\alpha }\otimes du^{\beta }$ can be also
represented equivalently in N-adapted form
\begin{eqnarray}
\mathbf{\mathring{g}} &=&\mathring{g}_{\alpha }(u)\mathbf{\mathring{e}}%
^{\alpha }\otimes \mathbf{\mathring{e}}^{\beta }=\mathring{g}%
_{i}(x)dx^{i}\otimes dx^{i}+\mathring{g}_{a}(x,y)\mathbf{\mathring{e}}%
^{a}\otimes \mathbf{\mathring{e}}^{a},  \label{primedm} \\
&&\mbox{ for }\mathbf{\mathring{e}}^{\alpha }=(dx^{i},\mathbf{e}^{a}=dy^{a}+%
\mathring{N}_{i}^{a}(u)dx^{i}),\mbox{ and }\mathbf{\mathring{e}}_{\alpha }=(%
\mathbf{\mathring{e}}_{i}=\partial /\partial y^{a}-\mathring{N}%
_{i}^{b}(u)\partial /\partial y^{b},\ {e}_{a}=\partial /\partial y^{a}).
\notag
\end{eqnarray}

%%%%%%%%%%%
% Table 1
%%%%%%%%%%%

%\vskip5pt
%\begin{table*}[h]
{\scriptsize
\begin{eqnarray*}
&&%
\begin{tabular}{l}
\hline\hline
\begin{tabular}{lll}
&  &  \\
& {\ \textsf{Table 1:Quasiperiodic cosmological flows \ }}ansatz \textit{for
constructing generic off-diagonal exact and parametric solutions} &  \\
& using the Anholonomic Frame Deformation Method, \textbf{AFDM}, &
\end{tabular}%
\end{tabular}
\\
&&{%
\begin{tabular}{lll}
\hline
diagonal ansatz: PDEs $\rightarrow $ \textbf{ODE}s &  & AFDM: \textbf{PDE}s
\textbf{with decoupling; \ generating functions} \\
radial coordinates $u^{\alpha }=(r,\theta ,\varphi ,t)$ & $u=(x,y):$ &
\mbox{  2+2
splitting, } $u^{\alpha }=(x^{1},x^{2},y^{3},y^{4}=t);$%
\mbox{  flow
parameter  }$\tau $ \\
LC-connection $\mathring{\nabla}$ & [connections] & $%
\begin{array}{c}
\mathbf{N}:T\mathbf{V}=hT\mathbf{V}\oplus vT\mathbf{V,}\mbox{ locally }%
\mathbf{N}=\{N_{i}^{a}(x,y)\} \\
\mbox{ canonical connection distortion }\mathbf{D}=\nabla +\mathbf{Z}%
\end{array}%
$ \\
$%
\begin{array}{c}
\mbox{ diagonal ansatz  }g_{\alpha \beta }(u) \\
=\left(
\begin{array}{cccc}
\mathring{g}_{1} &  &  &  \\
& \mathring{g}_{2} &  &  \\
&  & \mathring{g}_{3} &  \\
&  &  & \mathring{g}_{4}%
\end{array}%
\right)%
\end{array}%
$ & $\mathbf{\mathring{g}}\Leftrightarrow \mathbf{g}(\tau )$ & $%
\begin{array}{c}
g_{\alpha \beta }(\tau )=%
\begin{array}{c}
g_{\alpha \beta }(\tau ,x^{i},y^{a})\mbox{ general frames / coordinates} \\
\left[
\begin{array}{cc}
g_{ij}+N_{i}^{a}N_{j}^{b}h_{ab} & N_{i}^{b}h_{cb} \\
N_{j}^{a}h_{ab} & h_{ac}%
\end{array}%
\right] ,\mbox{ 2 x 2 blocks }%
\end{array}
\\
\mathbf{g}_{\alpha \beta }(\tau )=[g_{ij}(\tau ),h_{ab}(\tau )], \\
\mathbf{g}(\tau )=\mathbf{g}_{i}(\tau ,x^{k})dx^{i}\otimes dx^{i}+\mathbf{g}%
_{a}(\tau ,x^{k},y^{b})\mathbf{e}^{a}\otimes \mathbf{e}^{b}%
\end{array}%
$ \\
$\mathring{g}_{\alpha \beta }=\mathring{g}_{\alpha }(t)\mbox{ for FLRW }$ &
[coord.frames] & $g_{\alpha \beta }(\tau )=g_{\alpha \beta }(\tau ,r,\theta
,y^{4}=t)\mbox{ cosm. configurations}$ \\
&  &  \\
$%
\begin{array}{c}
\mbox{coord.tranfsorms }e_{\alpha }=e_{\ \alpha }^{\alpha ^{\prime
}}\partial _{\alpha ^{\prime }}, \\
e^{\beta }=e_{\beta ^{\prime }}^{\ \beta }du^{\beta ^{\prime }},\mathring{g}%
_{\alpha \beta }=\mathring{g}_{\alpha ^{\prime }\beta ^{\prime }}e_{\ \alpha
}^{\alpha ^{\prime }}e_{\ \beta }^{\beta ^{\prime }} \\
\mathbf{\mathring{g}}_{\alpha }(x^{k},y^{a})\rightarrow \mathring{g}_{\alpha
}(t),\mathring{N}_{i}^{a}(x^{k},y^{a})\rightarrow 0.%
\end{array}%
$ & [N-adapt. fr.] & $\left\{
\begin{array}{cc}
\mathbf{g}_{i}(\tau ,r,\theta ),\mathbf{g}_{a}(\tau ,r,\theta ,t), &
\mbox{
d-metrics } \\
N_{i}^{3}(\tau )=n_{i}(\tau ,r,\theta ,t),N_{i}^{4}=w_{i}(\tau ,r,\theta ,t),
&
\end{array}%
\right. $ \\
$\mathring{\nabla},$ $Ric=\{\mathring{R}_{\ \beta \gamma }\}$ & Ricci tensors
& $\widehat{\mathbf{D}},\ \widehat{\mathcal{R}}ic=\{\widehat{\mathbf{R}}_{\
\beta \gamma }\}$ \\
$~^{m}\mathcal{L[\mathbf{\phi }]\rightarrow }\mathbf{T}_{\alpha \beta }[F]$
& sources & $\ ^{e}\widehat{Y}_{\ \nu }^{\mu }(\tau )=diag[~\ _{h}\widehat{Y}%
(\tau ,x^{i})\delta _{j}^{i},\widehat{Y}(\tau ,x^{i},t)\delta _{b}^{a}]$ \\
trivial equations for $\mathring{\nabla}$-torsion & LC-conditions & $%
\widehat{\mathbf{D}}_{\mid \widehat{\mathcal{T}}\rightarrow 0}=\mathbf{%
\nabla }$ \\ \hline\hline
\end{tabular}%
}
\end{eqnarray*}%
}

%\caption{Off-diagonal ansatz, connections and sources for the AFDM}
%\label{tb1par}
%\end{table*}

In this paper, we study physically important cases when $\mathbf{\mathring{g}%
}$ is a Friedman--Lema\^{\i}tre--Robertson--Walker (FLRW) type metric, or a
Bianchi anisotropic metrics. For diagonalizable prime metrics, we can always
find a coordinate system when $\mathring{N}_{i}^{b}=0.$ Nevertheless, it is
convenient to construct exact solutions with nontrivial functions $\eta
_{\alpha }=(\eta _{i},\eta _{a}),\eta _{i}^{a},$ and nonzero coefficients $%
\mathring{N}_{i}^{b}(u)$ in order to avoid constructions with nonholonomic
deformations for singular coordinates.

\subsubsection{Cosmological canonical Ricci d-tensors, nonlinear symmetries
and LC-conditions}

For simplicity, we study locally anisotropic cosmological flows when the
coefficients of the geometric objects do not depend on the space coordinate $%
y^{3},$ i.e. using d-metric data (\ref{cosmf}) with $\omega =1.$ The GIF
entropic modified Einstein equations (\ref{entropfloweq}) are parameterized%
\begin{eqnarray}
\mathbf{R}_{1}^{1}(\tau ) &=&\mathbf{R}_{2}^{2}(\tau )=-\ \ _{h}\widehat{Y}%
(\tau )\mbox{
i.e. }\frac{g_{1}^{\bullet }g_{2}^{\bullet }}{2g_{1}}+\frac{\left(
g_{2}^{\bullet }\right) ^{2}}{2g_{2}}-g_{2}^{\bullet \bullet }+\frac{%
g_{1}^{\prime }g_{2}^{\prime }}{2g_{2}}+\frac{(g_{1}^{\prime })^{2}}{2g_{1}}%
-g_{1}^{\prime \prime }=-2g_{1}g_{2}\ _{h}\widehat{Y};  \label{eq1b} \\
\mathbf{R}_{3}^{3}(\tau ) &=&\mathbf{R}_{4}^{4}(\tau )=-\widehat{Y}(\tau )%
\mbox{  i.e. }\frac{\left( \overline{h}_{3}^{\ast }\right) ^{2}}{2\overline{h%
}_{3}}+\frac{\overline{h}_{3}^{\ast }\overline{h}_{4}^{\ast }}{2\overline{h}%
_{4}}-\overline{h}_{3}^{\ast \ast }=-2\overline{h}_{3}\overline{h}_{4}%
\widehat{Y};  \label{eq2b}
\end{eqnarray}%
\begin{eqnarray}
\mathbf{R}_{3k}(\tau ) &=&\frac{\overline{h}_{3}}{2\overline{h}_{3}}%
\overline{n}_{k}^{\ast \ast }+(\frac{3}{2}\overline{h}_{3}^{\ast }-\frac{%
\overline{h}_{3}}{\overline{h}_{4}}\overline{h}_{4}^{\ast })\frac{\overline{n%
}_{k}^{\ast }}{2\overline{h}_{4}}=0;  \label{eq3b} \\
2\overline{h}_{3}\mathbf{R}_{4k}(\tau ) &=&-\overline{w}_{k}[\frac{\left(
\overline{h}_{3}^{\ast }\right) ^{2}}{2\overline{h}_{3}}+\frac{\overline{h}%
_{3}^{\ast }\overline{h}_{4}^{\ast }}{2\overline{h}_{4}}-\overline{h}%
_{3}^{\ast \ast }]+\frac{\overline{h}_{3}^{\ast }}{2}(\frac{\partial _{k}%
\overline{h}_{3}}{\overline{h}_{3}}+\frac{\partial _{k}\overline{h}_{4}}{%
\overline{h}_{4}})-\partial _{k}\overline{h}_{3}^{\ast }=0.  \label{eq4b}
\end{eqnarray}

This system of nonlinear PDE (\ref{eq1b})--(\ref{eq4b}) can be written in an
explicit decoupled form. Let us consider $\overline{\alpha }_{i}=(\partial
_{t}\overline{h}_{3})\ (\partial _{i}\overline{\varpi }),\ \overline{\beta }%
=(\partial _{t}\overline{h}_{3})\ (\partial _{t}\overline{\varpi }),\
\overline{\gamma }=\partial _{t}\left( \ln |\overline{h}_{3}|^{3/2}/|%
\overline{h}_{4}|\right) ,$ where $\overline{\varpi }:={\ln |\partial _{t}}%
\overline{{h}}{_{3}/\sqrt{|\overline{h}_{3}\overline{h}_{4}|}|}$ is
considered as a generating function. In this work, we overline certain
coefficients for cosmological evolution following notations from \cite%
{vacaru19a}. For configurations with $\partial _{t}h_{a}\neq 0$ and $%
\partial _{t}\varpi \neq 0,$ we obtain such equations
\begin{equation}
\psi ^{\bullet \bullet }+\psi ^{\prime \prime }=2\ _{h}\widehat{Y};{%
\overline{\varpi }}^{\ast }\ \overline{h}_{3}^{\ast }=2\overline{h}_{3}%
\overline{h}_{4}\widehat{Y};\ \overline{n}_{i}^{\ast \ast }+\overline{\gamma
}\overline{n}_{i}^{\ast }=0;\overline{\beta }\overline{w}_{i}-\overline{%
\alpha }_{i}=0.  \label{cosmsimpl}
\end{equation}%
These equations can be integrated "step by step" for any (redefined)
generating function $\overline{\Psi }(\tau ,x^{i},t):=e^{\overline{\varpi }}$
and sources $\ _{h}\widehat{Y}(\tau ,x^{i})$ and $\widehat{Y}(\tau ,x^{k},t).
$

The system (\ref{cosmsimpl}) imposes certain conditions of nonlinear
symmetric on four functions ($\overline{h}_{3},\overline{h}_{4},\widehat{Y},%
\overline{\Psi }).$ We can redefine respective generating functions, $(%
\overline{\Psi }(\tau ),\widehat{Y}(\tau ))\iff (\overline{\Phi }(\tau ),%
\overline{\Lambda }(\tau ))$ if there are satisfied the equations
\begin{equation}
\overline{\Lambda }(\ \overline{\Psi }^{2})^{\ast }=|\widehat{Y}|(\overline{%
\Phi }^{2})^{\ast },\mbox{
or  }\overline{\Lambda }\overline{\ \Psi }^{2}=\overline{\Phi }^{2}|\widehat{%
Y}|-\int dt\ \overline{\Phi }^{2}|\widehat{Y}|^{\ast }.  \label{nsym1b}
\end{equation}
The vertical flow of effective cosmological constants $\overline{\Lambda }$
can be identified to certain models with $\Lambda $ from (\ref{lagrs}) if we
consider nonholonomic deformations of some classes of (anti) de Sitter
solutions in GR or a MGT.

The LC-conditions for zero torsion cosmological GIFs transform into a system
of 1st order PDEs with coefficients depending on $\tau$ and $t,$
\begin{equation}
\partial _{t}\overline{w}_{i}=(\partial _{i}-\overline{w}_{i}\partial
_{t})\ln \sqrt{|\overline{h}_{4}|},(\partial _{i}-\overline{w}_{i}\partial
_{t})\ln \sqrt{|\overline{h}_{3}|}=0,\partial _{k}\overline{w}_{i}=\partial
_{i}\overline{w}_{k},\partial _{t}\overline{n}_{i}=0,\partial _{i}\overline{n%
}_{k}=\partial _{k}\overline{n}_{i}.  \label{lccondit}
\end{equation}%
Such systems can be solved in explicit form for certain classes of
additional nonholonomic constraints on cosmological d--metrics and
N-coefficients, see (\ref{cosmf}).

\subsection{Integrability of KK quasiperiodic cosmological flow equations}

Integrating "step by step" the system of the \ nonlinear PDEs (\ref{eq1b})--(%
\ref{eq4b}) decoupled in the form (\ref{cosmsimpl}) (see similar details and
proofs in \cite{vacaru18tc,bubuianu18,bubuianu19,vacaru19a,gheorghiu16}), we
obtain such d--metric coefficients for (\ref{decomp31}) and/or (\ref{dme}),
\begin{equation}
\ g_{i}(\tau ) =e^{\ \psi (\tau ,x^{k})}%
\mbox{ as a solution of 2-d
Poisson eqs. }\psi ^{\bullet \bullet }+\psi ^{\prime \prime }=2~\ _{h}%
\widehat{Y}(\tau );  \notag
\end{equation}
\begin{eqnarray}
g_{3}(\tau ) &=&\overline{h}_{3}(\tau ,{x}^{i},t)=h_{3}^{[0]}(\tau
,x^{k})-\int dt\frac{(\overline{\Psi }^{2})^{\ast }}{4\widehat{Y}}%
=h_{3}^{[0]}(\tau ,x^{k})-\overline{\Phi }^{2}/4\overline{\Lambda }(\tau );
\notag \\
g_{4}(\tau ) &=&\overline{h}_{4}(\tau ,{x}^{i},t)=-\frac{(\overline{\Psi }%
^{2})^{\ast }}{4\widehat{Y}^{2}\overline{h}_{3}}=-\frac{(\overline{\Psi }%
^{2})^{\ast }}{4\overline{\widehat{Y}}^{2}(h_{3}^{[0]}(\tau ,x^{k})-\int dt(%
\overline{\Psi }^{2})^{\ast }/4\widehat{Y})}  \label{offdcosm} \\
&=&-\frac{[(\overline{\Phi }^{2})^{\ast }]^{2}}{4\overline{h}_{3}|\overline{%
\Lambda }(\tau )\int dt\widehat{Y}[\overline{\Phi }^{2}]^{\ast }|}=-\frac{[(%
\overline{\Phi }^{2})^{\ast }]^{2}}{4[h_{3}^{[0]}(x^{k})-\overline{\Phi }%
^{2}/4\overline{\Lambda }(\tau )]|\int dt\ \widehat{Y}[\overline{\Phi }%
^{2}]^{\ast }|}.  \notag
\end{eqnarray}%
Such d-metric coefficients are computed with respect to N-adapted bases
determined by N--connection coefficients computed by formulas
\begin{eqnarray}
N_{k}^{3}(\tau ) &=&\overline{n}_{k}(\tau ,{x}^{i},t)=\ _{1}n_{k}(\tau
,x^{i})+\ _{2}n_{k}(\tau ,x^{i})\int dt\frac{(\overline{\Psi }^{\ast })^{2}}{%
\widehat{Y}^{2}|h_{3}^{[0]}(\tau ,x^{i})-\int dt(\overline{\Psi }^{2})^{\ast
}/4\widehat{Y}|^{5/2}}  \notag \\
&=&\ _{1}n_{k}(\tau ,x^{i})+\ _{2}n_{k}(\tau ,x^{i})\int dt\frac{(\overline{%
\Phi }^{\ast })^{2}}{4|\overline{\Lambda }(\tau )\int dt\widehat{Y}[%
\overline{\Phi }^{2}]^{\ast }||\overline{h}_{3}|^{5/2}};  \label{nconcosm} \\
N_{i}^{4}(\tau ) &=&\overline{w}_{i}(\tau ,{x}^{i},t)=\frac{\partial _{i}\
\overline{\Psi }}{\ \overline{\Psi }^{\ast }}=\frac{\partial _{i}\ \overline{%
\Psi }^{2}}{(\overline{\Psi }^{2})^{\ast }}\ =\frac{\partial _{i}[\int dt\
\widehat{Y}(\ \overline{\Phi }^{2})^{\ast }]}{\widehat{Y}(\ \overline{\Phi }%
^{2})^{\ast }},\ \   \notag
\end{eqnarray}
There are considered different classes of functions in these formulas. The
values $h_{3}^{[0]}(\tau ,x^{k}),$ $\ _{1}n_{k}(\tau ,x^{i}),$ and $\
_{2}n_{k}(\tau ,x^{i})$ are integration functions encoding various possible
sets of (non) commutative parameters and integration constants (in general,
such constants run on a temperature like parameter $\tau $ for geometric
evolution flows). We can work equivalently with different generating data $(%
\overline{\Psi }(\tau ,{x}^{i},t),\widehat{Y}(\tau ,{x}^{i},t))$ or $(%
\overline{\Phi }(\tau ,{x}^{i},t),\overline{\Lambda }(\tau ))$ related via
nonlinear symmetries. These values are related by nonlinear differential /
integral transforms (\ref{nsym1b}), and respective integration functions. In
explicit form, such nonlinear symmetries involve certain topology/ symmetry
/ asymptotic conditions for some classes of exact / parametric cosmological
solutions.

With respect to coordinate \ frames, the coefficients (\ref{offdcosm}) and (%
\ref{nconcosm}) define generic off-diagonal cosmological solutions if the
corresponding anholonomy coefficients are not trivial. Such locally
cosmological solutions can not be diagonalized by coordinate transforms in a
finite spacetime region and its geometric flow evolutions. For the canonical
d-connection, the cosmological d-metrics are with nontrivial
nonholonomically induced d-torsion and N-adapted coefficients which can be
computed in explicit form if there are prescribed certain evolution
conditions, boundary values etc. We can generate as particular cases some
well-known cosmological FLRW, or Bianchi, type metrics if there are
considered data of type $(\overline{\Psi }(\tau ,t),\widehat{Y}(\tau ,t)),$
or $(\overline{\Phi }(\tau ,t),\overline{\Lambda }(\tau )),$ determined by
special classes of integration functions and certain frame/ coordinate
transforms to respective (off-) diagonal configurations $g_{\alpha \beta
}(\tau ,t).$ For cosmological Ricci soliton configurations, we can fix $\tau
=\tau _{0}.$

Let us discuss some nonholonomic transform and evolution properties of above
locally anisotropic cosmological solutions determined by effective
cosmological parameterizations sources (\ref{effsourc}). We have a system of
equations involving the evolution derivative $\partial _{\tau }.$ The v-part
of cosmologically vierbeinds under geometric flow depend on a time like
coordinate $y^{4}=t,$ with $\ \mathbf{e}_{\ \mu }^{\mu ^{\prime }}(\tau)=[%
\mathbf{e}_{\ 1}^{1^{\prime }}(\tau ,x^{k}),\mathbf{e}_{\
2}^{2^{\prime}}(\tau ,x^{k}),\mathbf{e}_{\ 3}^{3^{\prime }}(\tau ,x^{k},t),%
\mathbf{e}_{\ 4}^{4^{\prime }}(\tau ,x^{k},t)].$ There are considered also
coordinates $(x^{k},t),$ when the dependence on $y^{3}$ can be omitted for
configurations with Killing symmetry on $\partial _{3}.$ In the horizontal
part, the frame transforms for generating effective sources are of type $\
^{e}\widehat{Y}_{i}(\tau )=[\mathbf{e}_{\ i}^{i^{\prime }}(\tau )]^{2})[~\
^{tot}\widehat{Y}_{i^{\prime }1^{\prime }}(\tau )+\frac{1}{2}~\partial
_{\tau }\mathbf{g}_{1^{\prime }}(\tau )]=\ _{h}\widehat{Y}(\tau ,{x}^{k}).$
Similar transforms can be performed for v-components, $\ ^{e}\widehat{Y}%
_{a}(\tau )=[\mathbf{e}_{\ a}^{a^{\prime }}(\tau )]^{2})[~\ ^{tot}\widehat{Y}%
_{a^{\prime }a^{\prime }}(\tau )+\frac{1}{2}~\partial _{\tau }\mathbf{g}%
_{a^{\prime }}(\tau )]=\widehat{Y}(\tau ,{x}^{k},t).$ For generating locally
anisotropic cosmological solutions, we can prescribe any values for the
matter sources $\ ^{tot}\widehat{Y}_{\mu \nu }(\tau )$ in a geometric flow
model with cosmological evolution and quasiperiodic spacetime structure.
Considering N-adapted diagonal configurations and integrating on $\tau ,$ we
can compute a cosmological evolution flows of $\mathbf{g}_{\alpha ^{\prime
}}(\tau ,{x}^{k},t)$ encoding, for instance, nonholonomic and nonlinear
geometric diffusion process \cite{vacaru2000,vacaru2012}. Such geometric
constructions and physical models are performed with respect to a new system
of reference determined by $\mathbf{e}_{\ \mu }^{\mu ^{\prime }}(\tau
,x^{k},t).$ To elaborate realistic cosmological models, we have to prescribe
some locally anisotropic generating values $[\mathbf{e}_{\ \mu }^{\mu
^{\prime }}(\tau ,x^{k},t),\ ^{tot}\widehat{Y}_{\mu \nu }(\tau ,x^{k},t)]$
which are compatible with certain observational data, for instance, in
modern cosmology and dark matter and dark energy physics, see a series of
examples in \cite%
{vacaru18tc,bubuianu18,bubuianu19,vacaru19a,gheorghiu16,amaral17,aschheim18}.

\subsubsection{Quadratic line elements for off-diagonal cosmological
quasiperiodic flows}

We can work with different types of generating functions, for instance, $%
\overline{\Phi }$ and/or $\overline{\Psi }.$ Any coefficient $\overline{h}%
_{3}(\tau )=\overline{h}_{3}(\tau ,x^{k},t)=h_{3}^{[0]}(\tau ,x^{k})-%
\overline{\Phi }^{2}/4\overline{\Lambda }(\tau ),\overline{h}_{3}^{\ast
}\neq 0$ can be considered also as a generating function, for instance, for
locally anisotropic quasiperiodic cosmological configurations. Using the
second formula (\ref{offdcosm}), we express $\overline{\Phi }^{2}=-4%
\overline{\Lambda }(\tau )\overline{h}_{3}(\tau ,{r,\theta },t)$ and, using (%
\ref{nsym1b}), $(\overline{\Psi }^{2})^{\ast }=\int dt\widehat{Y}\overline{h}%
_{3}^{\ast }.$ Introducing such values into the formulas for $\overline{h}%
_{a}$ and $\widehat{Y}$ in (\ref{offdcosm}) and (\ref{nconcosm}), we
construct locally anisotropic cosmological solutions parameterized by \
d--metrics with N-adapted coefficients (\ref{cosmf}), for instance, as a
d-metric (\ref{decomp31}), {\small
\begin{eqnarray}
&&ds^{2}=e^{\ \psi (\tau ,x^{k})}[(dx^{1})^{2}+(dx^{2})^{2}]+
\label{gensolcosm} \\
&&\left\{
\begin{array}{cc}
\begin{array}{c}
\overline{h}_{3}[dy^{3}+(\ _{1}n_{k}+4\ _{2}n_{k}\int dt\frac{(\overline{h}%
_{3}^{\ast }{})^{2}}{|\int dy^{4}\ \widehat{Y}\overline{h}_{3}^{\ast }|\ (%
\overline{h}_{3})^{5/2}})dx^{k}] \\
-\frac{(\overline{h}_{3}^{\ast }{})^{2}}{|\int dt\ \ \widehat{Y}\overline{h}%
_{3}^{\ast }|\ \overline{h}_{3}}[dt+\frac{\partial _{i}(\int dt\ \widehat{Y}%
\ \overline{h}_{3}^{\ast }{})}{\ \ \ \widehat{Y}\overline{h}_{3}^{\ast }\ {}}%
dx^{i}], \\
\mbox{ or }%
\end{array}
&
\begin{array}{c}
\mbox{gener.  funct.}\overline{h}_{3}, \\
\mbox{ source }\ \widehat{Y},\mbox{ or }\overline{\Lambda }(\tau );%
\end{array}
\\
\begin{array}{c}
(h_{3}^{[0]}-\int dt\frac{(\overline{\Psi }^{2})^{\ast }}{4\ \widehat{Y}}%
)[dy^{3}+(_{1}n_{k}+\ _{2}n_{k}\int dt\frac{(\overline{\Psi }^{\ast })^{2}}{%
4\ \widehat{Y}^{2}|h_{3}^{[0]}-\int dy^{4}\frac{(\overline{\Psi }^{2})^{\ast
}}{4\widehat{Y}}|^{5/2}})dx^{k}] \\
-\frac{(\overline{\Psi }^{2})^{\ast }}{4\ \widehat{Y}^{2}(h_{3}^{[0]}-\int dt%
\frac{(\overline{\Psi }^{2})^{\ast }}{4\ \widehat{Y}})}[dt+\frac{\partial
_{i}\ \overline{\Psi }}{\ \overline{\Psi }^{\ast }}dx^{i}], \\
\mbox{ or }%
\end{array}
&
\begin{array}{c}
\mbox{gener.  funct.}\overline{\Psi }, \\
\mbox{source }\ \widehat{Y};%
\end{array}
\\
\begin{array}{c}
(h_{3}^{[0]}-\frac{\overline{\Phi }^{2}}{4\overline{\Lambda }}%
)[dy^{3}+(_{1}n_{k}+\ _{2}n_{k}\int dt\frac{[(\overline{\Phi }^{2})^{\ast
}]^{2}}{|\ 4\overline{\Lambda }\int dy^{4}\ \widehat{Y}(\overline{\Phi }%
^{2})^{\ast }|}|h_{4}^{[0]}(x^{k})-\frac{\overline{\Phi }^{2}}{4\overline{%
\Lambda }}|^{-5/2})dx^{k}] \\
-\frac{[(\overline{\Phi }^{2})^{\ast }]^{2}}{|\ 4\overline{\Lambda }\int
dy^{4}\ \widehat{Y}(\overline{\Phi }^{2})^{\ast }|(h_{3}^{[0]}-\frac{%
\overline{\Phi }^{2}}{4\overline{\Lambda }})}[dt+\frac{\partial _{i}[\int
dt\ \ \widehat{Y}(\overline{\Phi }^{2})^{\ast }]}{\ \ \widehat{Y}(\overline{%
\Phi }^{2})^{\ast }}dx^{i}],%
\end{array}
&
\begin{array}{c}
\mbox{gener.  funct.}\overline{\Phi } \\
\overline{\Lambda }(\tau )\mbox{ for }\ \widehat{Y}.%
\end{array}%
\end{array}%
\right.  \notag
\end{eqnarray}%
} Such solutions posses a Killing symmetry on $\partial _{3}.$ We can
re-write equivalently such linear quadratic elements in terms of $\eta $%
--polarization functions and describe geometric flows of certain target
locally anisotropic cosmological metrics $\ \widehat{\mathbf{g}}=[g_{\alpha
}=\eta _{\alpha }\mathring{g}_{\alpha },\ \eta _{i}^{a}\mathring{N}%
_{i}^{a}]. $ The primary cosmological data $[\mathring{g}_{\alpha },%
\mathring{N}_{i}^{a}]$ are also encoded when the cosmological flow evolution
is determined by generating and integration functions and generating sources.

\subsubsection{Off-diagonal Levi-Civita KK quasiperiodic cosmological
configurations}

From (\ref{gensolcosm}), we can extract and model entropic flow evolution of
cosmological spacetimes in GR encoding both electromagnetic interactions and
nontvrivial quasiperiodic structures under geometric evolution. The zero
torsion conditions are satisfied by a special class of generating functions
and sources when, for instance, $\overline{\Psi }(\tau )=\overline{\check{%
\Psi}}(\tau ,x^{i},t),$ when $(\partial _{i}\overline{\check{\Psi}})^{\ast
}=\partial _{i}(\overline{\check{\Psi}}^{\ast })$ and $\widehat{Y}(\tau
,x^{i},t)=\widehat{Y}[\overline{\check{\Psi}}]=\check{Y}(\tau ),$ or $%
\widehat{Y}=const.$ For such classes of quasiperiodic and flow evolving
generating functions and sources, the nonlinear symmetries (\ref{nsym1b})
are written%
\begin{equation*}
\overline{\Lambda }(\tau )\ \overline{\check{\Psi}}^{2}=\overline{\check{\Phi%
}}^{2}|\check{Y}|-\int dt\ \overline{\check{\Phi}}^{2}|\check{Y}|^{\ast },%
\overline{\check{\Phi}}^{2}=-4\overline{\Lambda }(\tau )\overline{\check{h}%
_{3}}(\tau ,{r,\theta },t),\overline{\check{\Psi}}^{2}=\int dt\ \check{Y}%
(\tau ,{r,\theta },t)\overline{\check{h}}_{3}^{\ast }(\tau ,{r,\theta },t).
\end{equation*}%
We conclude that $\overline{h}_{4}(\tau )=\overline{\check{h}}_{4}(\tau
,x^{i},t)$ can be considered also as generating function for GIF emergent
cosmological solutions. There are generated LC--configurations with some
parametric on $\tau $ functions $\overline{\check{A}}(\tau ,x^{i},t)$ (this
function is not related to an electromagnetic potential) and $n(\tau ,x^{i}),
$ when the N--connection coefficients are computed $\overline{n}_{k}(\tau )=%
\overline{\check{n}}_{k}(\tau )=\partial _{k}\overline{n}(\tau ,x^{i})$ and
$\overline{w}_{i}(\tau )=\partial _{i}\overline{\check{A}}(\tau )=\frac{%
\partial _{i}(\int dt\ \check{Y}\ \overline{\check{h}}_{3}^{\ast }{}{}])}{\
\ \check{Y}\ \overline{\check{h}}_{3}^{\ast }{}}=\frac{\partial _{i}%
\overline{\check{\Psi}}}{\overline{\check{\Psi}}^{\ast }}=\frac{\partial
_{i}[\int dt\ \ \check{Y}(\overline{\check{\Phi}}^{2})^{\ast }]}{\ \check{Y}(%
\overline{\check{\Phi}}^{2})^{\ast }}.$ In result, we construct new classes
of cosmological solutions for the KK theory and GR with electromagnetic
interactions defined as subclasses of solutions (\ref{gensolcosm}) with
quasiperiodic geometric flow evolution, {\small
\begin{eqnarray}
&&ds^{2}=e^{\ \psi (\tau ,x^{k})}[(dx^{1})^{2}+(dx^{2})^{2}]+
\label{lcsolcosm} \\
&&\left\{
\begin{array}{cc}
\begin{array}{c}
\overline{\check{h}}_{3}\left[ dy^{3}+(\partial _{k}\overline{n})dx^{k}%
\right] -\frac{(\ \overline{\check{h}}_{3}^{\ast }{}{}{})^{2}}{|\int dt\ \ \
\ \check{Y}\ \overline{\check{h}}_{3}^{\ast }{}|\ \ \overline{\check{h}}_{3}}%
[dt+(\partial _{i}\overline{\check{A}})dx^{i}], \\
\mbox{ or }%
\end{array}
&
\begin{array}{c}
\mbox{gener.  funct.}\overline{\check{h}}_{3}, \\
\mbox{ source }\ \ \check{Y},\mbox{ or }\overline{\Lambda };%
\end{array}
\\
\begin{array}{c}
(h_{3}^{[0]}-\int dt\frac{(\overline{\check{\Psi}}^{2})^{\ast }}{4\ \check{Y}%
})[dy^{3}+(\partial _{k}\overline{n})dx^{k}]-\frac{(\overline{\check{\Psi}}%
^{2})^{\ast }}{4\check{Y}^{2}(h_{3}^{[0]}-\int dt\frac{(\overline{\check{\Psi%
}}^{2})^{\ast }}{4\ \check{Y}})}[dt+(\partial _{i}\overline{\check{A}}%
)dx^{i}], \\
\mbox{ or }%
\end{array}
&
\begin{array}{c}
\mbox{gener.  funct.}\overline{\check{\Psi}}, \\
\mbox{source }\ \ \check{Y};%
\end{array}
\\
(h_{3}^{[0]}-\frac{\overline{\check{\Phi}}^{2}}{4\overline{\Lambda }}%
)[dy^{3}+(\partial _{k}\overline{n})dx^{k}]-\frac{[(\overline{\check{\Phi}}%
^{2})^{\ast }]^{2}}{|\ 4\overline{\Lambda }\int dt\ \check{Y}(\overline{%
\check{\Phi}}^{2})^{\ast }|(h_{3}^{[0]}-\frac{\overline{\check{\Phi}}^{2}}{4%
\overline{\Lambda }})}[dt+(\partial _{i}\overline{\check{A}})dx^{i}], &
\begin{array}{c}
\mbox{gener.  funct.}\ \overline{\check{\Phi}} \\
\mbox{effective }\overline{\Lambda }\mbox{ for }\ \ \check{Y}.%
\end{array}%
\end{array}%
\right.   \notag
\end{eqnarray}%
}

Such cosmological metrics are generic off-diagonal and define new classes of
solutions if the anholonomy coefficients are not zero for $N_{k}^{3}(\tau
)=\partial _{k}\overline{n}$ and $N_{i}^{4}(\tau )=\partial _{i}\overline{%
\check{A}}.$

\section{Exact solutions for KK quasiperiodic cosmological flows}

\label{s7}Cosmological models with entropic and quasiperiodic flow evolution
of locally anisotropic and inhomogeneous cosmological spacetimes are studied
in details in Refs. \cite{bubuianu18,vacaru19a} (for MGTs and entropic
gravity see respective Table 3 in those works). The geometric formalism and
technical results on exact solutions for KK quasiperiodic cosmological flows
are summarized below in Table 2.

\subsection{The AFDM for generating cosmological flow solutions}

We outline the key steps and results on application of the AFDM for
generating cosmological solutions with geometric flows and Killing symmetry
on $\partial _{3}.$ The nonholonomic deformation procedure is elaborated for
geometric flow evolution of a generating function $g_{3}(\tau )=\overline{h}%
_{3}(\tau ,x^{i},y^{3})$ (\ref{offdcosm}), a cosmological constant $%
\overline{\Lambda }(\tau ),$ and a source $\ \widehat{Y}(\tau )=\ \widehat{Y}%
(\tau ,x^{k},t)$ (\ref{effsourc}). There are constructed exact solutions of
the system of nonlinear PDEs for emergent cosmology (\ref{cosmsimpl}) which
can be typically parameterized in the form
\begin{eqnarray*}
ds^{2} &=&e^{\ \psi (\tau )}[(dx^{1})^{2}+(dx^{2})^{2}]+\overline{h}%
_{3}(\tau )[dy^{3}+(\ _{1}n_{k}+4\ _{2}n_{k}\int dt\frac{(\overline{h}%
_{3}^{\ast }{}(\tau ))^{2}}{|\int dt\ \ \widehat{Y}(\tau )\overline{h}%
_{3}^{\ast }(\tau )|\ (\overline{h}_{3}(\tau ))^{5/2}})dx^{k}] \\
&&-\frac{[\overline{h}_{3}^{\ast }{}(\tau )]^{2}}{|\int dt\ \ \widehat{Y}%
(\tau )(h_{3}{}^{\ast }(\tau ))|\ h_{3}(\tau )}[dt+\frac{\partial _{i}(\int
dt\ \widehat{Y}(\tau )\overline{h}_{3}^{\ast }{}(\tau ))}{\ \widehat{Y}(\tau
)\ \overline{h}_{3}^{\ast }{}(\tau )}dx^{i}],
\end{eqnarray*}
see also other type solutions in Table 2.

%%%%%%%%%%%
% Table 2
%%%%%%%%%%%
%\vskip5pt
%\begin{table*}[t]
{\scriptsize
\begin{eqnarray*}
&&%
\begin{tabular}{l}
\hline\hline
\begin{tabular}{lll}
& {\large \textsf{Table 2:\ Parameterization of quasiperiodic KK
cosmological flows }} &  \\
& Exact solutions of \ $\mathbf{R}_{\mu \nu }(\tau )=\ ^{eff}\yen _{\mu \nu
}(\tau )$ (\ref{entropfloweq}) transformed into a system of nonlinear PDEs (%
\ref{eq1b})-(\ref{eq4b}) &
\end{tabular}
\\
\end{tabular}
\\
&&%
\begin{tabular}{lll}
\hline\hline
$%
\begin{array}{c}
\mbox{d-metric ansatz with} \\
\mbox{Killing symmetry }\partial _{3}=\partial _{\varphi }%
\end{array}%
$ &  & $%
\begin{array}{c}
ds^{2}=g_{i}(\tau )dx^{i})^{2}+g_{a}(\tau )(dy^{a}+N_{i}^{a}(\tau
)dx^{i})^{2},\mbox{ for } \\
g_{i}=e^{\psi {(\tau ,x}^{k}{)}},\,\,\,\,g_{a}=\overline{h}_{a}(\tau ,{x}%
^{k},t),\ N_{i}^{3}=\overline{n}_{i}(\tau ,{x}^{k},t),\,\,\,\,N_{i}^{4}=%
\overline{w}_{i}(\tau ,{x}^{k},t),%
\end{array}%
$ \\
Effective matter sources &  & $\ ^{e}\widehat{Y}_{\ \nu }^{\mu }(\tau )=[~\
_{h}\widehat{Y}(\tau ,{x}^{k})\delta _{j}^{i},\widehat{Y}(\tau ,{x}%
^{k},t)\delta _{b}^{a}];x^{1},x^{2},y^{3},y^{4}=t$ \\ \hline
Nonlinear PDEs (\ref{cosmsimpl}) &  & $%
\begin{array}{c}
\psi ^{\bullet \bullet }+\psi ^{\prime \prime }=2~\ \ _{h}\widehat{Y}(\tau );
\\
\overline{\varpi }^{\ast }\ \overline{h}_{3}^{\ast }=2\overline{h}_{3}%
\overline{h}_{4}\widehat{Y}(\tau ); \\
\overline{n}_{k}^{\ast \ast }+\overline{\gamma }\overline{n}_{k}^{\ast }=0;
\\
\overline{\beta }\overline{w}_{i}-\overline{\alpha }_{i}=0;%
\end{array}%
$ for $%
\begin{array}{c}
\overline{\varpi }{=\ln |\partial _{t}}\overline{{h}}{_{3}/\sqrt{|\overline{h%
}_{3}\overline{h}_{4}|}|,}\overline{\alpha }_{i}=(\partial _{t}\overline{h}%
_{3})\ (\partial _{i}\overline{\varpi }), \\
\ \overline{\beta }=(\partial _{t}\overline{h}_{3})\ (\partial _{t}\overline{%
\varpi }),\ \overline{\gamma }=\partial _{t}\left( \ln |\overline{h}%
_{3}|^{3/2}/|\overline{h}_{4}|\right) , \\
\partial _{1}q=q^{\bullet },\partial _{2}q=q^{\prime },\partial
_{4}q=\partial q/\partial t=q^{\ast }%
\end{array}%
$ \\ \hline
$%
\begin{array}{c}
\mbox{ Generating functions:}\ h_{3}(\tau ,{x}^{k},t), \\
\overline{\Psi }(\tau ,x^{k},t)=e^{\overline{\varpi }},\overline{\Phi }(\tau
,{x}^{k},t); \\
\mbox{integration functions:}\ h_{3}^{[0]}(\tau ,x^{k}),\  \\
_{1}n_{k}(\tau ,x^{i}),\ _{2}n_{k}(\tau ,x^{i}); \\
\mbox{\& nonlinear symmetries}%
\end{array}%
$ &  & $%
\begin{array}{c}
\ (\overline{\Psi }^{2})^{\ast }=-\int dt\ \widehat{Y}\overline{h}_{3}^{\ast
},\overline{\Phi }^{2}=-4\overline{\Lambda }(\tau )\overline{h}_{3},%
\mbox{
see }(\ref{nsym1b}); \\
\overline{h}_{3}(\tau )=\overline{h}_{3}^{[0]}-\overline{\Phi }^{2}/4%
\overline{\Lambda }(\tau ),\overline{h}_{3}^{\ast }\neq 0,\overline{\Lambda }%
(\tau )\neq 0=const \\
\overline{\Lambda }(\ \overline{\Psi }^{2})^{\ast }=|\widehat{Y}|(\overline{%
\Phi }^{2})^{\ast },\mbox{
or  }\overline{\Lambda }\overline{\ \Psi }^{2}=\overline{\Phi }^{2}|\widehat{%
Y}|-\int dt\ \overline{\Phi }^{2}|\widehat{Y}|^{\ast }%
\end{array}%
$ \\ \hline
Off-diag. solutions, $%
\begin{array}{c}
\mbox{d--metric} \\
\mbox{N-connec.}%
\end{array}%
$ &  & $%
\begin{array}{c}
\ g_{i}=e^{\ \psi (\tau ,x^{k})}\mbox{ as a solution of 2-d Poisson eqs. }%
\psi ^{\bullet \bullet }+\psi ^{\prime \prime }=2~\ _{h}\widehat{Y}(\tau );
\\
\overline{h}_{4}(\tau )=-(\overline{\Psi }^{2})^{\ast }/4\widehat{Y}^{2}%
\overline{h}_{3},\mbox{ see }(\ref{offdcosm}); \\
\overline{h}_{3}(\tau )=h_{3}^{[0]}-\int dt(\overline{\Psi }^{2})^{\ast }/4%
\widehat{Y}=h_{3}^{[0]}-\overline{\Phi }^{2}/4\overline{\Lambda }(\tau ); \\
\\
\overline{n}_{k}(\tau )=\ _{1}n_{k}+\ _{2}n_{k}\int dt(\overline{\Psi }%
^{\ast })^{2}/\widehat{Y}^{2}|h_{3}^{[0]}-\int dt(\overline{\Psi }%
^{2})^{\ast }/4\widehat{Y}|^{5/2}; \\
\overline{w}_{i}(\tau )=\partial _{i}\ \overline{\Psi }/\ \partial _{t}%
\overline{\Psi }=\partial _{i}\ \overline{\Psi }^{2}/\ \partial _{t}%
\overline{\Psi }^{2}|.%
\end{array}%
$ \\ \hline
LC-configurations (\ref{lccondit}) &  & $%
\begin{array}{c}
\partial _{t}\overline{w}_{i}=(\partial _{i}-\overline{w}_{i}\partial
_{t})\ln \sqrt{|\overline{h}_{4}|},(\partial _{i}-\overline{w}_{i}\partial
_{4})\ln \sqrt{|\overline{h}_{3}|}=0, \\
\partial _{k}\overline{w}_{i}=\partial _{i}\overline{w}_{k},\partial _{t}%
\overline{n}_{i}=0,\partial _{i}\overline{n}_{k}=\partial _{k}\overline{n}%
_{i}; \\
\Psi =\overline{\check{\Psi}}(\tau ,x^{i},t),(\partial _{i}\overline{\check{%
\Psi}})^{\ast }=\partial _{i}(\overline{\check{\Psi}}^{\ast })\mbox{ and }
\\
\widehat{Y}(\tau ,x^{i},t)=\widehat{Y}[\overline{\check{\Psi}}]=\check{Y},%
\mbox{ or }\widehat{Y}=const.%
\end{array}%
$ \\ \hline
N-connections, zero torsion &  & $%
\begin{array}{c}
\overline{n}_{k}(\tau )=\overline{\check{n}}_{k}(\tau )=\partial _{k}%
\overline{n}(\tau ,x^{i}) \\
\mbox{ and }\overline{w}_{i}(\tau )=\partial _{i}\overline{\check{A}}(\tau
)=\left\{
\begin{array}{c}
\partial _{i}(\int dt\ \overline{\check{\yen }}\ \overline{\check{h}}%
_{3}^{\ast }])/\overline{\check{\yen }}\ \overline{\check{h}}_{3}^{\ast }{};
\\
\partial _{i}\overline{\check{\Psi}}/\overline{\check{\Psi}}^{\ast }; \\
\partial _{i}(\int dt\ \overline{\check{\yen }}(\overline{\check{\Phi}}%
^{2})^{\ast })/(\overline{\check{\Phi}})^{\ast }\overline{\check{\yen }};%
\end{array}%
\right. .%
\end{array}%
$ \\ \hline
$%
\begin{array}{c}
\mbox{polarization functions} \\
\mathbf{\mathring{g}}\rightarrow \overline{\widehat{\mathbf{g}}}\mathbf{=}[%
\overline{g}_{\alpha }=\overline{\eta }_{\alpha }\mathring{g}_{\alpha },\
\overline{\eta }_{i}^{a}\mathring{N}_{i}^{a}]%
\end{array}%
$ &  & $%
\begin{array}{c}
ds^{2}=\overline{\eta }_{i}(\tau ,x^{k},t)\mathring{g}%
_{i}(x^{k},t)[dx^{i}]^{2}+\overline{\eta }_{3}(\tau ,x^{k},t)\mathring{h}%
_{3}(x^{k},t)[dy^{3}+ \\
\overline{\eta }_{i}^{3}(\tau ,x^{k},t)\mathring{N}%
_{i}^{3}(x^{k},t)dx^{i}]^{2}+\overline{\eta }_{4}(\tau ,x^{k},t)\mathring{h}%
_{4}(x^{k},t)[dt+\overline{\eta }_{i}^{4}(\tau ,x^{k},t)\mathring{N}%
_{i}^{4}(x^{k},t)dx^{i}]^{2},%
\end{array}%
$ \\ \hline
$%
\begin{array}{c}
\mbox{ Prime metric defines } \\
\mbox{ a cosmological solution}%
\end{array}%
$ &  & $%
\begin{array}{c}
\lbrack \mathring{g}_{i}(x^{k},t),\mathring{g}_{a}=\mathring{h}_{a}(x^{k},t);%
\mathring{N}_{k}^{3}=\mathring{w}_{k}(x^{k},t),\mathring{N}_{k}^{4}=%
\mathring{n}_{k}(x^{k},t)] \\
\mbox{diagonalizable by frame/ coordinate transforms.}%
\end{array}%
$ \\
$%
\begin{array}{c}
\mbox{Example of a prime } \\
\mbox{ cosmological metric }%
\end{array}%
$ &  & $%
\begin{array}{c}
\mathring{g}_{1}=\frac{a^{2}(t)}{1-kr^{2}},\mathring{g}_{2}=a^{2}(t)r^{2},%
\mathring{h}_{3}=a^{2}(t)r^{2}\sin ^{2}\theta ,\mathring{h}%
_{4}=c^{2}=const,k=\pm 1,0; \\
\mbox{ any frame transform of a FLRW or a Bianchi metrics}%
\end{array}%
$ \\ \hline
Solutions for polarization funct. &  & $%
\begin{array}{c}
\eta _{i}(\tau )=e^{\ \psi (\tau ,x^{k})}/\mathring{g}_{i};\overline{\eta }%
_{4}(\tau )\mathring{h}_{4}=-\frac{4[(|\overline{\eta }_{3}\mathring{h}%
_{3}|^{1/2})^{\ast }]^{2}}{|\int dt\ \widehat{Y}[(\overline{\eta }_{3}%
\mathring{h}_{3})]^{\ast }|\ }; \\
\overline{\eta }_{3}=\overline{\eta }_{3}(\tau ,x^{i},t)%
\mbox{ as a generating
function}; \\
\overline{\eta }_{k}^{3}\ (\tau )\mathring{N}_{k}^{3}=\ _{1}n_{k}+16\ \
_{2}n_{k}\int dt\frac{\left( [(\overline{\eta }_{3}\mathring{h}%
_{3})^{-1/4}]^{\ast }\right) ^{2}}{|\int dt\ \widehat{Y}[(\overline{\eta }%
_{3}\mathring{h}_{3})]^{\ast }|\ };\ \overline{\eta }_{i}^{4}\ (\tau )%
\mathring{N}_{i}^{4}=\frac{\partial _{i}\ \int dt\ \widehat{Y}(\overline{%
\eta }_{3}\mathring{h}_{3})^{\ast }}{\ \widehat{Y}(\overline{\eta }_{3}%
\mathring{h}_{3})^{\ast }},%
\end{array}%
$ \\ \hline
Polariz. funct. with zero torsion &  & $%
\begin{array}{c}
\overline{\eta }_{i}(\tau )=\frac{e^{\ \psi (\tau ,x^{k})}}{\mathring{g}_{i}}%
_{i};\overline{\eta }_{4}(\tau )=-\frac{4[(|\overline{\eta }_{3}\mathring{h}%
_{3}|^{1/2})^{\ast }]^{2}}{\mathring{g}_{4}|\int dt\ \widehat{Y}[(\overline{%
\eta }_{3}\mathring{h}_{3})]^{\ast }|\ };\overline{\eta }_{3}(\tau )=%
\overline{\check{\eta}}_{3}(\tau ,{x}^{i},t) \\
\mbox{  as
a generating function};\overline{\eta }_{k}^{4}(\tau )=\partial _{k}%
\overline{\check{A}}(\tau \,,x^{i},t)/\mathring{w}_{k};\overline{\eta }%
_{k}^{3}(\tau )=(\partial _{k}\overline{n})/\mathring{n}_{k},%
\end{array}%
$ \\ \hline\hline
\end{tabular}%
\end{eqnarray*}%
}

\subsection{Cosmological metrics with entropic quasiperiodic flows}

We analyse two possibilities to transform the geometric flow modified
Einstein equations (\ref{entropfloweq}) into systems of nonlinear PDEs (\ref%
{eq1b})--(\ref{eq4b}) for which generic off-diagonal or diagonal solutions
depending in explicit form on a evolution parameter, a time like variable
and two space like coordinates can be constructed. In the first case, there
are considered entropic quasiperiodic sources determined by some additive or
general nonlinear functionals for effective matter fields. In the second
case, respective nonlinear functionals determining quasiperiodic solutions
for entropic configurations are prescribed for generating functions
subjected to nonlinear symmetries (\ref{nsym1b}). We also note that it is
possible to construct certain classes of locally anisotropic and
inhomogeneous cosmological solutions using nonlinear / additive functionals
both for generating functions and (effective) sources.

\subsubsection{Cosmological flows generated by additive functionals for
effective sources}

We consider $^{ad}\ \widehat{\mathbf{\Upsilon }}(\tau )=\ ^{e}\widehat{Y}%
(\tau ,x^{i},t)$ (\ref{effsourc}) for an additive source of components $\
\widehat{\mathbf{\Upsilon }}_{\mu \nu },\ ^{int}\widehat{\mathbf{\Upsilon }}%
_{\mu \nu },$ and $^{\chi }\widehat{\mathbf{\Upsilon }}_{\mu \nu }$ in (\ref%
{totsourc}) when
\begin{equation}
\ ^{ad}\ \widehat{Y}(\tau )=\ ^{ad}\ \widehat{Y}(\tau ,{x}^{i},t)=\ \widehat{%
Y}(\tau ,{x}^{i},t)+\ \ _{0}^{int}\widehat{Y}(\tau ,{x}^{i},t)+\ _{0}^{\chi }%
\widehat{Y}(\tau ,{x}^{i},t).  \label{adsourccosm}
\end{equation}%
To (\ref{lcsolcosm}), we associate an additive cosmological constant $\ ^{ad}%
\overline{\Lambda }(\tau )=\overline{\Lambda }(\tau )+\ ^{int}\Lambda (\tau
)+\ ^{\chi }\Lambda (\tau )$ corresponding to nonlinear symmetries (\ref%
{nsym1b}) for any component $\ \widehat{\mathbf{\Upsilon }},\ ^{int}\widehat{%
\mathbf{\Upsilon }},$ and $\ ^{\chi }\widehat{\mathbf{\Upsilon }}$ but one
general generating functions. For such conditions, the equation (\ref{eq2b})
transforms into ${\overline{\varpi }}^{\ast }\ \overline{h}_{3}^{\ast }=2%
\overline{h}_{3}\overline{h}_{4}\ \ ^{ad}\ \widehat{Y}(\tau )$ and can be
integrated on $y^{4}=t.$ Following the procedure summarized in Table 2, we
can integrate the system of nonlinear PDEs (\ref{cosmsimpl}) for
cosmological quasiperiodic flows determined by above type effective sources
and nonlinear symmetries. We obtain quadratic line elements {\small
\begin{eqnarray}
&&ds^{2}=e^{\ \psi (\tau ,x^{k})}[(dx^{1})^{2}+(dx^{2})^{2}]+\overline{h}%
_{3}(\tau )[dy^{3}+(\ _{1}n_{k}(\tau ,{x}^{i})+4\ _{2}n_{k}(\tau ,{x}^{i})
\label{cosmasdm} \\
&&\int dt\frac{[\overline{h}_{3}{}^{\ast }(\tau )]^{2}}{|\int dy^{4}\ \ \
^{ad}\overline{\Lambda }(\tau )\overline{h}_{3}{}^{\ast }(\tau )|[\overline{h%
}_{3}(\tau )]^{5/2}})dx^{k}]-\frac{[\overline{h}_{3}{}^{\ast }(\tau )]^{2}}{%
|\int dy^{4}\ \ ^{ad}\overline{\Lambda }(\tau )\overline{h}_{3}{}^{\ast
}(\tau )|\ \overline{h}_{3}(\tau )}[dt+\frac{\partial _{i}(\int dt\ \ ^{ad}%
\overline{\Lambda }(\tau )\ \overline{h}_{3}{}^{\ast }(\tau ))}{\ \ ^{ad}%
\overline{\Lambda }(\tau )\ \overline{h}_{3}{}^{\ast }(\tau )}dx^{i}].
\notag
\end{eqnarray}%
} We have to fix a sign of the coefficient $\overline{h}_{3}(\tau ,x^{k},t)$
which describes relativistic flow evolution with a generating function with
Killing symmetry on $\partial _{3}$ determined by sources $(\ _{h}\widehat{Y}%
(\tau ),\ ^{ad}\ \widehat{Y}(\tau )).$ Such entropic and quasiperiodic flow
solutions are of type (\ref{gensolcosm}) and can be re-written equivalently
with coefficients stated by other type functionals like $\ ^{ad}\overline{%
\Phi }(\tau ,{x}^{i},t)$ and $\ ^{ad}\overline{\Psi }(\tau ,{x}^{i},t).$

Let us discuss how we can extract from off-diagonal d-metrics (\ref{cosmasdm}%
) certain cosmological LC-configurations when there are imposed additional
zero torsion constraints (\ref{lccondit}). Such equations can be considered
as some anholonomy conditions restricting the respective classes of
generating functions $(\overline{\check{h}}_{3}(\tau ,x^{i},t),\overline{%
\check{\Psi}}(\tau ,x^{i},t)$ and/ or $\overline{\check{\Phi}}(\tau
,x^{i},t))$ for $\overline{n}(\tau ,x^{i}),\overline{\check{A}}(\tau ,{x}%
^{i},t))$ and sources $\ ^{as}\widehat{Y}(\tau )=\ \check{Y}(\tau )$ with
additive splitting of type(\ref{adsourccosm}) and $\ ^{as}\overline{\Lambda }%
(\tau )$ (\ref{lcsolcosm}), {\small
\begin{equation}
ds^{2}=e^{\ \psi (\tau )}[(dx^{1})^{2}+(dx^{2})^{2}]+\overline{\check{h}}%
_{3}(\tau )\left[ dy^{3}+(\partial _{k}\overline{n}(\tau ))dx^{k}\right] -%
\frac{[\overline{\check{h}}_{3}{}^{\ast }(\tau )]^{2}}{|\int dt\ ^{as}\check{%
Y}(\tau )\overline{\check{h}}_{3}{}^{\ast }(\tau )|\ \overline{\check{h}}%
_{3}(\tau )}[dt+(\partial _{i}\overline{\check{A}}(\tau ))dx^{i}].
\label{cosmadmlc}
\end{equation}%
}

In this subsection, we constructed two classes of generic flows off-diagonal
cosmological metrics of type (\ref{cosmasdm}) and/or (\ref{cosmadmlc}). Such
solutions define off-diagonal cosmological flows generated by entropic
quasiperiodic additive sources $\ ^{as}\widehat{Y}(\tau )$ and/or $\ ^{as}%
\check{Y}(\tau )$ when terms of type (\ref{adsourccosm}) encode and model
respectively contributions of gravitational electromagnetic fileds with
certain nonholonomically emergent flows for conventional dark matter fields
and effective entropic evolution sources. The type and values of such
generating additive sources can be can be prescribed in some forms which are
compatible with observational data of cosmological (and geometric/entropic)
evolution for dark matter distributions. Such configurations describe the
geometric evolution of possible quasiperiodic, aperiodic, pattern forming,
solitonic nonlinear wave interactions.

\subsubsection{Cosmological evolution for nonlinear entropic quasiperiodic
functionals for sources}

We can generate other classes of considering nonlinear quasiperiodic
functionals for effective sources, $\ ^{qp}\widehat{Y}(\tau )=\quad ^{qp}%
\widehat{Y}(\tau ,x^{i},t)=\ ^{qp}\widehat{Y}[\ \ \widehat{Y},\ \ _{0}^{int}%
\widehat{Y}+\ _{0}^{\chi }\widehat{Y}]$ (instead of additional functional
dependencies in (\ref{adsourccosm})) subjected to nonlinear symmetries (\ref%
{nsym1b}). Following the AFDM, see Table 2, we construct geometric flow
cosmological solutions of with nonlinear sources, {\small
\begin{eqnarray}
&&ds^{2}=e^{\ \psi (\tau )}[(dx^{1})^{2}+(dx^{2})^{2}]+\overline{h}_{3}(\tau
)[dy^{3}+(\ _{1}n_{k}(\tau )+4\ _{2}n_{k}(\tau )\int dt\frac{[\overline{h}%
_{3}{}^{\ast }(\tau )]^{2}}{|\int dt\ \ ^{qp}\widehat{Y}(\tau )\overline{h}%
_{3}{}^{\ast }(\tau )|[\overline{h}_{3}(\tau )]^{5/2}})dx^{k}]  \notag \\
&&-\frac{[\overline{h}_{3}{}^{\ast }(\tau )]^{2}}{|\int dt\ ^{qp}\widehat{Y}%
(\tau )\ \overline{h}_{3}{}^{\ast }(\tau )|\ \overline{h}_{3}(\tau )}[dt+%
\frac{\partial _{i}(\int dt\ \ ^{qp}\widehat{Y}(\tau )\ \overline{h}%
_{3}{}^{\ast }(\tau )])}{\ \ ^{qp}\widehat{Y}(\tau )\ \overline{h}%
_{3}{}^{\ast }(\tau )}dx^{i}].  \label{cosmnfdm}
\end{eqnarray}%
}

The equations (\ref{lccondit}) for extracting LC-configurations are solved
by d-metrics of type {\small
\begin{equation}
ds^{2}=e^{\ \psi (\tau )}[(dx^{1})^{2}+(dx^{2})^{2}]+\overline{\check{h}}%
_{3}(\tau )[dy^{3}+(\partial _{k}\overline{n}(\tau ))dx^{k}]-\frac{[%
\overline{\check{h}}_{3}{}^{\ast }(\tau )]^{2}}{|\int dt\ {\ ^{qp}}\check{Y}%
(\tau )\overline{h}_{3}{}^{\ast }(\tau )|\ \overline{\check{h}}_{3}(\tau )}%
[dt+(\partial _{i}\overline{\check{A}}(\tau ))dx^{i}].  \label{cosmnfdmlcn}
\end{equation}%
} For additive functionals for cosmological entropic and quasiperiodic
sources, the formulas (\ref{cosmnfdm}) and (\ref{cosmnfdmlcn}) transforms
respectively into quadratic linear elements (\ref{cosmasdm}) and (\ref%
{cosmadmlc}).

Finally, we note that additive and nonlinear functionals can be considered
for generating functions%
\begin{equation*}
\ ^{a}\overline{\Phi }(\tau )=\ ^{a}\overline{\Phi }(\tau ,{x}^{i},t)=\
\overline{\Phi }(\tau ,{x}^{i},t)+\ \ _{0}^{int}\overline{\Phi }(\tau ,{x}%
^{i},t)+\ _{0}^{\chi }\overline{\Phi }(\tau ,{x}^{i},t)\mbox{ or }^{qp}%
\overline{\Phi }(\tau )=\ \ ^{qp}\overline{\Phi }(\tau ,{x}^{i},t)=\ ^{qp}%
\overline{\Phi }[\overline{\Phi },\ \ _{0}^{int}\overline{\Phi },\
_{0}^{\chi }\overline{\Phi }],
\end{equation*}%
when the generation functions for sources are arbitrary ones or certain
additive/ nonlinear functionals. A number of such examples and respective
cosmological inflation and accelerating cosmology scenarios were elaborated
in our previous works \cite{bubuianu18,vacaru19a,ruchin13,vacaru09,
vacaru18tc,amaral17,aschheim18,rajpoot17,gheorghiu16} for other types of
Lagrangians and effective sources which are different from (\ref{lagrs}) and
(\ref{totsourc}). All such locally anisotropic cosmological metrics in MGTs,
geometric flow and GIF theories can not be characterised by
Bekenstein-Hawking or holographic type entropies. In \cite%
{ruchin13,gheorghiu16,bubuianu19,vacaru19b,vacaru19c,vacaru19d,vacaru19}, we
concluded that such classes of generalized cosmological and other type
physically important solutions can be characterized by respective
nonholonomic modifications of G. Perelman's W-entropy and associated
thermodynamical models.

\subsection{Computing W-entropy and cosmological geometric flow
thermodynamic values}

In subsection, we show how G. Pereman's W-entropy and associated
thermodynamic models with extensions to QGIFs of KK systems with
entanglement elaborated in sections \ref{s3}-\ref{s5} can be applied in
modern accelerating cosmology with quasiperiodic distributions modeling dark
energy and dark matter fields. For simplicity, we shall analyze an example
of cosmological flows determined by $\mathbf{q}_{3}(\tau )=\overline{h}%
_{3}(\tau ,x^{k},t)$ and $\ ^{ad}\overline{\Lambda }(\tau ).$ Similar
constructions can be performed for any type of cosmological solutions
parameterized in Table 2.

\subsubsection{An example of geometric thermodynamic model for cosmological
geometric flows}

Let us consider in explicit form how a GIF model can be elaborated for a
cosmological KK system for 5-d metrics $\mathbf{g}_{\underline{\alpha }%
\underline{\beta }}(\tau )=(\mathbf{g}_{\alpha \beta }(\tau ),\mathbf{A}%
_{\gamma }(\tau ))$ defined as a solution of generalized R. Hamilton
equations in canonical variables (\ref{canhamiltevol}) for an arbitrary
normalization function $\ \widehat{f}(\tau ).$ Such relativistic geometric
flow equations can be re-written equivalently in the form (\ref%
{cankkhamiltevol}). We shall work in N-adapted frames when the d-metrics are
parameterized in the form
\begin{equation}
\mathbf{g}_{\underline{\alpha }\underline{\beta }}(\tau )=diag[q_{i}(\tau ),%
\mathbf{q}_{3}(\tau ),\mathbf{g}_{4}(\tau )=-[\ _{q}N(\tau )]^{2},\mathbf{g}%
_{5}=1],  \label{solcosmex1}
\end{equation}%
when the 4-d component of such a d-metric is of type (\ref{decomp31}) and
determined by a family of locally anisotropic cosmological solutions of type
(\ref{cosmasdm}), i.e. by certain data for d-metric coefficients {\small
\begin{equation*}
q_{1}(\tau ) = q_{2}(\tau )=e^{\ \psi (\tau ,x^{k})},\mathbf{q}_{3}(\tau )=%
\overline{h}_{3}(\tau ,x^{k},t)\mbox{ is a generating function }; \lbrack \
_{q}N(\tau )]^{2} = \frac{[\overline{h}_{3}{}^{\ast }(\tau )]^{2}}{|\int
dy^{4}\ \ ^{ad}\overline{\Lambda }(\tau )\overline{h}_{3}{}^{\ast }(\tau )|\
\overline{h}_{3}(\tau )},
\end{equation*}%
} where $\psi $ is a solution of parametric 2-d Poisson equation with source
$\ _{h}\widehat{Y},$ see also formulas for nonlinear symmetries (\ref{nsym1b}%
), and N-connection coefficients
\begin{equation*}
n_{k}(\tau ) = \ _{1}n_{k}(\tau ,{x}^{i})+4\ _{2}n_{k}(\tau ,{x}^{i})\int dt%
\frac{[\overline{h}_{3}{}^{\ast }(\tau )]^{2}}{|\int dy^{4}\ \ \ ^{ad}%
\overline{\Lambda }(\tau )\overline{h}_{3}{}^{\ast }(\tau )|[\overline{h}%
_{3}(\tau )]^{5/2}}, w_{k}(\tau ) = \frac{\partial _{i}(\int dt\ \ ^{ad}%
\overline{\Lambda }(\tau )\ \overline{h}_{3}{}^{\ast }(\tau ))}{\ \ ^{ad}%
\overline{\Lambda }(\tau )\ \overline{h}_{3}{}^{\ast }(\tau )},
\end{equation*}%
where contributions from electromagnetic fields, possible quasiperiodic
configurations, emergent gravity effects are encoded via nonlinear
symmetries into $\ ^{ad}\overline{\Lambda }(\tau )$ and data for generating
sources $(\ ~\ _{h}\widehat{Y}(\tau ),\ \ ^{ad}\ \widehat{Y}(\tau )).$ Such
data determine cosmological flow solutions of type (\ref{gensolcosm}) and
can be re-written equivalently with coefficients stated by other type
functionals of the form $\ ^{ad}\overline{\Phi }(\tau ,{x}^{i},t)$ and $\
^{ad}\overline{\Psi }(\tau ,{x}^{i},t).$

The W-entropy (\ref{wfkk}) can be written in terms of of canonical
d-connection in 4-d, $\widehat{\mathbf{D}},$ for a corresponding class of
normalizing functions $\underline{f}=\widehat{f}(u)f(u^{5}),$
\begin{eqnarray*}
\underline{\mathcal{W}} &=&\int \underline{\mu }\sqrt{|\underline{\mathbf{g}}%
|}d^{5}\underline{u}[\tau (\ _{s}\underline{R}+|\ \ _{h}\underline{\mathbf{D}%
}\ \underline{f}|+|\ \ _{v}\underline{\mathbf{D}}\ \underline{f}|)^{2}+%
\underline{f}-5] \\
&=&\int \widehat{\mu }\sqrt{|\widehat{\mathbf{g}}|}\delta ^{4}udu^{5}[\tau
(\ _{s}\widehat{R}+|\ f(u^{5})\ _{h}\widehat{\mathbf{D}}\ \widehat{f}|+|\ f\
(u^{5})_{v}\widehat{\mathbf{D}}\widehat{\underline{f}}|)^{2}+\widehat{f}%
f(u^{5})-5].
\end{eqnarray*}%
The normalizing functions satisfy the condition $\ \int \underline{\mu }%
\sqrt{|\underline{\mathbf{g}}|}d^{5}\underline{u}=\int_{t_{1}}^{t_{2}}\int_{%
\Xi _{t}}\int_{u_{1}^{5}}^{u_{2}^{5}}\underline{\mu }\sqrt{|\underline{%
\mathbf{g}}|}d^{5}\underline{u}=1$, where $\ \underline{\mu }=\left( 4\pi
\tau \right) ^{-5/2}e^{-\underline{f}}=\ \widehat{\mu }e^{-f(u^{5})},\
\widehat{\mu }=\left( 4\pi \tau \right) ^{-5/2}e^{-\widehat{f}}.$ Such a
function determine the geometric normalization for a nonlinear diffusion
process on a temperature like parameter $\tau $ when relativistic dynamics
for any fixed $\tau _{0}$ is a cosmological KK type. A parametrization $%
\underline{f}=\widehat{f}(u)f(u^{5})$ can be chosen for respective
cosmological models if there are considered certain experimental data. Here
we note that the integration volume is of type
\begin{equation}
d^{5}\mathcal{V}ol[\underline{\mathbf{g}}(\tau )]:=\sqrt{|\widehat{\mathbf{g}%
}|}\delta ^{4}u=e^{\ \psi (\tau ,x^{k})}\sqrt{|\overline{h}_{3}|}\
_{q}N(\tau )dx^{1}dx^{2}[dy^{3}+n_{k}(\tau )dx^{k}][dt+w_{k}(\tau )dx^{2}]
\label{volfex}
\end{equation}%
for the functions $\psi ,\overline{h}_{3},\ _{q}N$ and $n_{k},w_{k}$ (with $%
k=1,2)$ are determined by the coefficients in (\ref{cosmasdm}).

The thermodynamic generating function (\ref{genfkk}) for the above
assumptions on this class of cosmological flow solutions is computed
\begin{equation}
\underline{\widehat{\mathcal{Z}}}[\underline{\mathbf{g}}(\tau )]=\int \left(
4\pi \tau \right) ^{-5/2}e^{-\widehat{f}}e^{-f(u^{5})}d^{5}\mathcal{V}ol[%
\widehat{f}(u)f(u^{5})+5/2],\mbox{ for }\underline{\mathbf{V}}\mathbf{,}
\label{zfunct}
\end{equation}%
In this subsection, we underline symbols in order to emphasize their KK
geometric nature and put "hats" in order to emphasize that the solutions and
computations are performed for the canonical d-connection. For such
d-metrics,  $_{s}\underline{R}=\ _{s}\widehat{R}=4\ \ ^{ad}\overline{\Lambda
}(\tau ),$ and $\underline{\mathbf{R}}_{\alpha \beta }=\ ^{ad}\overline{%
\Lambda }(\tau )\underline{\mathbf{g}}_{\alpha \beta },$ where $\ _{h}%
\widehat{Y}=\ _{h}\Lambda (\tau )=\ ^{ad}\overline{\Lambda }(\tau )$ is
approximated to a running flow constant $\tau $ under assumption for the
same normalization function $\underline{f}$ \thinspace\ (for simplicity,
chosen to satisfy the conditions $\underline{\mathbf{D}}_{\beta }\underline{f%
}=\widehat{\mathbf{D}}_{\beta }\underline{f}=0)$ on cosmological flows on
all directions on $\mathbf{V}.$ In result, we can compute the thermodynamic
values for geometric/cosmological evolution flows of KK systems (\ref%
{thvalkk}) determined by the class of cosmological solutions{\small \ }%
\begin{eqnarray}
\underline{\widehat{\mathcal{E}}}\  &=&-\tau ^{2}\int \left( 4\pi \tau
\right) ^{-5/2}e^{-\widehat{f}}e^{-f(u^{5})}d^{5}\mathcal{V}ol(\ _{s}%
\underline{R}+|\underline{\mathbf{D}}\underline{f}|^{2}\mathbf{\ }-\frac{5}{%
2\tau })=  \label{auxener} \\
&&-\tau ^{2}\int \left( 4\pi \tau \right) ^{-5/2}e^{-\widehat{f}%
}e^{-f(u^{5})}d^{5}\mathcal{V}ol[4\ \ ^{ad}\overline{\Lambda }(\tau )-\frac{5%
}{2\tau }],  \notag \\
\underline{\widehat{\mathcal{S}}}\  &=&-\int \left( 4\pi \tau \right)
^{-5/2}e^{-\widehat{f}}e^{-f(u^{5})}d^{5}\mathcal{V}ol\left[ \tau \left( \
_{s}\underline{R}+|\underline{\mathbf{D}}\underline{f}|^{2}\right) +%
\underline{f}-5\right] =  \label{auxentr} \\
&&-\int \left( 4\pi \tau \right) ^{-5/2}e^{-\widehat{f}}e^{-f(u^{5})}d^{5}%
\mathcal{V}ol[4\ \ ^{ad}\overline{\Lambda }(\tau )+\widehat{f}f(u^{5})-5],
\notag \\
\underline{\widehat{\eta }}\  &=&2\tau ^{4}\int \left( 4\pi \tau \right)
^{-5/2}e^{-\widehat{f}}e^{-f(u^{5})}d^{5}\mathcal{V}ol[|\ \underline{\mathbf{%
R}}_{\alpha \beta }+\underline{\mathbf{D}}_{\alpha }\ \underline{\mathbf{D}}%
_{\beta }\underline{f}-\frac{1}{2\tau }\underline{\mathbf{g}}_{\alpha \beta
}|^{2}]=  \notag \\
&&2\tau ^{4}\int \left( 4\pi \tau \right) ^{-5/2}e^{-\widehat{f}%
}e^{-f(u^{5})}d^{5}\mathcal{V}ol[(\ ^{ad}\overline{\Lambda }(\tau )-\frac{1}{%
2\tau })\underline{\underline{\mathbf{g}}}_{\alpha \beta }],  \notag
\end{eqnarray}%
where $\underline{\mathbf{g}}_{\alpha \beta }$ is a d-metric (\ref%
{solcosmex1}). Using such values, we can compute  the free energy and
relative entropy (\ref{relentr}),%
\begin{eqnarray*}
\underline{\widehat{\mathcal{F}}}\ (\ _{1}\underline{\mathbf{g}}) &=&%
\underline{\widehat{\mathcal{E}}}(\ _{1}\underline{\mathbf{g}})-\beta ^{-1}%
\underline{\widehat{\mathcal{S}}}(\ _{1}\underline{\mathbf{g}})\mbox{ and }%
\underline{\widehat{\mathcal{S}}}(\ _{1}\underline{\mathbf{g}}\shortparallel
\underline{\mathbf{g}})=\tau ^{-1}[\underline{\widehat{\mathcal{F}}}(\ _{1}%
\underline{\mathbf{g}})-\underline{\widehat{\mathcal{F}}}(\underline{\mathbf{%
g}})],\mbox{ where } \\
\underline{\mathcal{E}}(\ _{1}\underline{\mathbf{g}}) &=&-\tau ^{2}\int
\left( 4\pi \tau \right) ^{-5/2}e^{-\widehat{f}}e^{-f(u^{5})}d^{5}\mathcal{V}%
ol_{1}[4\ \ _{1}^{ad}\overline{\Lambda }(\tau )-\frac{5}{2\tau }]\mbox{ and }
\\
\underline{\mathcal{S}}(\ _{1}\underline{\mathbf{g}}) &=&-\int \left( 4\pi
\tau \right) ^{-5/2}e^{-\widehat{f}_{1}}e^{-f_{1}(u^{5})}d^{5}\mathcal{V}%
ol_{1}[4\ \ ^{ad}\overline{\Lambda }_{1}(\tau )+\widehat{f}%
_{1}f_{1}(u^{5})-5]
\end{eqnarray*}%
are computed for respective functionals $d^{5}\mathcal{V}ol[\ _{1}\underline{%
\mathbf{g}}(\tau ),\tau ,\ \underline{f}_{1}]$ and $d^{5}\mathcal{V}ol[%
\underline{\mathbf{g}}(\tau ),\tau ,\underline{f}],$ with respective
effective runnning cosmological constants, $\ \ _{1}^{ad}\overline{\Lambda }%
(\tau )$ and $\ \ ^{ad}\overline{\Lambda }(\tau ),$ and different
normailizing functions, $\ \underline{f}_{1}$ and $\underline{f}.$

Conventionally, such types of thermodynamic systems for cosmological flow
evolution are of type $\underline{\mathcal{A}}=\left[ \underline{\mathcal{Z}}%
,\ \underline{\mathcal{E}},\underline{\mathcal{S}},\underline{\eta }\right] $
(\ref{thvalkk}) when the geometric thermodynamic data are determined by (\ref%
{zfunct}) \ and respective (\ref{auxener}) and (\ref{auxentr}), i.e. $%
\underline{\widehat{\mathcal{A}}}=\left[ \underline{\widehat{\mathcal{Z}}},\
\underline{\widehat{\mathcal{E}}},\underline{\widehat{\mathcal{S}}},%
\underline{\widehat{\eta }}\right] ,$ when all integrals are computed for
volume forms of type (\ref{volfex}).

\subsubsection{An example of QGIF for a locally anisotropic cosmological
solution}

We can consider a conventional state density (\ref{statedens}) as a density
matric for quantum models using formula $\ \underline{\sigma }(\tau ,%
\underline{E})=\underline{Z}^{-1}e^{-\tau ^{-1}\underline{E}}.$ In this
subsection, we take $\underline{Z}=\underline{\widehat{\mathcal{Z}}}[%
\underline{\mathbf{g}}(\tau )]$ (\ref{zfunct}) following our previous
constructions are possible for the KK QGIF theory with a thermodynamic
generating function $\underline{\mathcal{Z}}\ [\underline{\mathbf{g}}(\tau
)] $ (\ref{genfkk}) and related statistical density $\underline{\rho }(\beta
,\underline{\mathcal{E}}\ ,\underline{\mathbf{g}})$ (\ref{statedens}) (such
values be used for defining $\ \underline{\sigma }_{\mathcal{A}}$ (\ref%
{aux01}) \ as a probability distribution). \

For the class of cosmological flows with solutions (\ref{cosmasdm}), we can
associate a quantum system $\underline{\mathcal{A}}$ when the density matrix%
\begin{equation}
\ \underline{\widehat{\rho }}_{\mathcal{A}}:=\underline{\widehat{\mathcal{Z}}%
}^{-1}e^{-\tau ^{-1}\underline{\widehat{\mathcal{E}}}\ }  \label{statdensex}
\end{equation}
is determined by values $\underline{\widehat{\mathcal{Z}}}[\underline{%
\mathbf{g}}(\tau )]$ (\ref{zfunct}) and $\underline{\widehat{\mathcal{E}}}[%
\underline{\mathbf{g}}(\tau )]$ (\ref{auxener}) (fixing normalization and
intergation functions and constants, we can approximate $\underline{\widehat{%
\mathcal{E}}}[\underline{\mathbf{g}}(\tau )]$ $\approx \underline{E}=const$%
). In result, we can compute the entanglement entropy (\ref{entangentr}) for
such a class of cosmological solutions
\begin{equation}
\ _{q}\underline{\widehat{\mathcal{S}}}(\ \underline{\widehat{\rho }}_{%
\mathcal{A}}):=Tr(\ \underline{\widehat{\rho }}_{\mathcal{A}}\ \log
\underline{\widehat{\rho }}_{\mathcal{A}}),  \label{entangentrex}
\end{equation}%
when $\ \underline{\rho }_{\mathcal{A}}$ is computed using formulas (\ref%
{statdensex}). This entanglement entropy $\ _{q}\underline{\widehat{\mathcal{%
S}}}(\ \underline{\widehat{\rho }}_{\mathcal{A}})$ is a QGIF version of the
G. Perelman thermodynamic entropy $\underline{\widehat{\mathcal{S}}}$ (\ref%
{auxentr}).

This class of cosmological flows are characterized also by relative KK GIF
entropy and monotonocity properties, see formulas (\ref{3relz}) and (\ref%
{3rels}). They involve the thermodynamic generating function (generalization
of (\ref{genfkk})),  $\ _{ABC}\widehat{\mathcal{Z}}[\underline{\mathbf{g}}%
(\tau ),\ _{1}\underline{\mathbf{g}}(\tau ),\ _{2}\underline{\mathbf{g}}%
(\tau )]=\int \ \ _{1}\int \ _{2}\int (4\pi \tau )^{-15/2}e^{-\ _{ABC}%
\underline{f}}d^{5}\mathcal{V}ol[\underline{\mathbf{g}}(\tau )]d^{5}\mathcal{%
V}ol[\ _{1}\underline{\mathbf{g}}(\tau )]d^{5}\mathcal{V}ol[\ _{2}\underline{%
\mathbf{g}}(\tau )](-\ _{ABC}\underline{f}+15),$ for $\mathbf{\underline{%
\mathbf{V}}\otimes \underline{\mathbf{V}}\otimes \ \underline{\mathbf{V}},}$
with volume forms of type (\ref{aux01}) with a normalizing function $\ _{ABC}%
\underline{f}(\ \underline{u},\ _{1}\underline{u},\ _{2}\underline{u}).$ On
such tensor products of KK cosmological manifolds, their flow evolution is
of type $\ _{ABC}\underline{\mathbf{g}}=\{\underline{\mathbf{g}}%
=[q_{1},q_{2},\mathbf{q}_{3}=\overline{h}_{3}(\tau ,x^{k},t),_{q}N,q_{5}=1],$
$_{1}\underline{\mathbf{g}}=[\ _{1}q_{1},\ _{1}q_{2},\ _{1}\mathbf{q}_{3}=\
_{1}\overline{h}_{3}(\tau ,x^{k},t),_{1q}N,\ _{1}q_{5}=1],$ $\ _{2}%
\underline{\mathbf{g}}=[\ _{2}q_{1},\ _{2}q_{2},\ _{2}\mathbf{q}_{3}=\ _{2}%
\overline{h}_{3}(\tau ,x^{k},t),_{2q}N,\ _{2}q_{5}=1]\},\ $where generating
functions $\overline{h}_{3}(\tau ),\ _{1}\overline{h}_{3}(\tau )$ and $\ _{2}%
\overline{h}_{3}(\tau ).$ For respective tensor products, we can consider a
canonical d--connection $\ _{ABC}\underline{\widehat{\mathbf{D}}}=\
\underline{\widehat{\mathbf{D}}}+\ _{B}\ \underline{\widehat{\mathbf{D}}}+\
_{C}\ \underline{\widehat{\mathbf{D}}}$ and respective scalar curvature $\
_{sABC}\underline{\widehat{R}}=\ _{s}\underline{\widehat{R}}+\ _{s1}%
\underline{\widehat{R}}+\ _{s2}\underline{\widehat{R}}.$ The resulting
entropy function for such a three partite system can be computed for similar
assumption as we introduced for (\ref{auxentr}) (on any term of the tensor
product)
\begin{eqnarray*}
\ _{ABC}\underline{\widehat{\mathcal{S}}}=\ \underline{\mathcal{S}}[%
\underline{\widehat{A}},\underline{\widehat{B}},\underline{\widehat{C}}]
&=&-\int \ \ _{1}\int \ _{2}\int (4\pi \tau )^{-15/2}e^{-\ _{ABC}\underline{f%
}}d^{5}\mathcal{V}ol[\underline{\mathbf{g}}(\tau ),\tau ]d^{5}\mathcal{V}%
ol[\ _{1}\underline{\mathbf{g}}(\tau ),\tau ]d^{5}\mathcal{V}ol[\ _{2}%
\underline{\mathbf{g}}(\tau ),\tau ] \\
&&\left[ \tau \left( \ _{s}\underline{\widehat{R}}+\ _{s1}\underline{%
\widehat{R}}+\ _{s2}\underline{\widehat{R}}+|\underline{\mathbf{D}}\ _{ABC}%
\underline{f}|^{2}\right) +\ _{ABC}\underline{f}-15\right] ,\mbox{ where } \\
\left[ \tau \left( \ ...\right) +\ _{ABC}\underline{f}-15\right]  &\approx
&[4\tau (\ \ ^{ad}\overline{\Lambda }(\tau )+\ \ ^{ad}\overline{\Lambda }%
_{1}(\tau )+\ \ ^{ad}\overline{\Lambda }_{2}(\tau ))+\ _{ABC}\underline{f}%
-15].
\end{eqnarray*}

We can compute the R\'{e}nyi entropy (\ref{renentr}) using the replica
method with $\ \underline{\widehat{\rho }}_{\mathcal{A}}$ (\ref{statdensex}%
). For an integer replica parameter $r$ (replica parameter) the R\'{e}nyi
entropy for the mentioned class of cosmological flow solutions (\ref%
{cankkhamiltevol}),
\begin{equation}
\ _{r}\ \underline{\widehat{\mathcal{S}}}(\underline{\widehat{\mathcal{A}}}%
):=\frac{1}{1-r}\log [tr_{\mathcal{A}}(\underline{\widehat{\rho }}_{\mathcal{%
A}})^{r}]  \label{renentrex}
\end{equation}%
for a KK QGIF\ system determined by the matrix $\underline{\widehat{\rho }}_{%
\mathcal{A}}$ and associated thermodynamic model $\underline{\widehat{%
\mathcal{A}}}=\left[ \underline{\widehat{\mathcal{Z}}},\ \underline{\widehat{%
\mathcal{E}}},\underline{\widehat{\mathcal{S}}},\underline{\widehat{\eta }}%
\right] .$ We can consider the same computational formalism elaborated for
an analytic continuation of $r$ to a real number with a well defined limit $%
\ _{q}\underline{\widehat{\mathcal{S}}}(\widehat{\underline{\rho }}_{%
\mathcal{A}})=\lim_{r\rightarrow 1}\ \ _{r}\ \underline{\widehat{\mathcal{S}}%
}(\underline{\widehat{\mathcal{A}}})$ and normalization $tr_{\mathcal{A}}(%
\underline{\widehat{\rho }}_{\mathcal{A}})$ for $r\rightarrow 1.$ For such
limits and cosmological flow d-metrices, the R\'{e}nyi entropy (\ref%
{renentrex}) reduces to the entanglement entropy (\ref{entangentrex}).

Finally, we note that the volume forms (\ref{volfex}) can be considered for
certain prime metrics (in particular, for the Minkowski 4-d and extended to
5-d metric) when all classical and quantum GIF integrals are well defined.
In result, we can always define a self-consistent QGIF model with
quasi-classical limits for which the GIF theory can be with classical
relativistic and nonlinear evolution. All above considered classical and
quantum thermodynamic values depend on respective classes of normalizing and
integration and generating functions. Explicit numerical values can be
obtained by fixing such functions in order to reproduce certain
observational cosmological data or to model them with a corresponding
quantum computing calculus.

\section{Summary and conclusions}

\label{s8} In this paper, we develop our research on emergent gravity and
matter field interactions and entanglement in the framework of the theory of
quantum geometric information flows (QGIFs). It is a partner in a series of
works \cite{bubuianu19,vacaru19b,vacaru19c,vacaru19d,vacaru19,vacaru19a}) on
geometric evolution theories of classical and quantum relativistic
mechanical systems and modified gravity theories, MGTs. We show how
(modified) Einstein-Maxwell, EM, and Kaluza--Klein, KK, theories can be
derived using a generalized Ricci flow formalism. It should be emphasized
that the KK geometric formulation in canonical nonholonomic and analogous
Hamilton variables allows new applications in geometric information flow,
GIF, theories and for elaborating quantum mechanical models.

We first introduce the geometric flow equations for EM and KK systems and
show how such constructions can be performed in nonholonomic, canonical and
analogous (mechanical like) Lagrange-Hamilton variables. Here, we note that
the KK gravity models can be derived as respective classes of nonholonomic
Ricci soliton configurations. This is consistent with our former results on
relativistic geometric flows and (non) commutative and/or supersymmetric
MGTs \cite{ruchin13,bubuianu19,gheorghiu16,rajpoot17}. The geometric flow
and (quantum) information methods are motivated also by a renewed interest
in modern literature on Ricci Yang-Mill gradient flows and gravity \cite%
{narayanan06,luescher09,weisz11,carosso18,bergner18,streets09}.

We then postulate certain classes of nonholonomic deformations of G.
Perelman F- and W-functionals \cite{perelman1} which allow us to prove the
geometric flow equations for KK theories (i.e. generalizations of R.
Hamilton and D. Friedan equations originally proposed for Riemannian metrics
and Levi-Civitat connectons in \cite{friedan2,hamilt1}). It has been
observed earlier \cite{vacaru09,bubuianu19,vacaru19a} and supported by new
findings in this work that the concept of W-entropy is more general than
those considered for solutions with area-horizon and holographic type
configurations using the standard approach involving the Bekenstein--Hawking
entropy and thermodynamics \cite%
{bekenstein72,bekenstein73,bardeen73,hawking75}. We elaborate on statistical
and geometric thermodynamic models for KK flows using covariant variables
and Hamilton variables for Perelman KK--functionals. This allows us to
develop certain alternative approaches to geometric flows and MGTs which are
different from the last two decades results on emergent gravity,
entanglement and information theory \cite%
{ryu06,raamsdonk10,faulkner14,swingle12,jacobson15,pastawski15,
solodukhin11,verlinde10,verlinde16}.

In this article, we formulated also an approach to the theory of classical
GIFs and QGIFs of KK systems. The standard concepts and methods of
information theory and quantum physics and gravity \cite%
{cover,nielsen,nielsen10,weedbrook11,hayashi17,watrous18,preskill,
witten18,aolita14,nishioka18,bao19} are generalized for the Shannon/von
Neumann/conditional/relative entropy determined for thermodynamic generating
functions and density matrices encoding KK GIFs and Perelman W-entropy. The
concept of entanglement and main properties (inequalities) are formulated
and studied for new classes of theories of QGIFs for KK systems.

We proved that the geometric flow equations for KK cosmological systems can
be decoupled and integrated in general forms for generating functions and
sources with quasiperiodic structure. There are studied various examples of
cosmological evolution flows with additive and nonlinear functionals for
effective quasiperiodic electromagnetic and other type sources. Such new
classes of cosmological solutions are characterized by nonlinear symmetries
but can not be described using the concept of Bekenstein--Hawking entropy.
We show how cosmological KK flows can be characterized by Perelman's
W-entropy and associated thermodynamic potentials and provide explicit
examples for computing such values and constructing cosmological QGIFs.

Finally, we note that further developments of our approach will involve
geometric information flows of noncommutative and nonassociative gravity and
quantum field theories involving generalized classes of exact solutions in
MGTs and quantum gravity models, see certain preliminary results in \cite%
{vacaru2000,vacaru09,vacaru18tc,bubuianu18}.

\vskip3pt

\textbf{Acknowledgments:} This research develops former programs partially
supported by IDEI, PN-II-ID-PCE-2011-3-0256, CERN and DAAD and extended to
collaborations with California State University at Fresno, the USA, and Yu.
Fedkovych Chernivtsi National University, Ukraine.

\end{document}